\documentclass[draft]{agujournal2019}
\usepackage{url} 
\usepackage{lineno}
\usepackage{amsmath}
\usepackage{float}
\usepackage[inline]{trackchanges} 
\usepackage{soul}

\draftfalse

\journalname{JGR: Planets}

\begin{document}

\title{Observed Latitudinal, Longitudinal and Temporal Variability of Io's Atmosphere Simulated by a Purely Sublimation Driven Atmosphere}

%
%




\authors{A.-C. Dott\affil{1}, J. Saur\affil{1}, S. Schlegel\affil{1}, D.F. Strobel\affil{2}, K. de Kleer\affil{3}, I. de Pater \affil{4}}


\affiliation{1}{Institute of Geophysics and Meteorology, University of Cologne, Cologne, Germany}
\affiliation{2}{Department of Earth and Planetary Sciences, The Johns Hopkins University, Baltimore, MD, USA}
\affiliation{3}{California Institute of Technology, Pasadena, CA, USA}
\affiliation{4}{University of California, Berkeley, CA, USA}




\correspondingauthor{Anne-Cathrine Dott}{anne-cathrine.dott@uni-koeln.de}



\begin{keypoints}
\item We model Io's surface temperature including thermal inertia and solar illumination based on the exact celestial geometry
\item A purely solar illumination driven atmosphere can reproduce, nearly completely, the observed global-scale characteristics of Io's atmosphere
\item Io's inclination causes hemispheric seasons with northern summer in perihelion and northern winter in aphelion
\end{keypoints}

%
%

%
%


\begin{abstract}
How much of Io's SO$_2$ atmosphere is driven by volcanic outgasing or sublimation of SO$_2$ surface frost is a question with a considerable history. We develop a time dependent surface temperature model including thermal inertia and the exact celestial geometry to model the radiation driven global structure and temporal evolution of Io's atmosphere.
We show that many observations can be explained by assuming a purely sublimation driven atmosphere. We find that a thermal diffusivity $\alpha=2.41\times10^{-7}$ m$^2$s$^{-1}$ yields an averaged atmospheric SO$_2$ column density decreasing by more than one order of magnitude from the equator to the poles in accordance with the observed spatial variations of Io's column densities. Our model produces a strong day-night-asymmetry with modeled column density variations of almost two orders of magnitude at the equator as well as a sub-anti-Jovian hemisphere asymmetry, with maximum dayside column densities of $3.7\times10^{16}$ cm$^{-2}$ for the sub-Jovian and $8.5\times10^{16}$ cm$^{-2}$ for the anti-Jovian hemisphere. Both are consistent with the observed temporal and large-scale longitudinal variation of Io's atmosphere.
We find that the diurnal variations of the surface temperature affect the subsurface structure up to a depth of 0.6m. Furthermore, we quantify seasonal effects with Io having a northern summer close to perihelion and a northern winter close to aphelion. Finally, we found that at Io's anomalous warm polar regions a conductive heat flux of at least 1.2 Wm$^{-2}$ is necessary to reach surface temperatures consistent with observations.

\end{abstract}

\section*{Plain Language Summary} 
Io is the innermost of Jupiter's four Galilean moons and the most volcanically active body in our Solar System. It has a fairly dense SO$_2$ atmosphere with column densities on the order of $\sim$$10^{16}$ cm$^{-2}$. The huge number of erupting volcanoes on Io's surface constantly inject SO$_2$ into its atmosphere. The volcanic ejecta cover Io's surface with yellowish SO$_2$ surface frost. The gas injection as well as the sublimation of the surface frost are considered to be the main causes of Io's atmosphere. However, what controls the spatial and temporal structure of Io's atmosphere and to what proportion the atmosphere is driven by volcanoes or sublimation, is still not fully clear. Since the sublimation of surface frost strongly depends on the surface temperature, we developed a surface temperature model that takes thermal inertia and the exact celestial geometry into account in order to calculate the corresponding sublimation atmosphere. Using this model, we show that the SO$_2$ column density not only depends on the latitude, but also has a diurnal, seasonal and longitudinal variation. With the simulations we show that many observations of Io's atmosphere can be well explained by an atmosphere that is purely sublimation driven.

\section{Introduction}
Io's widespread surface volcanism was first predicted by \citeA{Peale1979} right before Voyager arrived at the innermost Galilean moon. In contrast to other bodies in the Solar System, the ultimate source of Io's mostly SO$_2$ atmosphere (\citeA{Peale1979}; \citeA{lellouch_1990}) is  volcanism. It has both small and large scale, spatial and temporal variability (e.g., \citeA{bagenaldols2020}; \citeA{depater2020review}). The atmosphere is observed to have equatorial SO$_2$ column densities in the range of $0.7-16\times10^{16}$ cm$^{-2}$, based upon different observation techniques, with significantly decreasing values towards the poles (\citeA{Feldman2000}; \citeA{McGrath2000}; \citeA{StrobelWolven2001}; \citeA{Jessup2004}; \citeA{Feaga2009}; \citeA{dePater_2020}). Observations also show that the atmosphere has column densities that are higher by up to a factor of 10 on the anti-Jovian hemisphere, than on the sub-Jovian (\citeA{Jessup2004}; \citeA{Spencer2005}; \citeA{Feaga2009}; \citeA{Tsang_2012}; \citeA{Giono_Roth2021}). In the early 2000s, especially the latitudinal variation of the atmosphere was often explained by the global distribution of volcanoes (e.g., \citeA{StrobelWolven2001}). The debate whether Io's atmosphere is primarily driven by direct particle injection from volcanoes or the sublimation from SO$_2$ surface frost has evolved through time due to research by, e.g., \citeA{Jessup2004}, \citeA{SaurStrobel2004}, \citeA{Feaga2009}, \citeA{Tsang_2012} and \citeA{lellouch_2015}. Viewed on smaller scales, \citeA{lellouch2007} suggest that the direct SO$_2$ particle injection do not dramatically increase the atmosphere's column densities above the plumes. 
Since the solar illumination decreases with increasing latitude, the idea arises that these large scale spatial structures can also be caused by the varying surface temperature and the thermal inertia of Io's surface materials. The corresponding sublimation atmosphere would then show the effect of thermal inertia as well (e.g., \citeA{Tsang_2012}; \citeA{retherford2019}).\\
Looking at hemispheric differences, Io's atmosphere is also observed to be more extended to higher latitude regions on the anti-Jovian hemisphere compared to the sub-Jovian \cite{lellouch2007}.  Later observations by, e.g., \citeA{Feaga2009} support that finding and state that the longitudinal dependence of the SO$_2$ abundance may result from a higher number of volcanoes on the anti-Jovian hemisphere compared to the sub-Jovian. More recently, \citeA{davies_2024} found that the number of active volcanoes per area unit in the polar region is similar to those of lower latitudes, but the polar volcanoes are smaller in terms of thermal emission compared to the large equatorial volcanoes.\\
Additional to the spatial variation, Io's atmosphere is also observed to vary temporally: First, the atmosphere collapses during the daily 2 hour eclipse by Jupiter (\citeA{SaurStrobel2004}; \citeA{Roth2011}; \citeA{Tsang2016}; \citeA{dePater_2020}) and secondly, it varies as a function of Jupiter's (and Io's) distance to the Sun (\citeA{Tsang_2012}; \citeA{Tsang2013}). Since this variability, especially the short-term variation during Io's eclipse, can not be explained by the presence of active/non-active volcanoes, the sublimation of Io's SO$_2$ surface frost was realized to be the dominant atmospheric source, although the exact proportion is still not clear. Nevertheless, using eclipse observations naturally obtained mainly on the sub-Jovian hemisphere, e.g., \citeA{Tsang2016} or \citeA{dePater_2020}, suggest that Io's volcanoes contribute on the order of tens of percents to the global SO$_2$ atmosphere and only have a pronounced effect if there is an eruption. To summarize, the global scale column density varies on two different timescales: Io's rotational period of $\sim$42 hrs and Jupiter's orbital period around the sun of $\sim$12 years. The (local) minima are reached during eclipse and shortly before sunrise or in aphelion, respectively.\\ 
Previous model work on Io's atmosphere was done by, e.g., \citeA{Walker2010} and \citeA{Walker_2012}. The authors focused on modeling the surface temperature and atmospheric variations using a direct simulation Monte Carlo method. They include complex surface characteristics and dynamic atmospheric processes such as inhomogeneous frost distributions and variable flow velocities, respectively. In the present work, we develop a simple time dependent surface temperature model with only one free parameter, i.e., Io's thermal inertia. Our model considers exact celestial geometry and therefore is able to calculate Io's surface and subsurface temperatures at any point in time and for each astronomical position of Jupiter, Io and the sun. Our simplified model aims to investigate to what degree such a plain sublimation model can explain the large-scale structure and variability of Io's surface temperature and SO$_2$ column density on short (diurnal) and large (Jovian year) timescales. A parameter study determines the thermal inertia values that fit observations from the past decades the best.

\section{The Time-Dependent Temperature Model}\label{Model}
With our model we investigate how well observed structures of Io's SO$_2$ atmosphere can be explained with an atmosphere that is purely sublimation driven. Since the amount of sublimated SO$_2$ surface frost highly depends on the surface temperature, we develop a simplified and time dependent surface temperature model. We include the most important heat production and loss processes as well as heat conduction with a subsurface layer characterized through its thermal inertia. First, we use the insolation $F\left(|S\left(t\right)|\right)$ as the most important heat source. It depends on the distance $|S|$ between Io and the sun which in turn is dependent on the time $t$. The local heat production rate $P_{sol}$ due to insolation also depends on the albedo $A_{vis}$ and the zenith angle $\xi$ which is dependent on $t$, the latitude $\theta\in[-90^{\circ},90^{\circ}]$ and the longitude $\varphi\in[0^{\circ},360^{\circ})$. We define the depth below the surface as $z=R_{Io}-r$, with $r$ the radial component. All radiative interaction (incoming and outgoing) is assumed to occur only at the surface, i.e. $z=0$, which is represented in our description with the $\delta$-function $\delta(z)$. Therefore, $P_{sol}$ is given as

\begin{equation}\label{Psol}
    P_{sol}=F\left(|S\left(t\right)|\right)(1-A_{vis})\cos\xi\left(t,\theta,\varphi\right)\delta(z).
\end{equation}

For completeness and similar to previous works we also include the thermal radiation from Jupiter $P_{rad,Jup}$ as well as the sunlight reflected from Jupiter $P_{ref,Jup}$. The contribution from Jupiter can be written as

\begin{equation}\label{P_rad_Jup}
    P_{rad,Jup}=\sigma\varepsilon_J T_J^4\left(\frac{R_J}{d_{(Io,J)}}\right)^2(1-A_{IR})\cos\xi_2\left(t,\theta,\varphi\right)\delta(z),
\end{equation}
    
\begin{equation}\label{P_ref_Jup}
    P_{ref,Jup}=F_{J}A_{J}\left(\frac{R_J}{d_{(Io,J)}}\right)^2\cdot\zeta\cdot(1-A_{vis})\cos\xi_2(t,\theta,\varphi)\delta(z).
\end{equation}

and is illustrated in Figure \ref{T_ref_rad}. Here, $\sigma$ is the Stefan-Boltzmann constant, $A_{IR}=0.5$ \cite{Nash1979} is Io's albedo in the infrared, $\varepsilon_J=0.9$ \cite{Morrison1980}, $T_J=165$ K \cite{JupiterFactsheet}, $R_J=71\,492$ km and $A_{J}=0.54$ \cite{JupiterFactsheet} are the emissivity, surface temperature, radius and albedo of Jupiter and $d_{(Io,J)}$ is the distance between Io and Jupiter. To calculate the heat flux due to the thermal radiation from Jupiter reaching Io's surface, we include Io's albedo in the infrared as well as a second zenith angle $\xi_2$, which is the zenith angle with respect to Jupiter. Due to Io's size compared to Jupiter and its close distance to the planet we consider the thermal radiation to come only from the direction that is pointing towards Jupiter`s center.
Additionally, the reflected sunlight from Jupiter is calculated using a factor $\zeta$, which describes the portion of Jupiter that is illuminated by the sun as seen from Io. At the equator and the sub-Jovian point the additional production terms given in equations (\ref{P_rad_Jup}) and (\ref{P_ref_Jup}) contribute approximately 0.55 Wm$^{-2}$ (for parameters given above and in Table \ref{Initvalues}) to the total heat production at the surface, which is on the same order as the internal heating rate (see equation (\ref{P_int})).\\
Secondly, as the only internal heat source, we include Io's tidal heating $P_{int}$. In general, the internal heat flux is assumed to be in the range of 1.5-4 Wm$^{-2}$ (\citeA{McEwen2004}; \citeA{Rathbun2004}; \citeA{Veeder2004_polar}). In the simulation we use a constant and uniformly distributed total internal heat flux of 2.4 Wm$^{-2}$. Simulations suggest that 80\% of Io's internal heat is transported by magmatic processes \cite{Steinke2020}. \citeA{Veeder_2015} use Galileo NIMS data to constrain that $\sim$50\% of Io's internal heat flux can be attributed to volcanic hot spots. In the first parts of our study we assume 20\% of the total internal heat flux, and therefore only 0.48 Wm$^{-2}$, to be transported by heat conduction. The internal heat production term only acts on the lower limit of the modeling domain, i.e. $z=d$. At this lower boundary the internal heat flux can be written as

\begin{equation}\label{P_int}
    P_{int}=0.48 \text{ Wm}^{-2}\delta(z-d).
\end{equation}
The most important heat loss term is the thermal radiation 

\begin{equation}\label{loss}
    L=\sigma\epsilon T^4\left(z,t,\theta,\varphi\right)\delta(z),
\end{equation}
where $\varepsilon=0.9$ (cf. \citeA{Morrison1980}) is Io's emissivity and $T$ is Io's surface temperature.\\
In addition to the sources and sinks detailed in equation (\ref{Psol}) - (\ref{loss}), we assume that heat is only transported within the subsurface layer through conduction with a thermal diffusivity $\alpha$. We assume that the thermal diffusivity is spatially constant near the surface. The influence of the heat fluxes of the surface extends only several meters in the subsurface layer (as shown in Section \ref{sec_SkinDepth}). This extent is much smaller than the latitudinal and azimuthal scales of the surface which are the order of a fraction of Io's radius; therefore, the heat flux is effectively only radial (see \ref{Appendix_A}). With this assumption,  our temperature model can be written as

\begin{equation}\label{energy_equation}
    \frac{\partial T\left(z,t,\theta,\varphi\right)}{\partial t}=\alpha\frac{\partial T^2\left(z,t,\theta,\varphi\right)}{\partial z^2}+\frac{1}{\rho c_p}\left(P_{sol}+P_{int}-L+P_{rad,Jup}+P_{ref,Jup}\right).
\end{equation}
The complete derivation of equation (\ref{energy_equation}) can be found in \ref{Appendix_A}. Heat conduction is controlled by the thermal diffusivity $\alpha$, which depends on the thermal conductivity $\kappa$, the mass density $\rho$ and the specific heat capacity $c_p$ of the subsurface matter and is given as 
\begin{equation}\label{Alpha}
    \alpha=\frac{\kappa}{\rho c_p}.
\end{equation}
It physically describes the efficiency of heat transfer through Io's interior and is related to thermal inertia. 
The analysis and the findings of this work are expressed in terms of thermal diffusivity since it is the essential variable in equation (\ref{energy_equation}) controlling the temperature on Io's surface. In the literature, the thermal inertia $\Gamma=\sqrt{\kappa \rho c_p}$ is frequently used. In Table \ref{Initvalues} upper part we show the model parameters for our reference model, where we list for comparison both thermal diffusivity and thermal inertia (see Section \ref{TIvsTD} for the direct relation between the thermal diffusivity and thermal inertia). The zenith angles $\xi$ and $\xi_2$, the distance between Io and the Sun $S$ as well as the distance between Io and Jupiter $d_{(Io,Jup)}$ are calculated using SPICE kernels \cite{SPICE} to provide a model that enables us to consider the exact celestial geometry (e.g. Io's inclination and the exact timing and positions of the eclipse by Jupiter) automatically. We use the latest satellite ephemeris JUP365 and the planetary and lunar ephemerides DE440 \cite{park_2021}. \\
The lower boundary condition of our model was set at a depth of $z_{\text{low}} = 2$ m. This value was determined in a parameter study for our selections of thermal diffusivity values $\alpha$ (Table \ref{Initvalues} lower part) such that at the lower boundary $z_{\text{low}}$ no significant temperature changes due to the sun's diurnal cycle occurs anymore (see Section \ref{sec_SkinDepth} and Equation (\ref{skindepth})). Due to the absence of solar driven heat flux at the lower boundary $z_{\text{low}}$ only conductive heat flux from Io's tidal heating is present.  We assume the latter heat flux to be constant and uniformly distributed over all latitudes and longitudes in most of our studies (except Section \ref{pole_problem}). The outer boundary condition is determined by the heat production and loss terms as they only act on Io's surface ($z=0$) (cf. Equation (\ref{energy_equation})). The boundary conditions can be found in \ref{Appendix_A}. This set of equations (Equations (\ref{energy_equation}), (\ref{innerbc}) and (\ref{outerbc})) is then solved numerically in our work. Each simulation is resolved with the same resolution for every thermal inertia assumptions.\\
Since a main objective of this work is to investigate Io's atmosphere, we calculate the corresponding SO$_2$ column density of the sublimation driven part $N_{sub}$ according to \citeA{Wagman_1979} assuming a hydrostatic, isothermal atmosphere in instantaneous thermal and vapor pressure equilibrium with a pure SO$_2$ ice underlying surface:
\begin{equation}\label{Wagman_eq}
    N_{sub}(t,\theta,\varphi)[m^{-2}]=\frac{1.1516\times10^{16}\exp\left(-\frac{4510}{T(t,\theta,\varphi)[K]}\right)}{m_{SO_2}[kg]\cdot g_0[ms^{-2}]},
\end{equation}
with $T$ the surface temperature, $m_{SO_2}$ the mass of an SO$_2$ molecule and $g_0=1.81$ m\,s$^{-2}$ Io's surface gravitational acceleration. In addition to that, we assume the atmospheric SO$_2$ column density to be completely controlled by the surface SO$_2$ sublimation temperature with no additional SO$_2$ source included in the simulations. \\

\begin{table}
 \caption{Model Parameters that are used to create a reference case for our simulations and range of values of a parameter study regarding the thermal diffusivity. We determine a thermal diffusivity that fits the modeled and observed column densities well for all latitudes and longitudes and call this the default value.}
 \centering
 \begin{tabular}{l c c}
 \hline
  \textbf{Parameters of Reference Model}  &    \\
 \hline
   Mass density $\rho$ [kg\,m$^{-3}$] & & 2096 \cite{Leone_2011}   \\
   Specific heat capacity $c_p$ [J\,kg$^{-1}$K$^{-1}$] & & 290 \cite{Leone_2011}   \\
   Thermal conductivity $\kappa$ [Wm$^{-1}$K$^{-1}$] & & 1.46 \cite{Leone_2011}   \\
   Thermal diffusivity $\alpha_0$ [m$^2$s$^{-1}$] & & $2.41\times10^{-6}$ \cite{Leone_2011}   \\
   Thermal inertia $\Gamma$ [Jm$^{-2}$s$^{-\frac{1}{2}}$K$^{-1}$] (MKS) & & 942\\
   \hline
   Albedo Io (visible) $A_{vis}$ & & 0.62 \cite{IoFactsheet}   \\
   Albedo Io (IR) $A_{IR}$ & & 0.5 \cite{Nash1979}   \\
   Albedo Jupiter $A_{J}$ & &  0.54 \cite{JupiterFactsheet}   \\
 \hline
 \hline
  \textbf{Parameter Study} & Range of values & Default \\
  \hline
  Thermal diffusivity $\alpha$ [m$^2$s$^{-1}$] & $1.61-96.4\times10^{-7}$ & $2.41\times10^{-7}$   \\
  \hline
 \end{tabular}
 \label{Initvalues}
\end{table}

\section{Results}
In this section we present a parameter study where we investigate the effects the thermal inertia of the surface has on Io's SO$_2$ atmosphere (assuming it is completely controlled by surface temperature).  In Section \ref{Basics} we discuss the effects of  Io's eclipse by Jupiter and changes in the thermal inertia on the diurnal temperature and SO$_2$ column density. To demonstrate the influence of thermal inertia not only on diurnal variations but also on global characteristics meaning the latitudinal and longitudinal variations, we present two extreme cases for extremely high and low thermal inertia values in Section \ref{LatVariation}. By comparing the simulations with several observations taken in the last decades we constrain a thermal diffusivity that fits the observed structure of Io's atmosphere best. In Section \ref{sec_SkinDepth} we estimate the "Depth of solar influence" (DOSI) that is defined as the depth from where the diurnal temperature variation is not significant anymore and discuss its dependence on the thermal diffusivity. Additionally, we also discuss an alternative variable to estimate the depth of the solar influence, i.e. the skin depth. We take a look at the long term variations of the surface temperature and the SO$_2$ column density over timescales of Jovian years using the exact celestial geometry in Section \ref{LongTermVariation}. By using also the exact body related geometry, e.g. the inclination,  we show that we can expect hemispheric seasons on Io depending on Jupiter's position in its orbit (Section \ref{SeasonalEffects}). Finally, we discuss whether or not our model could also explain Io's anomalous warm poles \cite{Rathbun2004} assuming a higher or non-uniformly distributed internal heat flux (Section \ref{pole_problem}).

\subsection{Thermal inertia effects on the diurnal temperature and column density}\label{Basics}
\begin{figure}[H]
    \centering
    \includegraphics[scale=0.8]{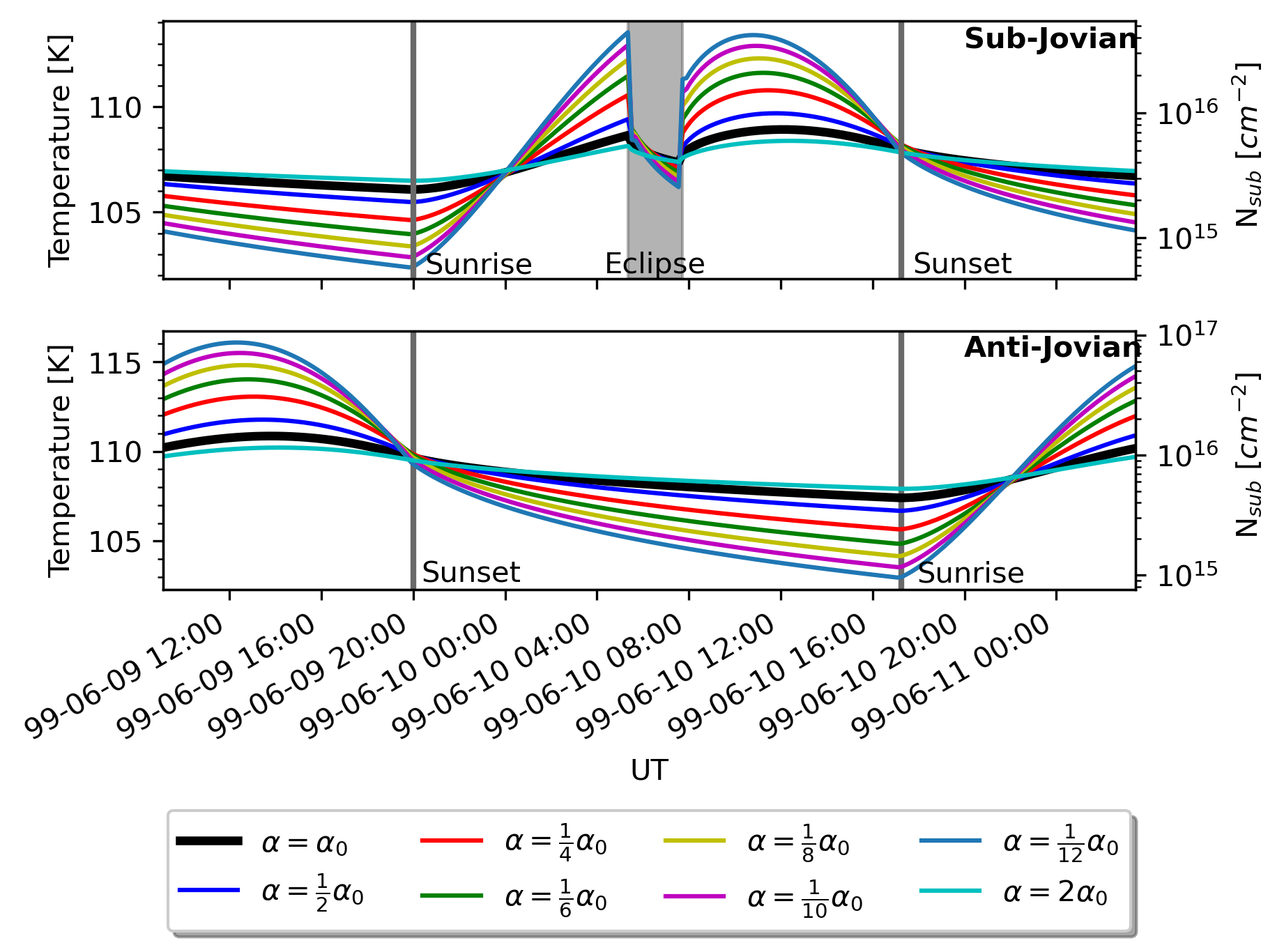}
    \caption{Diurnal equatorial surface temperatures and sublimation driven SO$_2$ column densities as a function of time elapsed over 42 hour Io rotation period for different values of the thermal diffusivity at the sub- and anti-Jovian point. Key Io local solar times are annotated on the plot. As reference value for the thermal diffusivity we assume $\alpha_0=2.41\times10^{-6}$m$^2$s$^{-1}$ and perform simulations varying $\alpha$ between $\frac{1}{12}\alpha_0$ and $2\alpha_0$. Remarkable is the eclipse by Jupiter which only affects the Sub-Jovian hemisphere due to the locked rotation of Io (cf. Figure \ref{fig:Io_orbit}) and which causes a drop in surface temperature within a few minutes. As a consequence, the surface temperature and atmospheric SO$_2$ column density is enhanced on the anti-Jovian hemisphere.}
    \label{Param_Study_Ts}
\end{figure}

\begin{figure}[H]
    \centering
    \includegraphics[scale=0.11]{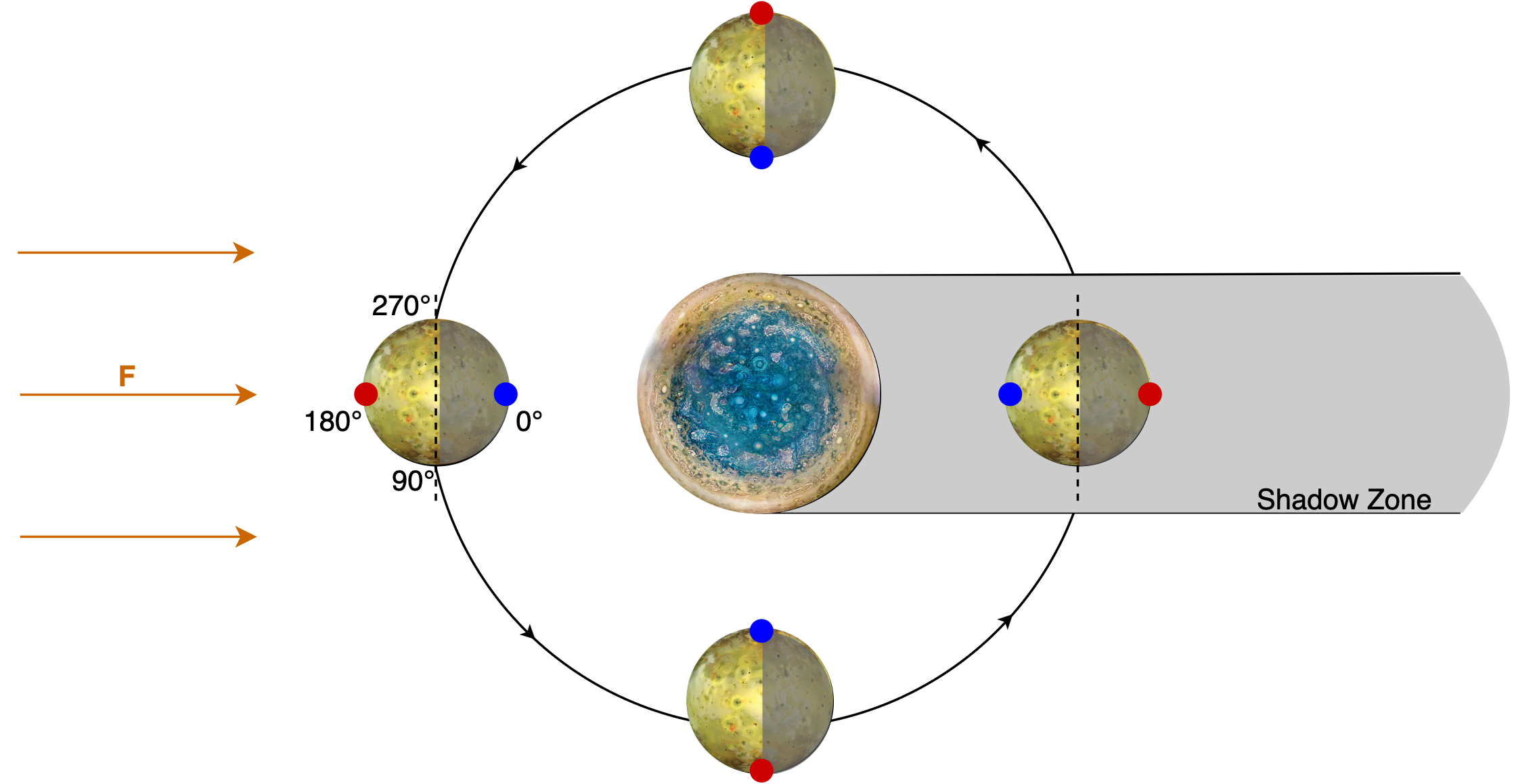}
    \caption{Top view of Io's orbit around Jupiter to illustrate the effect of Io's eclipse by Jupiter. Remarkable here is that due to the moon's locked rotation around the planet the anti-Jovian hemisphere is Io's nightside whenever Io is in eclipse by Jupiter. As a consequence, the anti-Jovian hemisphere receives 2 hrs more sunlight compared to the sub-Jovian hemisphere, which has a significant effects on Io's atmosphere.}
    \label{fig:Io_orbit}
\end{figure}

\begin{figure}[H]
    \centering
    \includegraphics[scale=0.8]{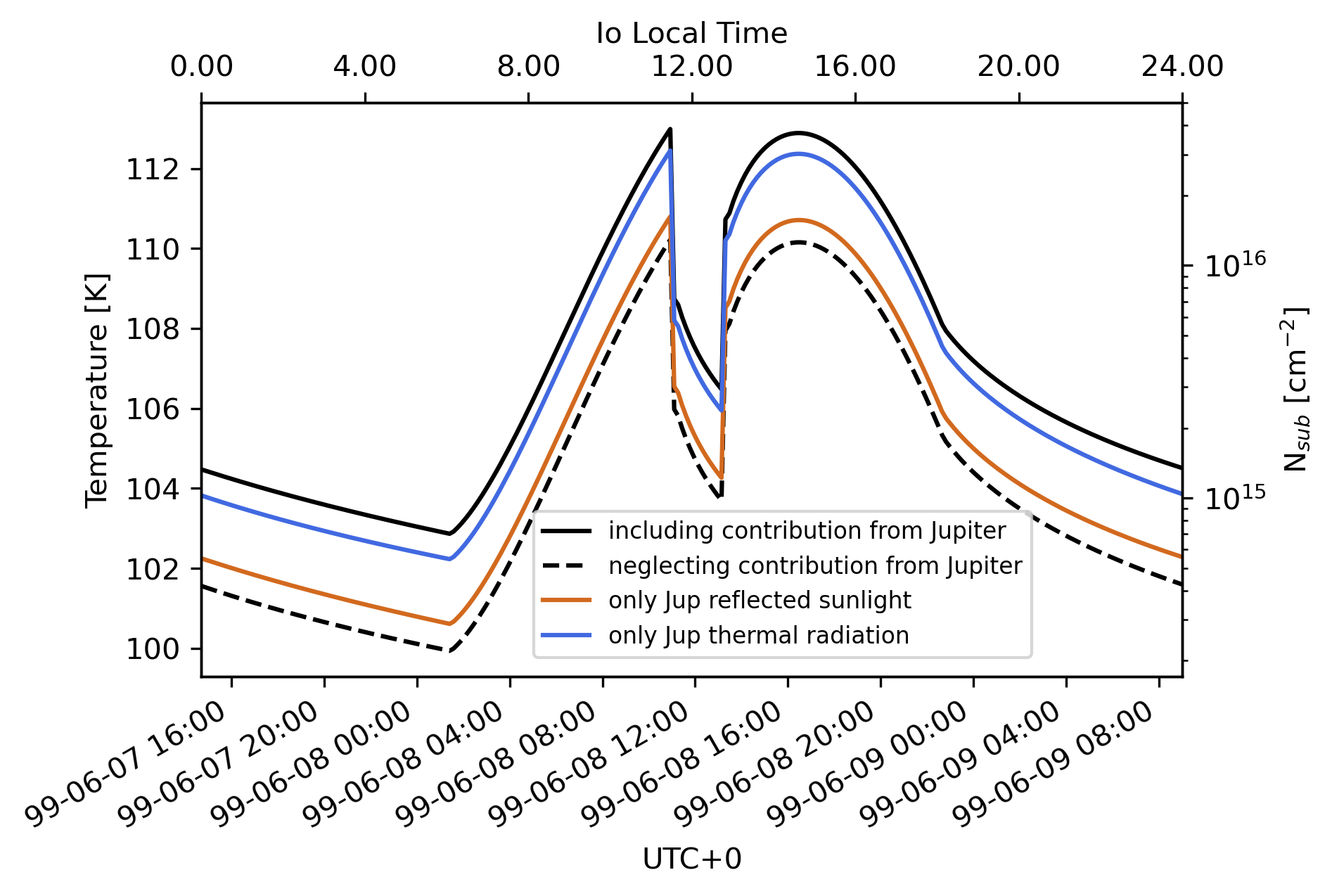}
      \caption{Diurnal equatorial surface temperatures and sublimation driven SO$_2$ column densities as a function of time including the contribution of the thermal radiation from Jupiter $P_{rad,Jup}$ and the sunlight reflected from Jupiter $P_{ref,Jup}$ (solid line) and neglecting them (dashed line) at the sub-Jovian point. The surface temperature differs by $\sim$3 K with largest influence of $P_{rad,Jup}$ and $P_{ref,Jup}$ during the night. The corresponding SO$_2$ column density is higher by a factor of $\sim$ 3.5 when including the contribution from Jupiter. Around 22\,\% of that enhancement is due to the reflected sunlight by Jupiter (red line) and 78\,\% is due to the thermal radiation from Jupiter (blue line).}
    \label{T_ref_rad}
\end{figure}

With the temperature model presented in Section \ref{Model} we are able to calculate Io's surface temperature $T$ as a function of time $t$, latitude $\theta$, longitude $\varphi$ for different values of thermal diffusivity. We begin with a study  how the thermal diffusivity  controls the surface temperature in response to diurnal variations. Therefore we present the equatorial surface temperature and column density as a function of time during one Io day ($\sim$42 hrs) and for different thermal diffusivity values in Figure \ref{Param_Study_Ts}. The top panel shows the diurnal variation at the sub-Jovian hemisphere (0° W), and the bottom panel shows the same but for the anti-Jovian hemisphere (180° W). Here, we present one Io day in 1999 when Jupiter was near its perihelion just as an example. The effect of the varying heliocentric distance on the surface temperature and atmosphere is discussed in Sections \ref{LongTermVariation} and \ref{SeasonalEffects}. \\
In our parameter study we start with a reference value for the thermal diffusivity $\alpha_0=2.41\times10^{-6}$ m$^2$s$^{-1}$ \cite{Leone_2011}. Here we choose the values for the lowest depth provided by \citeA{Leone_2011} since we expect the DOSI to be on the order of $\sim$1 m (see Section \ref{sec_SkinDepth}). As shown by \citeA{dePater_2020} the thermal inertia also may vary as a function of depth with values of 50 Jm$^{-2}$s$^{-\frac{1}{2}}$K$^{-1}$ (MKS in the following) directly at the surface and $\sim$320 MKS at a depth of a cm and below. We assume the thermal diffusivity to be constant as a function of depth and vary it between $\frac{1}{12}\alpha_0$ and $2\alpha_0$ (see Table \ref{Initvalues}). The heat transfer through Io's interior is faster for a higher thermal diffusivity resulting in a weaker diurnal variation. Observations of Io's thermal radiation made with Galileo's photopolarimeter–radiometer show that Io's daytime surface temperature varies between 90 K minimum near the terminators and in the polar regions and 130 K maximum close to the subsolar point \cite{Rathbun2004}. The reference thermal diffusivity $\alpha_0$(black line) already leads to a diurnal surface temperature variation that is unrealistically low with minimum and maximum temperatures of 106 K and 109 K for the sub-Jovian and 107.5 K and 111 K for the anti-Jovian hemisphere. Nonetheless, we also added the diurnal variations resulting from simulations assuming even higher thermal diffusivity values of $2\alpha_0$ (cyan line) to show that the resulting temperature and column density variation is even lower compared to the reference case. We determine minimum and maximum values $\alpha_{\text{min}}=\frac{1}{12}\alpha_0$ and $\alpha_{\text{max}}=\frac{1}{8}\alpha_0$ of the thermal diffusivity that lead to a diurnal variation that corresponds to observations for the sub- and anti-Jovian hemisphere, especially when looking at the corresponding column density.\\
Accounting for the eclipse at each point on the surface individually, for the sub-Jovian hemisphere the eclipse around 12:00 local time (LT) has a pronounced effect with a sharp temperature decrease of up to 7.5 K within $\sim$10 min (for the respective values of $\alpha$) corresponding to an atmosphere that collapses by more than one order of magnitude. The drop in surface temperature is expected to occur within minutes (e.g., \citeA{dePater_2002}) with a reacting sublimation atmosphere that collapses on comparable timescales (e.g., \citeA{cruikshank_2010}). As a consequence, the simulations at times of very rapid changes of insolation and surface temperature do not include adjustment of the atmospheric column densities to equilibrium state in Equation (\ref{Wagman_eq}). We give more details on our model limitations in Appendix \ref{A3-Limitations}. However, since this work is about investigating the variability of Io's atmosphere on significantly longer timescales than $\sim$ 10-15 min we can assume the atmospheric temperature is the same as the surface temperature \cite{Wagman_1979}. The observed column densities decrease by a factor of 5$\pm$2 and on similar timescales as found in the simulation (\citeA{Tsang_2012}; \citeA{Tsang2016}; \citeA{Roth2011}). Observations at millimeter wavelengths in \citeA{dePater_2020} do not show a significant drop in disk-averaged column densities during Io being in eclipse. Instead, they show a much lower fractional coverage of Io in eclipse compared to Io being in sunlight and are assumed to probe Io's subsurface at a depth of 1-2 cm, so do not show the temperature variation directly at the surface. 
Since it is not possible to distinguish between a high column density with low fractional coverage or vice versa with the data used in \citeA{Tsang2016}, the findings of \citeA{dePater_2020} may agree with a column density decreasing by a factor of 5 in the mid-infrared, nonetheless. However, the weaker temperature decrease of $\sim$3 K described in \citeA{dePater_2020} is consistent with our simulations evaluated at a respective depth of $\sim$3 cm, since the strength of the temperature decrease during eclipse naturally decreases with increasing depth due to the diminishing influence of the sun with increasing depth.Although their inferred column density decrease match our simulations, \citeA{Tsang2016} found the brightness temperature to decrease up to 22 K in eclipse in two set of observations obtained on two different days in November 2013, which would require a relatively low thermal inertia. In general our model is consistent with larger thermal inertia values causing a weaker temperature drop in eclipse. \citeA{Walker_2012} determined a best fit thermal inertia of a frost-covered surface of 200 $\pm$ 50 MKS, which is close to our default value of 298 MKS and fits to the model assumption that Io is uniformly covered with SO$_2$ surface frost. Therefore, the observations by \citeA{Tsang2016} could be explained by either a two-layer model as described in \citeA{dePater_2020} or by a non-uniform surface frost coverage. Additionally, \citeA{tsang_2015} in another study did not see any significant column density increase shortly after eclipse egress, implying that the atmosphere has not collapsed in Jupiter's shadow. Whether the sharp temperature decrease found by \citeA{Tsang2016} is attributable to either a low thermal inertia surface (layer), a non-uniform surface frost coverage or, unknown time-variability, or how the apparently contradicting observations might be reconciled otherwise, may remain subject of future observations and simulations.\\
To explain the eclipse effect in more detail, Figure \ref{fig:Io_orbit} illustrates a top view of Io's orbit around Jupiter. Because of Io's locked rotation around the planet, the anti-Jovian hemisphere is always on Io's nightside when Io is in eclipse so that Jupiter's shadow has no direct impact on the anti-Jovian hemisphere with regard to its surface temperature. As a consequence, this hemisphere is exposed to sunlight $\sim$ 2 hrs longer during each Io day, i.e. every 48 hrs, compared to the sub-Jovian hemisphere. Therefore, the eclipse effect is also the reason for the surface temperature to be generally higher on the anti-Jovian hemisphere compared to the sub-Jovian as was shown before by, e.g., \citeA{Walker_2012}.\\
In addition to that, the surface is heated due to thermal radiation and the sunlight reflected from Jupiter at Io's sub-Jovian hemisphere. Figure \ref{T_ref_rad} shows the diurnal variation of the surface temperature and corresponding column density including and neglecting both effects. The surface temperature increases by $\sim$3 K due to the thermal radiation and the sunlight reflected by Jupiter, which therefore have only a very small influence on the surface temperature compared to the solar radiation. However, the corresponding sublimation driven SO$_2$ column densities increase by a factor of $\sim$3.5 at times of maximum influence, which is during the night, when the solar illumination vanishes. Here the thermal radiation from Jupiter has the larger influence with a temperature increase of $\sim$2.3 K compared to the reflected sunlight, which leads to a temperature increase of $\sim$0.7 K.

\subsection{Interpreting the parameter study with regard to thermal inertia}\label{TIvsTD}
Since both the thermal diffusivity $\alpha=\kappa\left(\rho c_p\right)^{-1}$ as well as the thermal inertial $\Gamma=\sqrt{\kappa\rho c_p}$ depend on the thermal conductivity $\kappa$, we can relate both quantities for a given mass density $\rho$ and specific heat capacity $c_p$ through
\begin{equation}\label{Gamma}
    \Gamma=\rho c_p\sqrt{\alpha}.
\end{equation}
In our parameter study, we perform simulations by assuming thermal diffusivity values in the range of $\frac{1}{12}\alpha_0\leq\alpha\leq2\alpha_0$ with $\alpha_0=2.41\times10^{-6}$m$^2$s$^{-1}$. Using the values of \citeA{Leone_2011} for the mass density $\rho$ and the specific heat capacity $c_p$ (see Table \ref{Initvalues}) this corresponds to thermal inertia values varying between 272 and 1334 MKS which is comparable to the range inferred as reasonable by \citeA{Tsang_2012}. In Section \ref{Basics} we determine minimum and maximum values $\alpha_{\text{min}}=\frac{1}{12}\alpha_0$ and $\alpha_{\text{max}}=\frac{1}{8}\alpha_0$ that lead to a diurnal surface temperature and column density variation in agreement with observations. For the given values of $\rho$ and $c_p$ from Table \ref{Initvalues}, this corresponds to thermal inertia values of 272 to 333 MKS, which agrees well with the thermal inertia (320 MKS) derived at millimeter wavelengths \cite{dePater_2020}.

\subsection{Latitudinal variation of a purely sublimation driven atmosphere}\label{LatVariation}
\begin{figure}[H]
    \centering
    \includegraphics[scale=0.35]{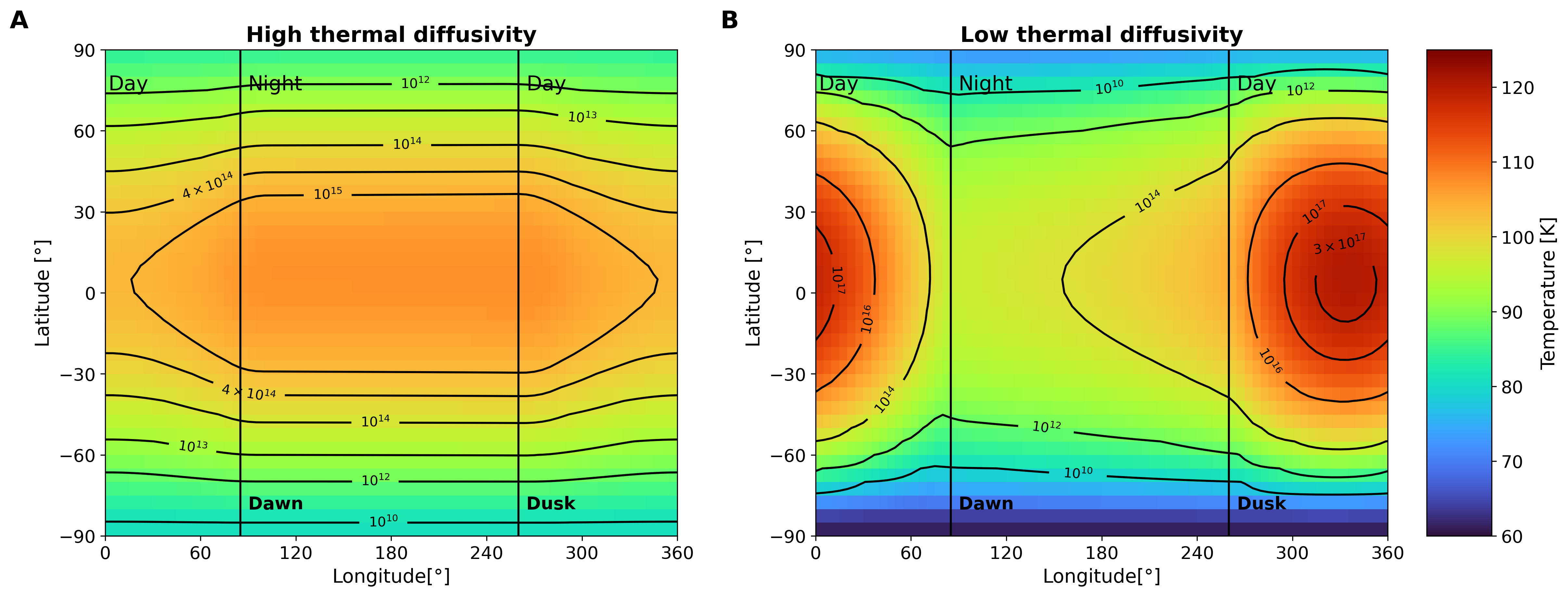}
      \caption{Global surface temperature and column density maps, assuming that the sub-Jovian longitude is the subsolar longitude, so the sub-Jovian hemisphere is illuminated by the sun. The column density is represented as contour lines and given in cm$^{-2}$, the surface temperature is color-coded and given in K. \textbf{(A)} For a high thermal diffusivity ($50\,\alpha_0$ with $\alpha_0=2.41\times10^{-6}$m$^2$s$^{-1}$) the atmosphere is clearly band structured with highest column densities around the equator decreasing towards the poles. There is no collapse of the atmosphere on the night side. On the contrary, here the night side is even warmer than the day side due to the eclipse effect (see Section \ref{Basics}). \textbf{(B)} For a low thermal diffusivity $\left(\frac{1}{50}\,\alpha_0\right)$, the atmosphere is mainly centered around the subsolar point with column densities decreasing towards the poles as well as towards the terminators. For identifying the surface temperature and column density patterns the relative distance of Io to Jupiter is not relevant at this point of our work, therefore the date of simulation is arbitrary.}
    \label{extreme_thermal_inertia}
\end{figure}

In the following we present complete maps of Io's surface temperature as well as the corresponding SO$_2$ column density as a function of latitude and longitude. We focus on two extreme cases with an extremely high and an extremely low thermal inertia to separate the effect of the thermal inertia on the surface temperature from other effects. Once this is shown, we determine a default thermal diffusivity by finding the value that fits modeled and observed column densities well for all latitudes and longitudes. At the end of this section we compare the simulations with observations.\\
In Figure \ref{extreme_thermal_inertia} we present an exceptionally high and exceptionally low case of the thermal inertia to explain its effects on the latitudinal structure of Io's surface temperature and correspondingly the SO$_2$ atmosphere. For the purposes of identifying surface temperature and column density patterns at this point the relative distance of Io to Jupiter is inconsequential, so the date of simulation is arbitrary. The left panel shows a map of Io at a certain point in time assuming a thermal diffusivity of $50\alpha_0$. Due to the extremely high thermal diffusivity the conductive heat transport into the subsurface of Io is very efficient. Therefore, the surface does not heat up very fast when exposed to sunlight and does not cool strongly on the night side due to the conductive heating from within Io. Consequently, the atmosphere is clearly band structured with the largest column densities around the equator. Since the solar zenith angle increases for higher latitudes the amount of heat absorbed by the surface decreases, which results in steadily decreasing surface temperatures and column densities towards the poles. Additionally, there is no or nearly no diurnal variation of the surface temperature or column density from dawn to dusk. The maps shown in Figure \ref{extreme_thermal_inertia} both result from simulations when Io's subsolar longitude corresponds to its sub-Jovian longitude, so the sub-Jovian hemisphere is illuminated by the sun and the anti-Jovian hemisphere is the night side. The enhanced column densities on the anti-Jovian hemisphere from 90 to 270° W are only attributed to the eclipse effect, mentioned in Section \ref{Basics} and in e.g., \citeA{Walker_2012}. \\
In contrast to that, the right panel presents a map assuming a thermal diffusivity of $\frac{1}{50}\alpha_0$. The extremely low thermal diffusivity causes the surface heat to remain in the very shallow subsurface structures and is not transported to deeper structures. Therefore, the surface temperature is only dependent on the solar zenith angle and increases quickly and strongly when exposed to sunlight. So the atmosphere is centered around the subsolar point with decreasing column densities towards the poles and the terminators. \\
Of particular note is that the thermal diffusivity strongly influences the peak column density. Due to the slower heat transfer the surface heats up much faster for a low thermal diffusivity, resulting in peak surface temperatures of 107.5 K for $50\alpha_0$ and much larger values of 122 K for $\frac{1}{50}\alpha_0$, respectively. The peak surface temperature extends over a longitudinal range of 180° for the high thermal inertia, whereas the peak temperature for the low thermal inertia is reached only in a small region close to the subsolar point. The corresponding peak column densities of $4.5\times10^{15}$ and $4\times10^{17}$cm$^{-2}$ differ by more than two orders of magnitude. \\
We conclude that the choice of thermal inertia has strong influence on the latitudinal structure of Io's atmosphere, which we will quantify further in the following Section.

\begin{figure}[H]
    \centering
    \includegraphics[scale=0.4]{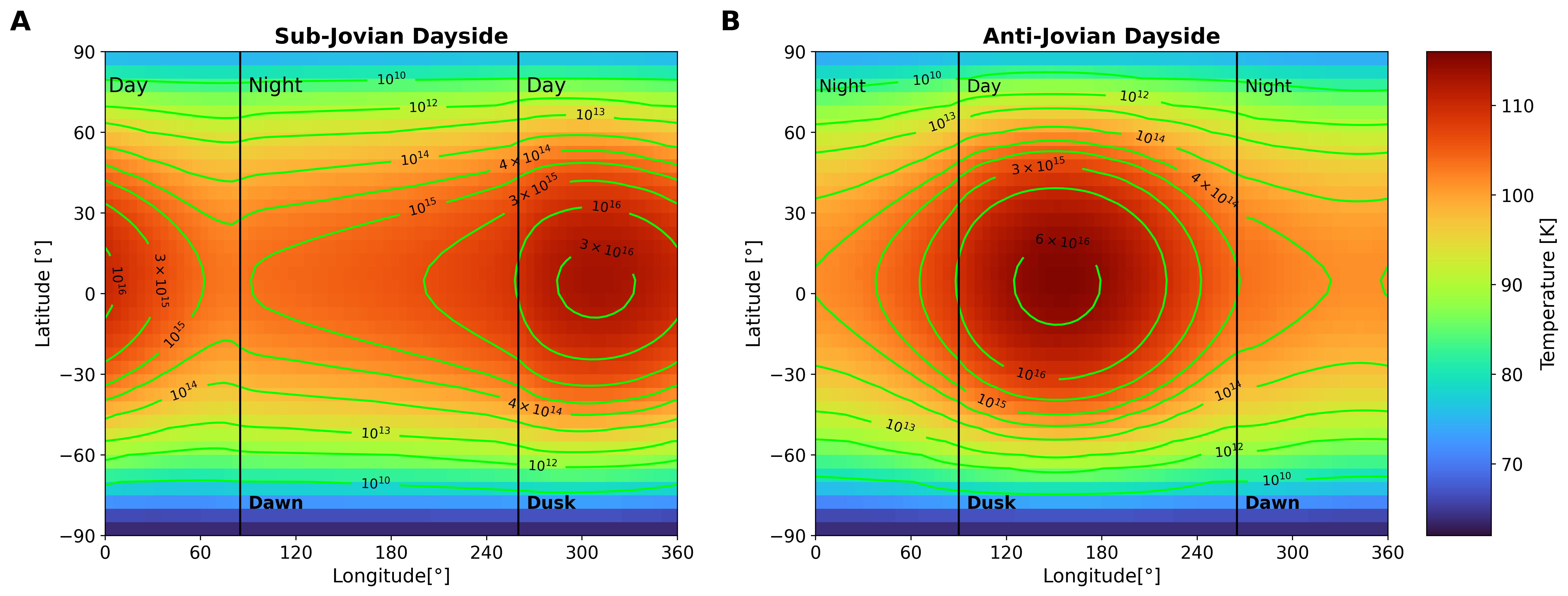}
    \caption{Global surface temperature and column density map assuming a thermal diffusivity of $0.1\alpha_0$ with $\alpha_0=2.41\times10^{-6}$m$^2$s$^{-1}$. We distinguish here between \textbf{(A)} the sub-Jovian and \textbf{(B)} the anti-Jovian hemisphere being Io's dayside. In both cases the atmosphere is mostly centered on Io's dayside as expected from observations and the model. The significant collapse of the atmosphere from dusk to dawn is controlled by thermal inertia with column densities decreasing by more than one order of magnitude. The anti-Jovian hemisphere has generally higher diurnal surface temperatures and column densities due to the eclipse by Jupiter.}
    \label{best_alpha_map}
\end{figure}

\begin{figure}[H]
    \centering
    \includegraphics[scale=0.65]{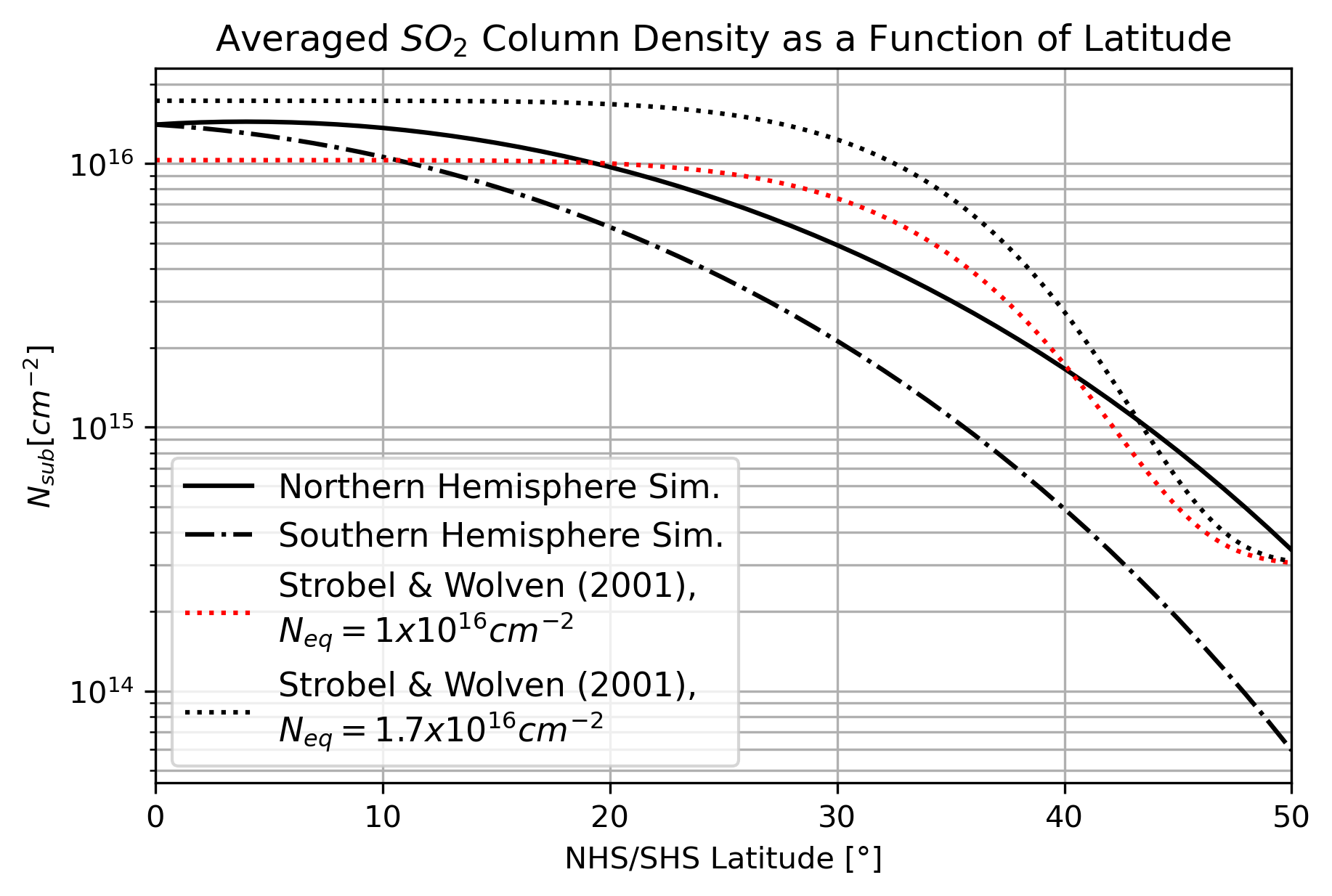}
    \caption{Column density as a function of latitude for Io's northern (solid line) and southern (dashed-dot line) hemispheres assuming a thermal diffusivity of $0.1\alpha_0$ with $\alpha_0=2.41\times10^{-6}$m$^2$s$^{-1}$ in each case. The simulated northern hemisphere column density is higher than the south due to Io's inclination causing hemispheric seasons, and this contributes to the strong agreement of the northern hemisphere results of \citeA{StrobelWolven2001} which are also obtained during perihelion. See the main text for a detailed summary.}
    \label{comp_strobel_wolven}
\end{figure}

The surface temperature and column density shown in Figure \ref{extreme_thermal_inertia} do not represent the observed structure of the atmosphere and only serve to demonstrate the effect of different thermal inertia assumptions on the latitudinal and longitudinal variation of the atmosphere separated from other physical effects. Now we further constrain the thermal diffusivity by comparing model results with observations of surface temperature and column densities for $\alpha$ in the range of $\frac{1}{12}\alpha_0\leq\alpha\leq2\alpha_0$ by taking into account Io's surface temperature at the sub- and anti-Jovian hemisphere as well as its diurnal variation. By comparing the model results also with observations of the latitudinal column density variation (e.g. \citeA{StrobelWolven2001}) and excluding temperature variations with unrealistic low and high diurnal variations we then choose our default value through quantitative comparison. We find 0.1$\alpha_0$, which compares to a thermal inertia of $\Gamma=298$ MKS. It is in a similar range of the best-fit thermal inertia values determined by \citeA{Tsang_2012} (150-300 MKS) and \citeA{Walker_2012} (200$\pm$50 MKS). The large uncertainty in our values is due to the large uncertainty of the mass density and specific heat capacity directly at Io's surface. \\
The surface temperature and column density map resulting from our simulation with a thermal diffusivity of 0.1$\alpha_0$ agrees well with the observations and is shown in Figure \ref{best_alpha_map}. 
At the equator on the subsolar point the solar heat flux is $\sim$ 9 times larger than the internal heat flux of 2.4 Wm$^{-2}$. The solar heat flux on the surface decreases with latitude. At a latitude of about 83 degree the solar heat flux becomes similar to the internal heat flux and further towards the pole the surface temperature is mainly controlled by the internal heat flux, which is why a slightly changing albedo towards the poles has no influence on the surface heating. For now, we will concentrate only on an extended equatorial region up to 60° latitude since Io's high latitude and polar regions are observed to have temperatures around 90 K \cite{Rathbun2004} and are therefore significantly warmer compared to what is expected from a surface heating due to tidal heating with the given total heat flux of 2.4 Wm$^{-2}$ only. \\
Figure \ref{best_alpha_map} shows two "snapshots" taken during one day in 1999, when Jupiter was near its perihelion. To highlight the sub-anti-Jovian hemisphere asymmetry, the left panel shows a time of day when the subsolar longitude occurs at sub-Jovian longitude, so the sub-Jovian hemisphere is fully illuminated by the sun whereas the right panel shows a time of day when the subsolar longitude occurs at anti-Jovian longitude, so the anti-Jovian hemisphere is the dayside. The maximum dayside column densities of 3.7$\times$10$^{16}$ cm$^{-2}$ and $8.5\times10^{16}$ cm$^{-2}$ respectively for the sub-Jovian and anti-Jovian hemisphere lie well with observed values that are in the range of $1-20\times10^{16}$ cm$^{-2}$ (\citeA{Feldman2000}; \citeA{Jessup2004}; \citeA{Spencer2005}; \citeA{Feaga2009}; \citeA{Tsang_2012}; \citeA{Jessup2014}). The sublimation atmosphere is observed to collapse during Io's eclipse by Jupiter because of the dwindling solar insolation. This is why Io's atmosphere is also assumed to collapse on the nightside with significantly decreasing column densities. As it is hard to observe Io's nightside from Earth, there are no measured values of the surface temperature or column density from dusk to dawn. According to our simulation, the column density decreases by one order of magnitude from $10^{16}$ cm$^{-2}$ at the dusk terminator to $0.4-1\times10^{15}$ cm$^{-2}$ at the dawn terminator depending on whether we investigate the sub-Jovian or the anti-Jovian hemisphere to be the dayside. \\
Additional to this longitudinal column density variation, the atmosphere is latitudinally structured as well. To quantify the equator-to-pole decrease, Figure \ref{comp_strobel_wolven} provides a direct comparison to the column density as a function of latitude determined by \citeA{StrobelWolven2001}, who processed reflected solar Lyman-$\alpha$ data measured by the Hubble Space Telescope (HST) when Jupiter was close to perihelion. In addition to the observationally derived column densities (dotted lines), we present the dayside averaged SO$_2$ column density as a function of latitude resulting from the simulation (solid and dashed dotted lines). \citeA{StrobelWolven2001} use two different equatorial column densities of $1\times10^{16}$ cm$^{-2}$ and $1.7\times10^{16}$ cm$^{-2}$ that are nearly constant between 0 and 20° latitude and found them to decrease between 20 and 50° to $3\times10^{14}$ cm$^{-2}$. Since our calculations consider Io's inclination of $\sim$3° (tilt of Io's spin axis with respect to Jupiter's orbital plane \cite{JupiterFactsheet}), our model shows a northern-southern hemisphere asymmetry, with higher column densities over the northern hemisphere during the time of the observation in the year, because Io's northern summer season is close to perihelion (discussed in Section \ref{SeasonalEffects}). Therefore, our results are split in one curve representing the latitudinal decrease on the northern hemisphere (solid) and one curve for the southern hemisphere (dashed dotted). Due to a poor telescope resolution at very high latitudes, \citeA{StrobelWolven2001} assume the column density to stay constant from 50° latitude polewards. As expected and observed in later observation, however, the column density would decrease even further (e.g., \citeA{Feaga2009}). For comparison purposes, in Figure \ref{comp_strobel_wolven} we only present our model results obtained between 0 and 50$^{\circ}$ latitude.\\
This comparison shows that our simulations yield dayside averaged column density values that agree with the ones from \citeA{StrobelWolven2001}. Thus, a plain sublimation generated atmosphere reproduces the observed latitudinal decrease of the column density which is more than one order of magnitude.

\subsubsection{Diurnal variations of the surface temperature and column density}\label{diurnal_variations}
\begin{figure}[H]
    \centering
    \includegraphics[scale=0.4]{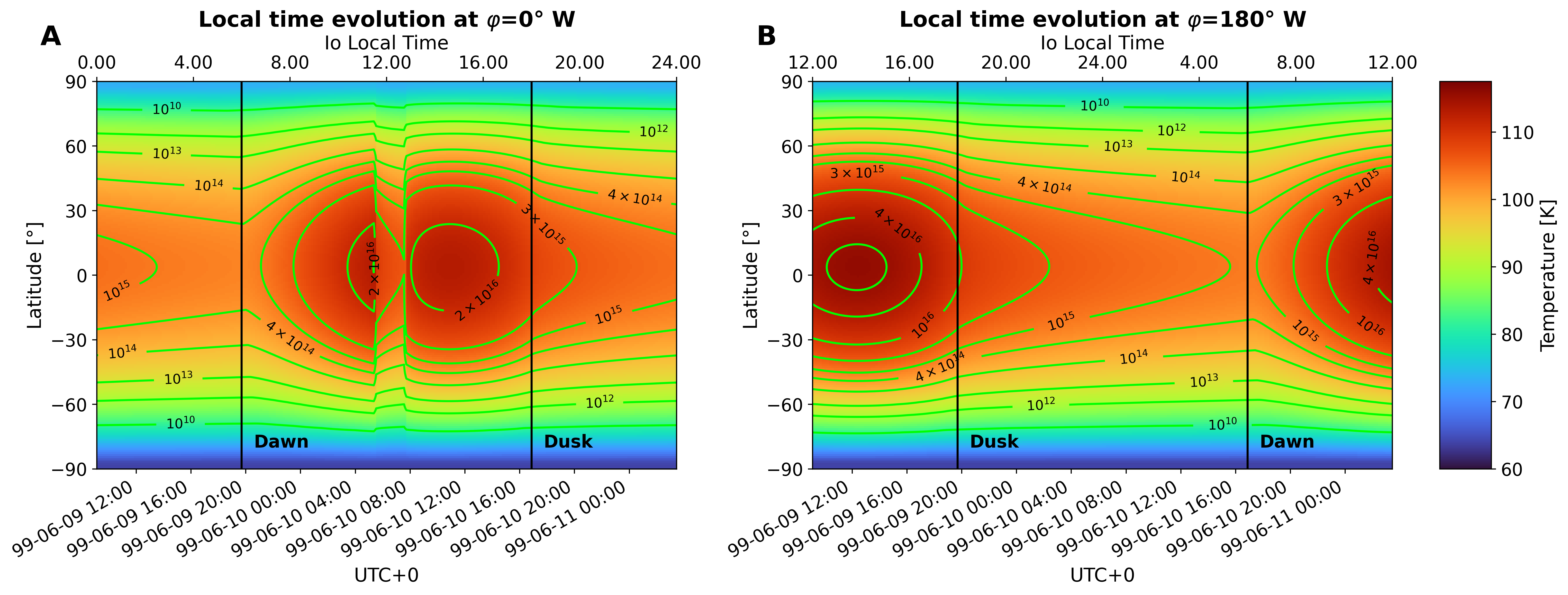}
    \caption{Diurnal variations of the surface temperature and column density for all latitudes at \textbf{(A)} the sub-Jovian ($\varphi$= 0° W) and \textbf{(B)} the anti-Jovian longitude ($\varphi$= 180° W) assuming $\alpha=0.1\alpha_0$ with $\alpha_0=2.41\times10^{-6}$m$^2$s$^{-1}$. The eclipse effect generates an atmosphere that has a more extended atmosphere with maximum column densities that are higher by a factor of four on the anti-Jovian hemisphere, compared to the sub-Jovian.} 
    \label{map_one_day}
\end{figure}

So far, we only presented surface temperature and column density maps at one specific time, meaning one specific solar time, but for all latitudes and longitudes (Section \ref{LatVariation}) as well as the diurnal variation at one specific point on Io's surface (cf. Section \ref{Basics}). In the following, we focus on the diurnal variation at one specific longitude but for all latitudes. Therefore, Figure \ref{map_one_day} shows the surface temperature (color-coded) and corresponding SO$_2$ column density (contour lines) as a function of latitude and time for both, the sub- ($\varphi=$ 0° W, panel A) and anti-Jovian longitude ($\varphi=$ 180° W, panel B). We assume a thermal diffusivity of $0.1\alpha_0$ as our default model to investigate the effect of the eclipse by Jupiter on Io's atmosphere. \\
For comparison, we again show the diurnal column density variation for one day in June 1999, when Jupiter was near perihelion. Figure \ref{map_one_day} A refers to the sub-Jovian longitude, where the eclipse effect is clearly visible. After passing the dawn terminator, the column density increases and reaches its morning maximum of $4\times10^{16}$ cm$^{-2}$ shortly before eclipse ingress. At 11:30 LT Io goes into eclipse by Jupiter for $\sim$2 hrs and no longer receives sunlight. Consequently, the surface temperature decreases, the atmosphere collapses and column densities decrease by one order of magnitude from $4\times10^{16}$ cm$^{-2}$ before eclipse ingress to $3.4\times10^{15}$ cm$^{-2}$ in eclipse, which compares well to previously modeled and observed values (\citeA{SaurStrobel2004}; \citeA{Tsang_2012}; \citeA{Tsang2016}). \citeA{dePater_2020} found the atmosphere to collapse and recover within 10 min after eclipse ingress and egress. As described in Section \ref{Basics} the simulation matches that well. Additionally, \citeA{Jessup2014} observed Io's low latitude dayside surface temperature to be $0.6\pm0.1$K larger at 10.00 LT than at 13.00 LT. The authors did not relate their observations to the eclipse effect, but the simulated surface temperatures show a similar trend before and during the eclipse phase: it decreases from 110 K at 10.00 LT (before eclipse) to 106 K at 12.30 LT (before eclipse egress). At the dusk terminator (18.00 LT) the surface temperature and column density starts to decrease slowly throughout the night controlled by the thermal inertia of the subsurface materials. In total, the equatorial surface temperature decreases by 6 K from 108.6 to 102.8 K from dusk to dawn, which corresponds to the column density decreasing by almost one order of magnitude from $7.4\times10^{15}$ cm$^{-2}$ to $7\times10^{14}$ cm$^{-2}$.\\
Figure \ref{map_one_day} B presents the diurnal variation of the surface temperature and column density at the anti-Jovian longitude at $\varphi=$ 180° W. It highlights the hemispheric differences between the sub and anti-Jovian side. The anti-Jovian hemisphere does not experience the eclipse by Jupiter (cf. Figure \ref{fig:Io_orbit}). Therefore, the surface temperature does not decrease until the afternoon LT and the atmosphere does not collapse during the day. Due to the enormous size of Jupiter and the small distance of Io from the planet, the eclipse affects the whole sub-Jovian hemisphere, leading to in general lower surface temperature and SO$_2$ column densities here. This effect is well observed in HST data (\citeA{Jessup2004}; \citeA{Feaga2009}). According to our simulation, the eclipse effect causes a temperature difference of 4 K at the equator between both hemispheres, which corresponds to maximum column densities that are higher by a factor of 4 on the anti-Jovian compared to the sub-Jovian longitude. \\
The diurnal maximum surface temperature occurs not exactly at noon where the solar heat flux is strongest. At the sub-Jovian longitude the diurnal maximum is reached shortly before noon because the following eclipse causes the surface to cool down. In contrast to that, at the anti-Jovian longitude, the maximum surface temperature is reached after noon at around 14.00 LT which is due to the thermal inertia of the subsurface materials. Furthermore, the eclipse by Jupiter lowers the total diurnal variation which is 10.5 K at the sub-Jovian but 12.3 K on the anti-Jovian longitude. This fits well with the model results of \citeA{Tsang_2012} who find diurnal variations of 10 K and 12 K, respectively, for their best-fit model. \citeA{Walker_2012} find the diurnal variation to be $\sim$14 K on the sub-Jovian and $\sim$17 K on the anti-Jovian hemisphere. This slightly enhanced diurnal variation compared to our results is attributable to a lower thermal inertia value ($\Gamma=200 \pm 50$MKS) assumed by \citeA{Walker_2012} as discussed in Section \ref{Basics}. \\
Lastly, the atmosphere is not only observed to be denser on the anti-Jovian hemisphere, but also more extended to high latitudes (\citeA{lellouch2007}; \citeA{Feaga2009}). Looking at the northern hemisphere, \citeA{Feaga2009} found the atmosphere to have its $10^{16}$ cm$^{-2}$ "column-density-boundary" at 30° latitude on the anti-Jovian hemisphere, and at 15° lat on the sub-Jovian. In our simulation the $10^{16}$ cm$^{-2}$ boundary is more at 20° latitude at the sub-Jovian longitude, and at around 30° latitude at the anti-Jovian. Therefore, the simulated atmosphere is slightly more extended on the sub-Jovian hemisphere than found in \citeA{Feaga2009}, but still clearly more expanded on the anti-Jovian hemisphere when comparing to the sub-Jovian. This effect is also attributed to the fact that the anti-Jovian hemisphere experiences $\sim$2 hrs more of sunlight each day compared to the sub-Jovian. 

\subsection{Equatorial Skin Depth}\label{sec_SkinDepth}
\begin{figure}[H]
    \centering
    \includegraphics[scale=0.75]{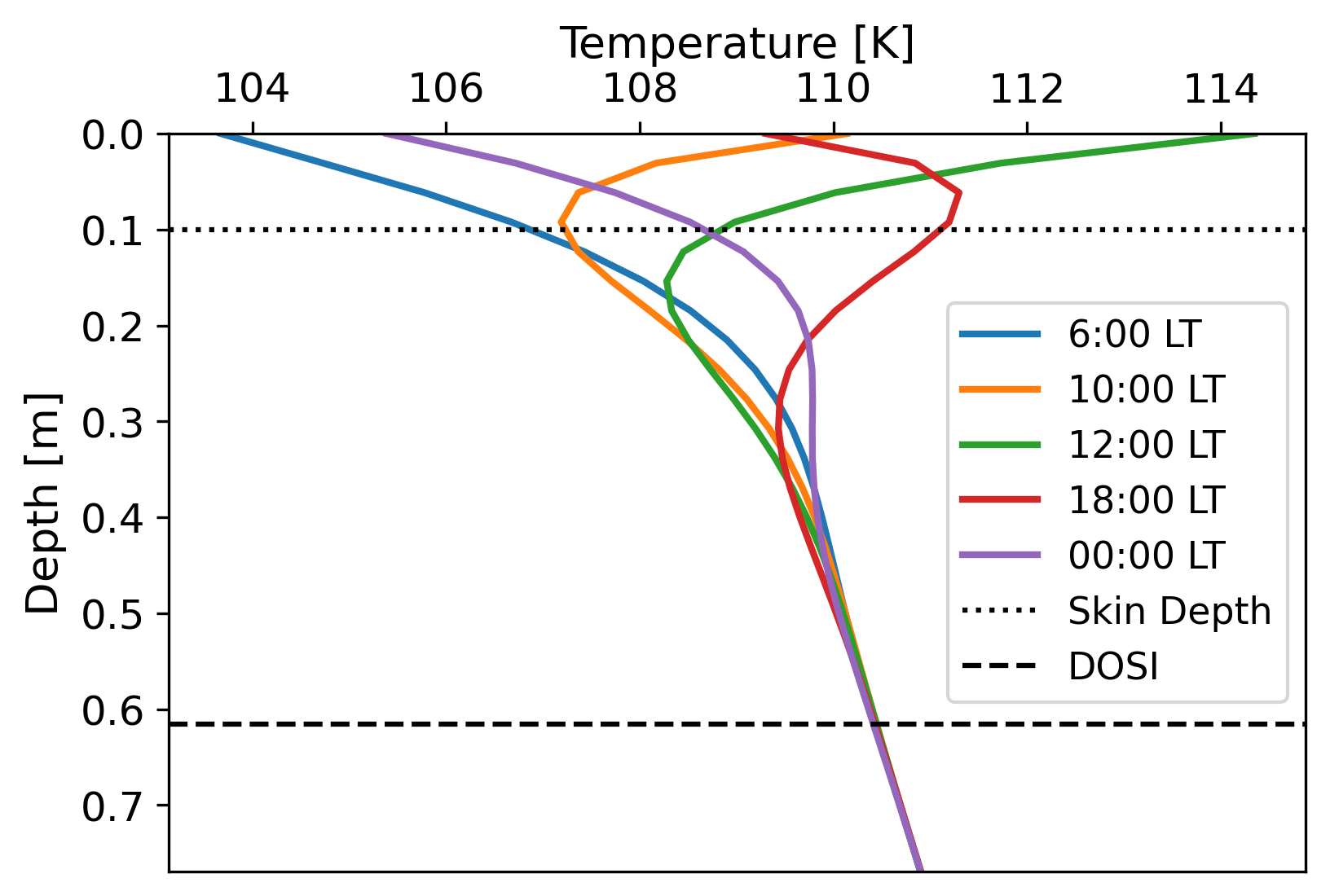}
    \caption{Temperature as a function of depth for different times during one Io day. The heating of the surface by insolation influences the subsurface up to a depth of $\sim$0.6 m (dashed line). This depth is referred to as DOSI (depth of solar influence). Further below, the temperature of the subsurface is controlled by internal heat sources only. The skin depth (dotted line, see Section \ref{sec_SkinDepth}) is at 0.1 m.}
    \label{plot_skin_depth}
\end{figure}

Additional to the surface temperatures, our model also provides values of subsurface temperatures. 
To characterize the heat diffusion through Io's subsurface layer, we define the following two measures:\\ 
First, we introduce a skin depth $d$. Applying an Ansatz of the form 
\begin{equation}
    T(z,t) = T_{(z=0)} \exp\left[i\left(\frac{t}{2\pi\Tilde{T}_{Io}}\right)\right]\exp\left(-\frac{z}{k}\right)
\end{equation}
within the simplified heat equation (Equation \ref{energy_equation}) without production and loss terms with k the complex wave number $k=\sqrt{\frac{i\omega}{\alpha}}$ and assume that the skin depth requires the surface temperature to decrease by a factor of $\frac{1}{e}$ (at depth $z=d$), the skin depth is given as
\begin{align}\label{skindepth}
    d=\sqrt{\frac{\Tilde{T}_{Io}\alpha}{\pi}}
\end{align}
with $\Tilde{T}_{Io}$ is the length of one Io day in units of seconds and $\alpha$ is the thermal diffusivity (see e.g., \citeA{dePater_2015_plansci}). Hence, the skin depth depends on thermal diffusivity $\alpha$ and decreases for decreasing $\alpha$. For the parameter study setup (Section \ref{Basics}) the skin depth varies between $d\approx0.1$ m for the lowest and $d\approx0.5$ m for highest thermal diffusivity value.\\
Figure \ref{plot_skin_depth} presents the vertical temperature profiles at the equator, on the anti-Jovian hemisphere ($\varphi=$ 180° W), assuming a thermal diffusivity of $0.1\alpha_0$ and for one Io day in June 1999, where Jupiter was at perihelion. The shape of the vertical temperature profiles is mainly determined by the heat production at Io's surface, the thermal diffusivity and the internal heat flux. In addition to the skin depth, we also determine the depth beyond which the diurnal temperature variability is not significant any more. This depth is referred to as DOSI (depth of solar influence) in the following. It is determined by making sure that the temperature difference between the different temperature profiles does not exceed 0.1 K here. This is 1 \% of the 10 K diurnal temperature variation at the surface and therefore considered to be adequate to determine the DOSI. For $\alpha=0.1\alpha_0$, the cyclic heating and cooling of the surface influences the subsurface temperatures up to a depth of $\sim$0.6 m (dashed line) in our simulation. Beyond that depth, the temperature gradient is determined by Io's internal heating.
It is worthwhile to point out that at the dusk terminator the shallow subsurface is hotter than the surface (see \citeA{dePater_2015_plansci}). This causes a conductive heat flux towards the surface, which heats the surface although there is no direct heating by the sun. Therefore the surface temperature and column density do not decrease instantaneously after sunset but decrease with a temporal delay lasting throughout the night time.\\
The simulated skin depth (dotted line) is at 0.1 m. Comparing the skin depth and the DOSI (dashed line) shows that the DOSI is higher by a factor of 6 compared to the skin depth. In order to not miss any significant diurnal temperature changes in analyzing the simulation results this fact is used to estimate the modeling domain boundary in the $z-$direction.\\[0.3cm]

\subsection{Long-term Column Density Variation}\label{LongTermVariation}
\begin{figure}[H]
    \centering
    \includegraphics[scale=0.7]{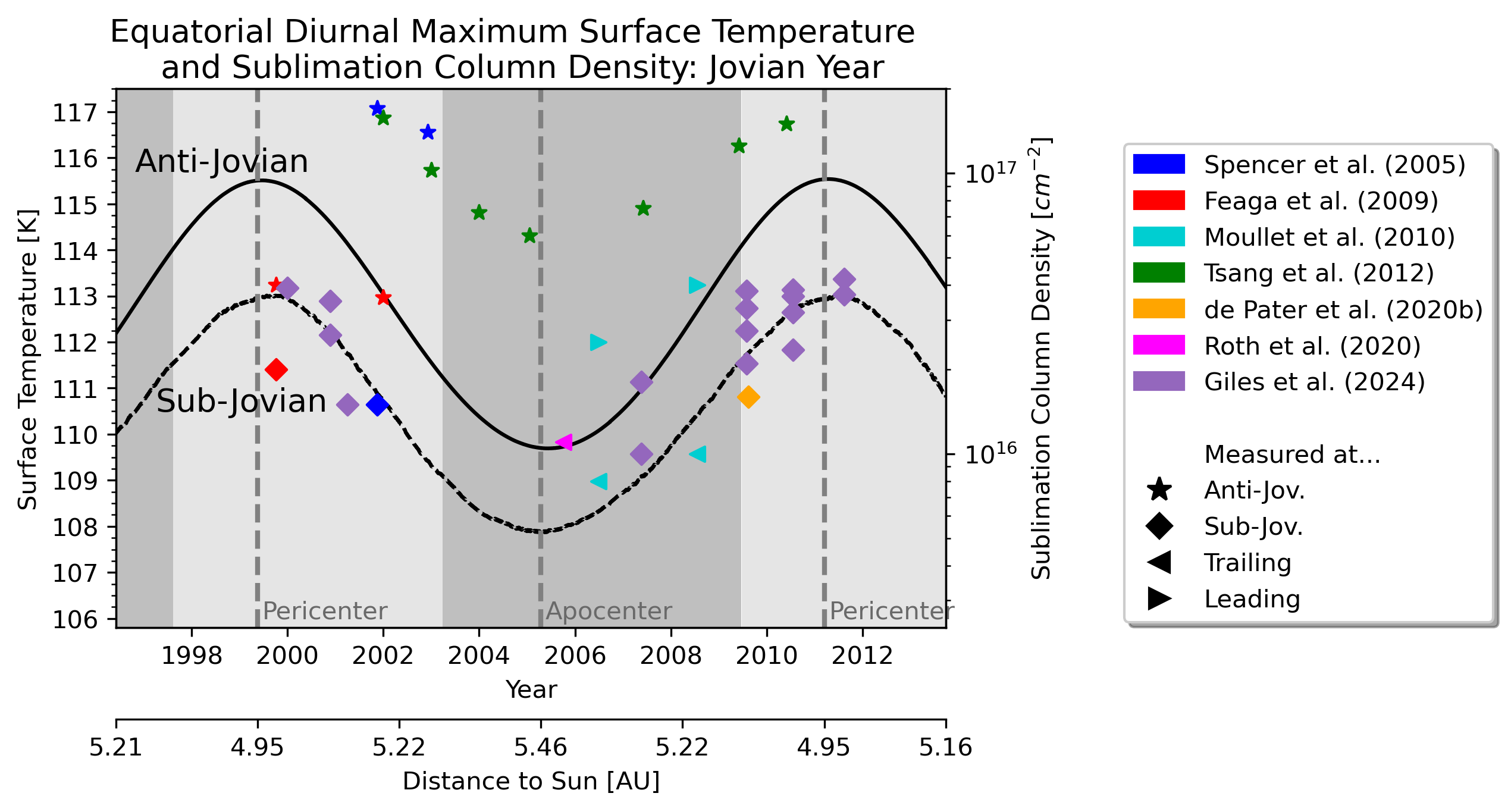}
    \caption{Modeled diurnal maximum surface temperature and corresponding sublimation column density as a function of time between June 1996 and September 2013. Added are also the pericenter/apocenter locations (dashed lines) and observed column densities published by various authors for comparison. Io's northern hemisphere summer and northern hemisphere winter seasons are indicated in light and dark gray.}
    \label{Jovian_year_overview}
\end{figure}

In the last decades, observations of Io's atmosphere have been performed regularly. This Section compares our simulations to observations obtained between June 1996 and September 2013 covering one Jovian year. Therefore, figure \ref{Jovian_year_overview} presents the maximum diurnal surface temperature as well as the corresponding sublimation column density over the time span from June 1996 to September 2013 at the sub-Jovian ($\varphi$= 0° W, dashed line) as well as at the anti-Jovian point ($\varphi$= 180° W, solid line). Also added are the times of perihelion and aphelion (dashed gray lines) that are very close to Io's summer and winter solstices. Io's northern summer and northern winter seasons are indicated in light and dark gray.\\
From aphelion to perihelion Io's distance to the sun decreases from 5.65 AU to 4.95 AU, which corresponds to change of the solar illumination of 45.9 Wm$^{-2}$ to 55.9 Wm$^{-2}$. As expected because of the smaller heat flux in aphelion, the surface temperature decreases from mid 1999 to 2005 by 5 K in the simulation. With regard to Jupiter's global thermal emission that is also included as a heat source at Io's surface (see Equation (\ref{P_rad_Jup}))there is no significant variation expected on timescales of a Jovian year. On the one hand this is due to the fact, that Jupiter emits more heat than it receives from insolation, meaning that it is mostly heated from internal heat sources that are assumed to have only very minor temporal variability on respective timescales (\citeA{smoluchowski_1967}; \citeA{guillot_2004}). On the other hand, further characteristics such as Jupiter's band structure and northern polar CH$_4$ emission, that are representative for the thermal structure, undergo relatively strong variations (up to 37\% on the case of CH$_4$ emissions) over one Jovian year (\citeA{antunano_2019}; \citeA{sinclair_2023}). However, those variations are not attributable to the seasonal cycle and occur only locally, leading to the conclusion that we can assume the global thermal emission of Jupiter to be constant over one Jovian year.\\
In the following, we compare our simulation with a set of observations labeled as stars, diamonds and triangles in Figure \ref{Jovian_year_overview} (for measurements on the anti-Jovian, sub-Jovian, trailing and leading hemisphere) and published values for atmospheric densities by various groups are color coded. 
\begin{subsubsection}{Comparison to UV-IR observations} 
\citeA{Feaga2009} use a set of Lyman-$\alpha$ images of Io's dayside atmosphere obtained with the HST Space Telescope Imaging Spectrograph when Io approaches and then passes perihelion between October 1997 and December 2001 (marked red in Figure \ref{Jovian_year_overview}). The authors found the average dayside atmosphere to be densest on the anti-Jovian hemisphere with column densities of $5\times10^{16}$ cm$^{-2}$. In late 1999, the observations show the averaged sub-Jovian hemispheric column densities to be in the range of $\sim$$2\times10^{16}$ cm$^{-2}$ and lower by a factor of 2 compared to the anti-Jovian hemisphere with column densities of $\sim$$4\times10^{16}$ cm$^{-2}$. Looking at data from 2002 to 2004, \citeA{Spencer2005} (labeled in blue in Figure \ref{Jovian_year_overview}) obtained disk-integrated 19-$\mu$m spectra taken with the TEXES high spectral resolution mid-infrared spectrograph and found column densities in the range of $1.5-18\times10^{16}$ cm$^{-2}$ depending on whether observing the sub- or anti-Jovian hemisphere. Compared to \citeA{Feaga2009} their values for the anti-Jovian hemisphere are significantly larger, although the decreasing surface temperature when approaching aphelion would lead to the assumption that the column densities should decrease. Our simulations suggest column densities in the range of $6-10.5\times10^{15}$ cm$^{-2}$ (sub-Jovian) and $1.5-3\times10^{16}$ cm$^{-2}$ (anti-Jovian) during that period of time. Comparing our simulation to the work of \citeA{Feaga2009} it is noticeable that it matches the observed column densities quite well considering that our values are diurnal maxima. The simulated column densities also match the observed values at the sub-Jovian hemisphere found by \citeA{Spencer2005}, although the simulation suggests column densities lower by a factor of 6 on the anti-Jovian hemisphere.
Covering almost one Jovian year, \citeA{Tsang_2012} also evaluated 19 $\mu$m spectra of Io’s atmosphere from the TEXES mid-infrared high spectral resolution spectrograph (green). They found the anti-Jovian column densities to increase by factor of 3 from $5.5\times10^{16}$ cm$^{-2}$ in aphelion to $1.6\times10^{17}$ cm$^{-2}$ in perihelion. The annual variation resulting from the simulations show a column density increasing by almost a factor of 10 from $\sim$$10^{16}$ cm$^{-2}$ at aphelion up to $\sim$$10^{17}$ cm$^{-2}$ at perihelion. The simulation underestimates the observed column density found by \citeA{Tsang_2012} by between 35-75\%. More recently, \citeA{giles_2024} expanded the data set analyzed in \citeA{Spencer2005}, \citeA{Tsang_2012} and \citeA{Tsang2013} by 19-$\mu$m IRTF/TEXES data obtained throughout an additional Jovian year from 2014 to 2023. They found the seasonal pattern observed from 2001 to 2013 to repeat in the following Jovian year. At the anti-Jovian hemisphere the data are well in agreement with a surface bond Albedo of 0.56$^{+0.04}_{-0.03}$, a thermal inertia of 250$^{+100}_{-90}$ MKS and some additional constant (volcanic) component of 0.74$^{+0.09}_{-0.11}$$\times10^{17}$ cm$^{-2}$, with column densities approaching $\sim$$2.1\times10^{17}$ cm$^{-2}$ in perihelion and $\sim$$7\times10^{16}$ cm$^{-2}$ in aphelion. The maximum and minimum surface temperatures are modeled to be 116.5 and 110.4 K shortly after Jupiter being at pericenter and apocenter, respectively. Similar to the \citeA{Tsang_2012} data, our model, which uses a slightly higher surface albedo and thermal inertia (0.62 and 298 MKS), underestimates these column density values by a facor of 2 - 6 depending on whether comparing at perihelion or aphelion, while our minimum and maximum surface temperature values are almost identical to the ones found by \citeA{giles_2024}. Thus our model would need to assume either a lower thermal inertia or an additional constant volcanic component to fit the data with higher agreement. For the first time, \citeA{giles_2024} investigated the seasonal variability also on the sub-Jovian hemisphere, shown as purple diamonds in Figure \ref{Jovian_year_overview}. In our plot the selected times of observations between 2014 and 2023 are matched to the previous Jovian year, which is sufficient because the seasonal pattern of 2001 - 2013 repeats with almost equal column density values as reported in \citeA{giles_2024}. The authors found the column density and surface temperature to decrease from $3.5\times10^{16}$ cm$^{-2}$ and 111.9 K in perihelion to $0.9\times10^{16}$ cm$^{-2}$ and 106.4 K in aphelion assuming a constant component of 0.06$^{+0.02}_{-0.06}$$\times10^{17}$ cm$^{-2}$. In contrast to the anti-Jovian hemisphere our simulation underestimates the column density values found in \citeA{giles_2024} by only a factor of 1.8 for both, perihelion and aphelion, where the surface temperature values almost fully agree again. However, \citeA{giles_2024} derived a comparable albedo of 0.66$^{+0.02}_{-0.16}$ but a thermal inertia of 80$^{+420}_{-20}$ MKS, which has large error bars but is significantly lower compared to our value of 298 MKS. Overall, our simulations show an atmosphere that is significantly thinner when Jupiter's distance to the Sun is maximum consistent with the observations of, e.g., \citeA{moullet_2010}, \citeA{Tsang_2012} and \citeA{giles_2024}.
\end{subsubsection}
\begin{subsubsection}{Comparison to millimeter observations}
Column densities derived from observations at millimeter wavelengths by \citeA{moullet_2010} are shown in cyan in Figure \ref{Jovian_year_overview}. The authors present disk-resolved observations obtained with the Submillimeter Array. Since millimeter observations usually do not cover the sub- or anti-Jovian but the trailing and leading hemispheres, those observations are highlighted differently and are shown as triangles in figure \ref{Jovian_year_overview}. \citeA{moullet_2010} evaluated data from mid 2006 and mid 2008 when Jupiter was moving from its apocenter towards the pericenter. The authors also observe the SO$_2$ column density to increase between the two dates by a factor of $\sim$1.5 for both, the trailing and leading hemisphere. The trailing hemisphere seems to have column densities that are lower by a factor of $\sim$3-4 compared to the leading hemisphere. This difference is similar to the sub- and anti-Jovian hemisphere asymmetry. \citeA{moullet_2010} found the atmosphere to have column densities of $8-10\times10^{15}$ cm$^{-2}$ on the trailing and $2.5-4\times10^{16}$ cm$^{-2}$ on the leading hemisphere. Since the simulation and the millimeter observation cover different hemispheres, a direct quantitative comparison of the modeled and the observed data is not fully possible, however, the simulations match well the observations from mid 2008 when comparing the anti-Jovian and leading as well as the sub-Jovian and trailing hemisphere. Two additional millimeter and well matching observations were published by \citeA{dePater_2020} and \citeA{roth_2020}. \citeA{dePater_2020} concentrated on the sub-Jovian hemisphere and performed observations with the Atacama Large (sub)Millimeter Array in 2018. The authors found the SO$_2$ atmosphere to have column densities of $1.6\times10^{16}$ cm$^{-2}$, which compares well with our simulation results for the sub-Jovian point in late 2009 where Jupiter had the same distance to the sun. Focusing on the anti-Jovian to trailing hemisphere, \citeA{roth_2020} (magenta) observed Io with the Northern Extended Millimetre Array in late 2016 and early 2017 where the sun distance was $\sim$5.45 AU. The observed column densities of $1.1\times10^{16}$ cm$^{-2}$ also match the simulation results for the same distance to the sun in late 2005 almost perfectly. 
\end{subsubsection}\\
In summary, our simulated sublimation-only atmosphere shows some slight differences when comparing with SO$_2$ gas abundances derived from mid-UV and mid-IR data, but matches well with the sub-millimeter and FUV data. It represents the general trend of an atmosphere well varying by more than one order of magnitude over the course of a Jovian year. In particular, the observed sub-Jovian column density abundances (Diamonds in Figure \ref{Jovian_year_overview}) agree well with our simulated sublimation atmosphere implying a high sublimation component on this hemisphere according to our model. The lacking SO$_2$ abundances at the anti-Jovian hemisphere could be balanced by assuming either a (longitudinally varying) lower thermal inertia compared to the sub-Jovian hemisphere or an additional constant (volcanic) component of $0.75-1.1\times10^{17}$ cm$^{-2}$, which is consistent with the constant component derived in \citeA{giles_2024}.

\subsection{Seasonal effects}\label{SeasonalEffects}
\begin{figure}[H]
    \centering
    \includegraphics[scale=0.38]{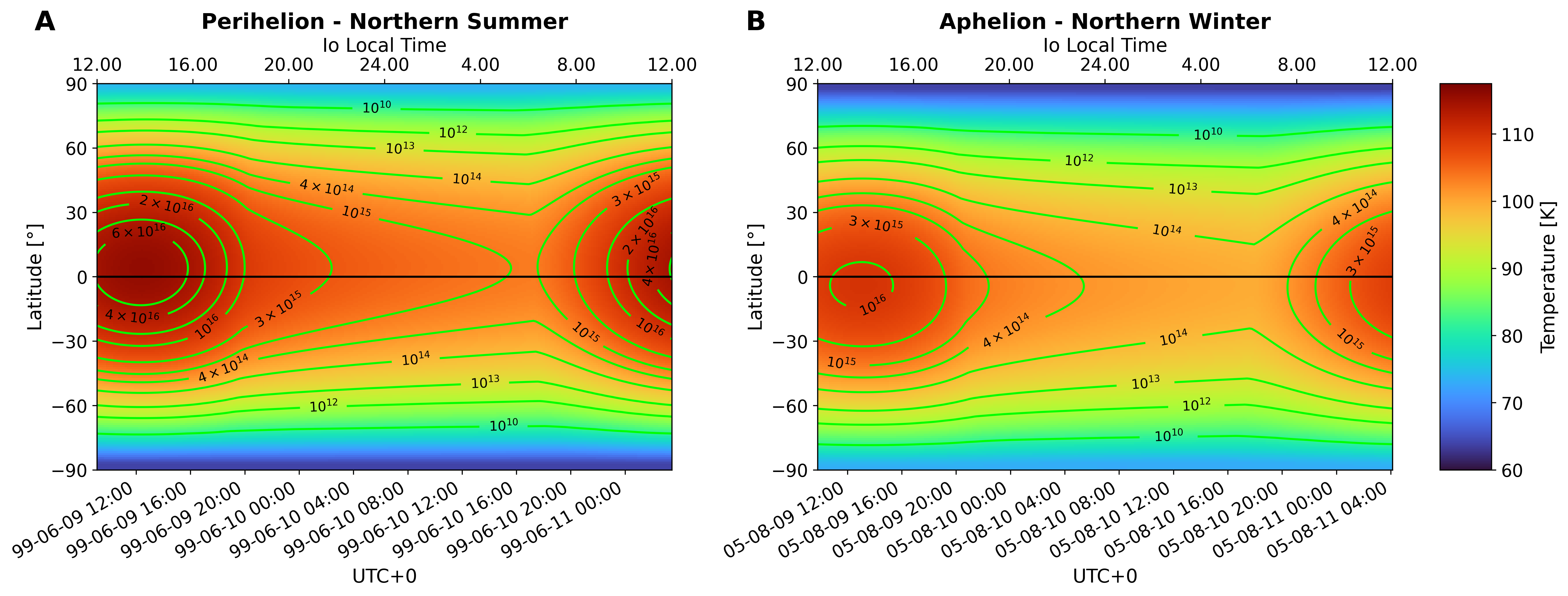}
    \caption{Diurnal surface temperature and column density variation at 
    $\varphi=$ 180° W (anti-Jovian). \textbf{(A)} One day during the northern summer season where Jupiter is near perihelion. \textbf{(B)} One day during southern summer, half a Jovian year ($\sim$6 years) later when Jupiter is near aphelion.}
    \label{seasons}
\end{figure}

Since we use the exact celestial geometry in our model, we are able to consider Io's inclination as well as Jupiter's (and Io's) distance to the sun over timescales of Jovian years. Figure \ref{seasons} shows the diurnal surface temperature and column density variation at one longitude on the anti-Jovian hemisphere ($\varphi=$ 180° W) on a day when Jupiter is near perihelion in 1999 (left panel) and on a day when Jupiter is near aphelion (right panel) 6 years later, in 2005. \\
The simulation indicates that the maximum surface temperature at Jupiter's perihelion is 6 K higher than at aphelion, which corresponds to sublimation column densities that differ by almost one order of magnitude. This effect is visible and was discussed in Figure \ref{Jovian_year_overview}.\\
In Figure \ref{seasons}, the column density contours in the low latitude region illustrate that the surface temperature and column density variation is not fully symmetric around the equator, but shifted towards the north around perihelion (left panel) and towards the south around the aphelion (right panel) because of Io's inclination. As a consequence, the northern hemisphere is $\sim$4 K hotter on average per day compared to the southern hemisphere near perihelion and vice versa near aphelion. Observations showed that near perihelion the column densities are modestly higher at northern mid-latitudes than at the same southern latitudes. The effect was not yet attributed to the inclination of Io \cite{Jessup2004}. Since these characteristics occur in a cycle of one Jovian year, these "seasonal effects" lead to Io having a winter and a summer hemisphere. 

\subsection{Io's warm poles}\label{pole_problem}
\begin{figure}[H]
    \centering
    \includegraphics[scale=0.6]{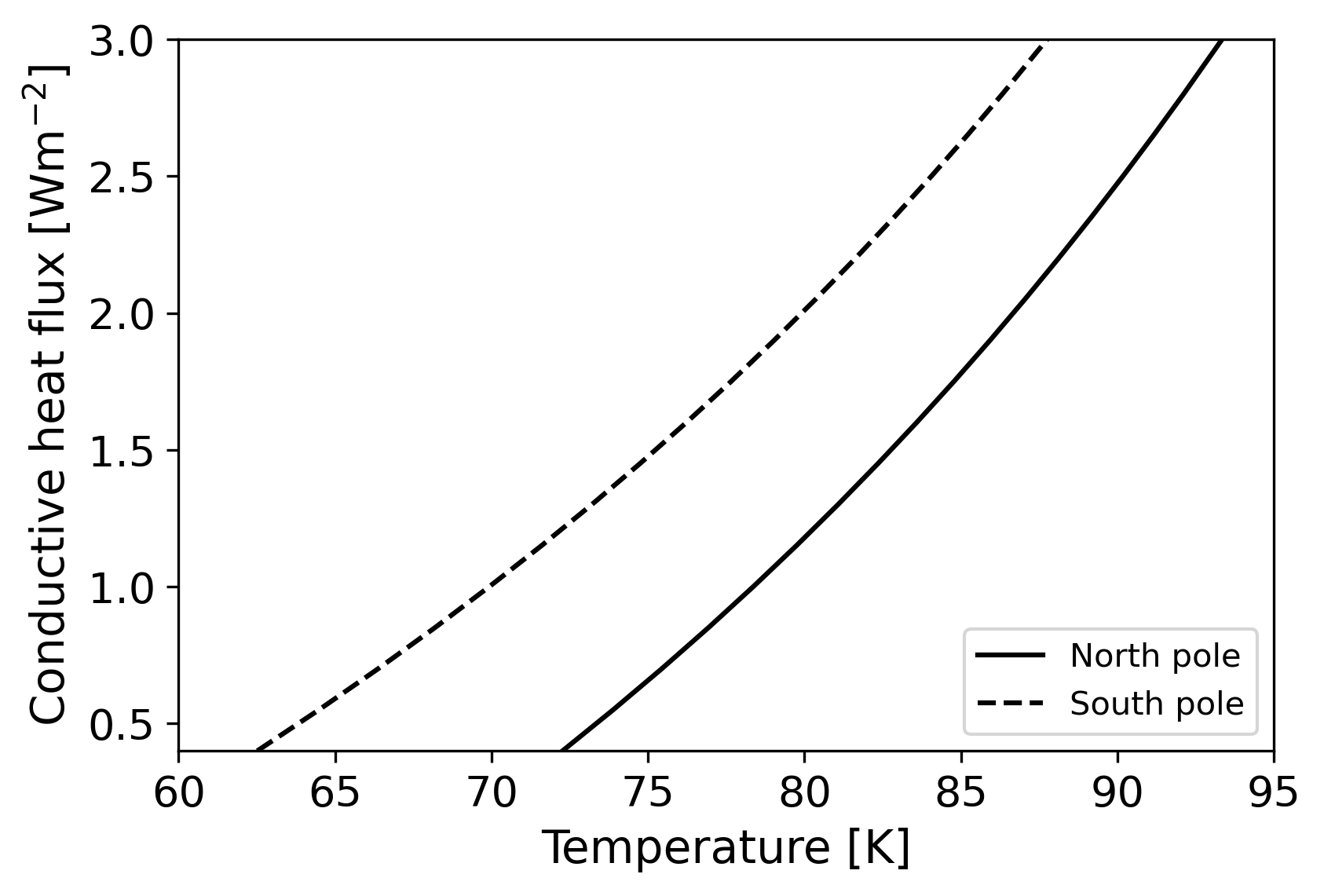}
    \caption{Polar surface temperature (default model) during northern summer for different conductive heat fluxes. To reach a surface temperature of 80 K a conductive heat flux of 1.2 Wm$^{-2}$ at the north pole and 2.1 Wm$^{-2}$ at the south pole is required for the selected season.}
    \label{Param_study_pole_heatflux}
\end{figure}

The simulations presented in this work can explain qualitatively in all cases, and quantitatively in many cases the longitudinal, latitudinal and temporal variation of Io's atmosphere quite well up to the mid-latitudes. Since the high latitudes from 60° to the poles are observed to have dayside surface temperatures of around 75-90 K  \cite{Rathbun2004} the simulations do not fit to observed values. Here, our modeled surface temperature values are in the range of 55-60 K with corresponding column densities of $10^9$ cm$^{-2}$ or less. \\
Close to the poles the zenith angle is close to 90°, therefore solar illumination and other external radiation sources are not sufficient to raise the surface temperature to values on the order of 75-90 K. In general, it is not known how Io's internal (conductive) heat flux is distributed. Due to the polar solar illumination geometry, the albedo does not affect the surface heating due to insolation at Io's poles significantly, so that even with an albedo of $A=0$ our simulation yields polar surface temperatures of $\sim$ 64 K on the winter hemisphere (cf. Equation \ref{Psol}).
This is why we can conclude that a uniformly distributed internal heat flux of 0.48 Wm$^{-2}$ (cf. Section \ref{Model}) does not match observations and there has to be an additional, most likely internal, heat flow. Assuming for a basic estimate that there is no surface heating from external heat sources and an equilibrium between internal heat production and heat loss at the poles, a conductive heat flux 2.1 Wm$^{-2}$ leads to polar surface temperatures of 80 K, which is 4 four times our initially assumed value of 0.48 Wm$^{-2}$. This calculation is confirmed by a parameter study presented in Figure \ref{Param_study_pole_heatflux} which shows the simulated surface temperature of Io's north and south pole during northern summer resulting from our default model with varying conductive heat fluxes between 0.4 and 3 Wm$^{-2}$. As expected, conductive heat fluxes of 1.2 Wm$^{-2}$ for the summer and 2.1 Wm$^{-2}$ for the winter hemisphere are necessary to reach a polar surface temperature of 80 K.\\
Although \citeA{Davies2023} found that the thermal emission of the volcanoes located near Io's north pole is higher than those at the south pole, the higher surface temperature of the northern polar in our simulation is only attributed to Io's seasons as discussed in Section \ref{SeasonalEffects}. \citeA{Veeder_2015} concluded that $\sim$50 \% of Io's internal heat, which would be 1.2 Wm$^{-2}$ for our assumptions, may not be attributed to known hot spots. Comparing that to the parameter study presented in Figure \ref{Param_study_pole_heatflux}, it might well be possible that at least in the polar regions 50\% of Io's internal heat flux is conductive. \\ 

\begin{figure}[H]
    \centering
    \includegraphics[scale=0.36]{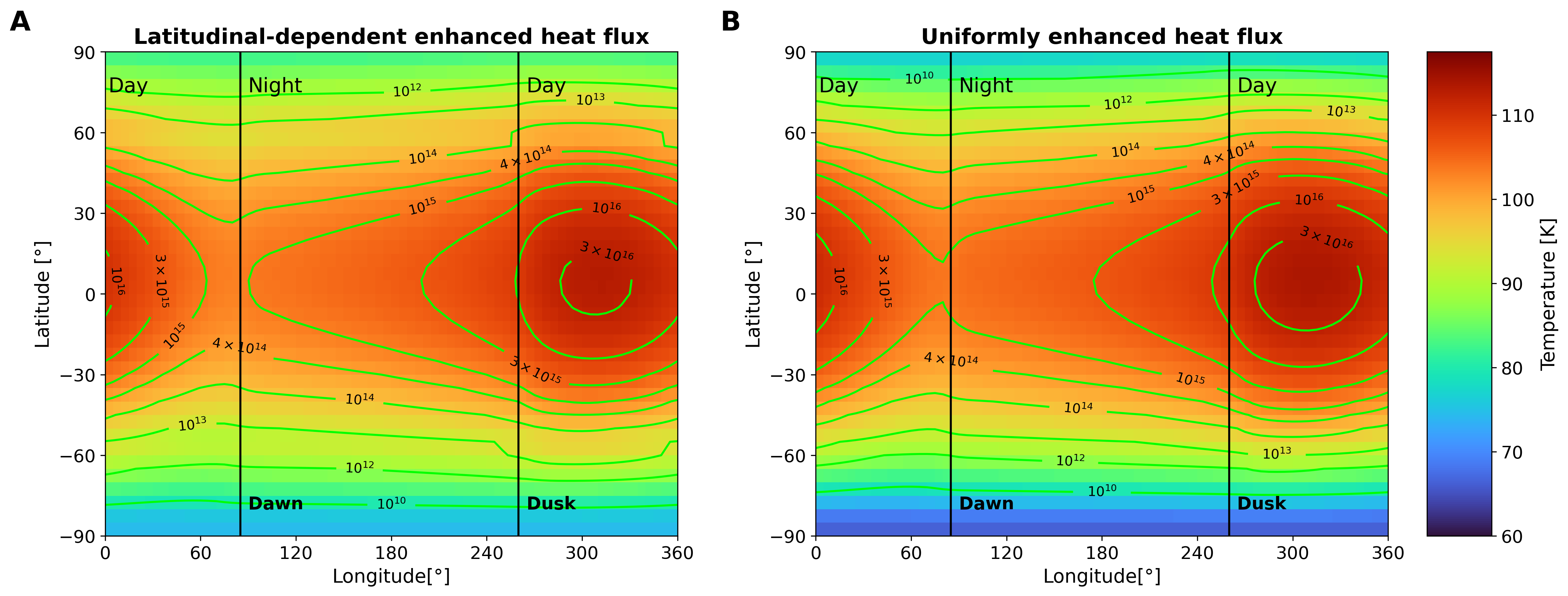}
    \includegraphics[scale=0.44]{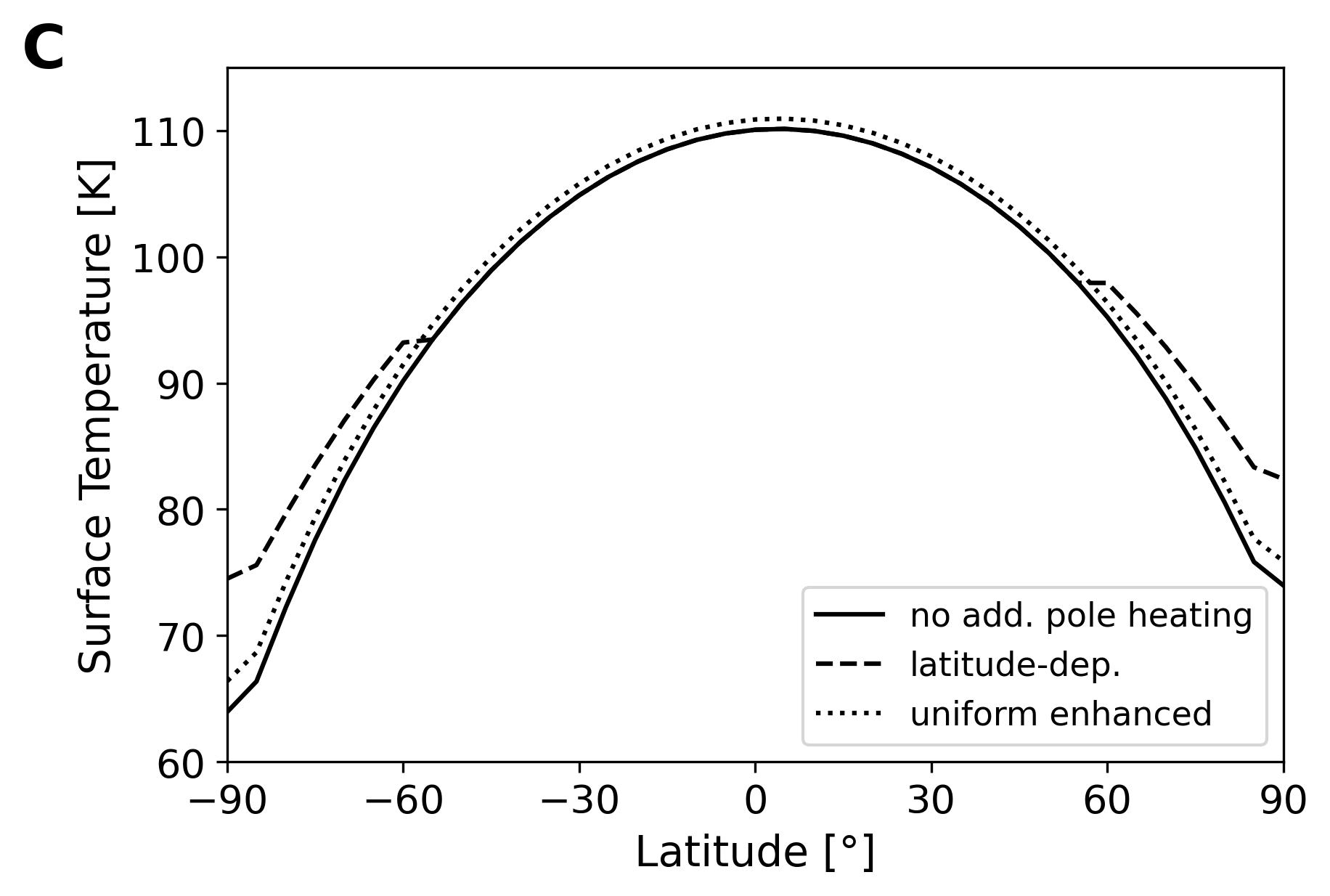}
    \includegraphics[scale=0.44]{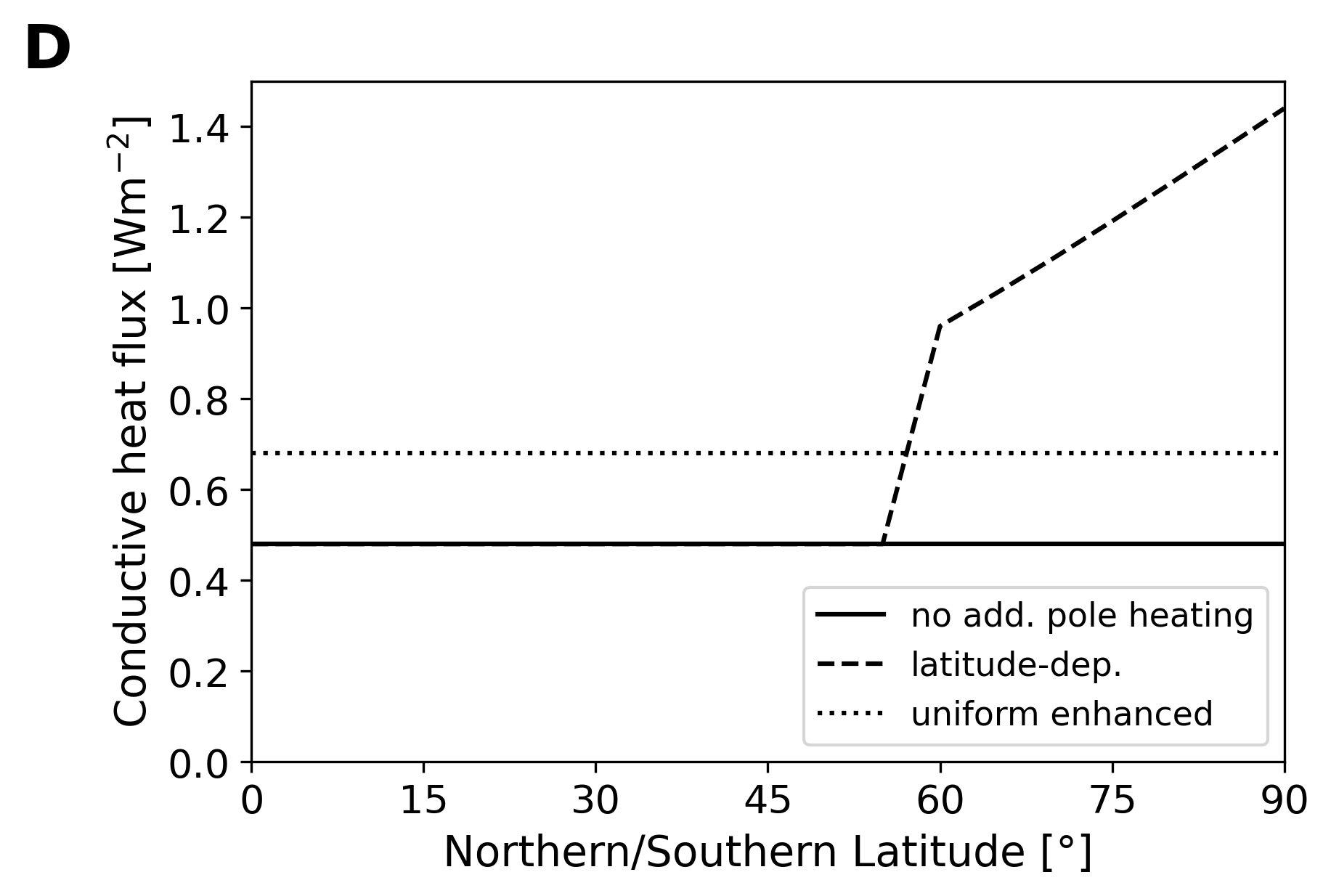}
    \caption{(top) Global surface temperature and column density map of Io's surface with the subsolar longitude occuring at $\varphi$=0$^\circ$ and the sub-Jovian hemisphere being fully illuminated. The simulations were performed for Jupiter (and Io) at perihelion. The internal heat fluxes were assumed to be \textbf{(A)} not uniformly distributed and \textbf{(B)} uniformly distributed but higher by 1 Wm$^{-2}$ leading to a global heat flow of 3.4 Wm$^{-2}$. (bottom) Corresponding \textbf{(C)} latitudinal variation of the sub-solar surface temperature at $\varphi=0$° (sub-Jovian hemisphere, dayside) for $\alpha=0.1\alpha_0$ with $\alpha_0=2.41\times10^{-6}$m$^2$s$^{-1}$ and \textbf{(D)} latitudinal variation of the conductive heat flux assuming no additional pole heating (bold line), a latitude-dependent heat flux (dashed line) and a uniformly enhanced heat flux (dotted line).}
    \label{warm_poles_plot}
\end{figure}

\citeA{Veeder2004_polar} suggested that volcanically active polar regions lead to an additional heat flux of $\sim$0.6 Wm$^{-2}$ to Io's global heat budget of $\sim$2.5 Wm$^{-2}$ and therefore could explain Io's comparatively warm poles. In the following we investigate whether our simulations lead to a comparable result. Therefore, we discuss two different scenarios: First, we assume a conductive heat flux that is not uniformly distributed, while the global heat flux stays at 2.4 Wm$^{-2}$ corresponding to $\sim$1$\times10^{14}$ W in total. In the simulations presented up to this point, we assumed 80\% of Io's internal heat to be lost through volcanic eruptions at all latitudes even though the volcanoes are observed to be predominantly located at low latitudes (e.g., \citeA{de_Kleer_2019_AJ}). We now assume a higher percentage of the internal heat flux to be conductive at the high latitudes. This scenario is referred to as "latitude-dependent" case in the following. The conductive heat flux is assumed to increase polewards from 0.48 Wm$^{-2}$ around the equator up to 1.44 Wm$^{-2}$, which corresponds to the assumptions that 60\% of the global heat flux is transported through conduction in the polar regions ($>$ 60° latitude). Secondly, we perform simulations assuming a total internal heat flux that is higher by 1 Wm$^{-2}$ but still uniformly distributed leading to a global heat flux of 3.4 Wm$^{-2}$ 
in total, and in the range of the total heat flux of 3$\pm$1 Wm$^{-2}$ derived by \citeA{Veeder2004_polar}. The conductive heat flux is then assumed to be still 20\% of the total internal heat flux, or 0.68 Wm$^{-2}$.
The simulation results for both cases as well as the surface temperature as a function of latitude for Io local noon at the prime meridian are presented in Figure \ref{warm_poles_plot}. To gain a better understanding of the different scenarios we also added the conductive heat fluxes as a function of latitude in Figure \ref{warm_poles_plot}.
\subsubsection{Uniformly enhanced heat flow}
In the equatorial and low latitude regions, the enhanced internal heat flux in the uniform case does not affect the surface temperature significantly since the total heat flow from solar radiation dominates. Looking at the poles, the simulation suggests that a uniformly distributed additional heat flux of 1 Wm$^{-2}$ is not enough to increase the surface temperature close to the observed values. To increase the simulated surface temperature in the polar regions with a uniformly distributed heat flux within the observed range of 75-90 K, it needs to be increased to 10.5 Wm$^{-2}$ and $4.4\times10^{14}$ W in total, under the condition that we still assume only 20\% of that heat flux to be conductive. This highly enhanced internal heat flux exceeds the expected range of internal heat due to tidal heating of $1.5-4$ Wm$^{-2}$ (\citeA{McEwen2004}; \citeA{Rathbun2004}; \citeA{Veeder2004_polar}) by more than a factor of 2 and it would also affect the equatorial regions such that the surface temperature and corresponding column densities would increase to unrealistic values; therefore, we consider this scenario to be less likely.
\subsubsection{Latitudinal-dependent enhanced heat flow}
In contrast to that, the enhanced heat flux at the higher latitudes for the latitude-dependent case scenario leads to simulated surface temperatures in the range of 75-85 K depending on whether looking at the northern or southern (summer or winter) hemisphere. These values are in the range of the observed 75-90 K \cite{Rathbun2004}.\\
In the equatorial regions we still assume 80\% of the internal heat to be lost through non-conductive processes. The surface temperature values here and in the low latitude regions are therefore the same or comparable to our default case and the uniformly enhanced case. In both cases, the latitudinal-dependent and the uniformly enhanced case, the diurnal variation of the surface temperature in the very high latitude to polar regions decreases to a minimum or vanishes completely, due to the missing solar insolation. \\
We can conclude at this point that the observed polar surface temperature according to our model is consistent with larger heat fluxes at the poles compared to the equator.
\subsubsection{Internal heating patterns}
The distribution of Io's internal (conductive) heat flux is assumed to be strongly dependent on whether the dissipation concentrates in the deep mantle or asthenosphere. Deep mantle heating models show a maximum heat flux in the polar regions and a minimum around the equator (\citeA{Segatz_1988}; \citeA{Hamilton2013}; \citeA{deKleer2019}). In contrast, the models of asthenospheric heating predict the minimum heat flux at the poles (\citeA{Segatz_1988}; \citeA{Hamilton2013}; \citeA{deKleer2019}; \citeA{kervazo_2022}). Another fact that is likely to change the heat flux pattern is whether or not Io has a subsurface magma ocean. If so, it is to be expected from models that, comparable to the asthenospheric heating, the heat flux is maximum in the equatorial region (\citeA{Tyler_2015}; \citeA{Hamilton2013}; \citeA{deKleer2019}; \citeA{Matsyama_2022}). In a recent work by \citeA{davies_2024}, the authors obtained a new global map of Io's volcanic thermal emission and point out a slight preference for a global magma ocean model over a heat flow model dominated by shallow asthenospheric tidal heating. \citeA{Khurana_2011} argue that Galileo magnetometer data show an electromagnetic induction signal from a global conductive layer of $\sim$50 km thickness and a rock-melt fraction of 20\% indicating that Io possesses a subsurface magma ocean. However, later works showed that the measured perturbations can also be caused by plasma interaction with a longitudinally asymmetric atmosphere and a magma ocean is not necessarily needed to explain the data \cite{Bloecker_2018}. Furthermore, an analysis of HST images from auroral spot oscillations does not reveal a phase shift that would be expected if induction in a highly conductive near-surface layer is present \cite{Roth_2017}. More recently \cite{park_2025} used Juno data to determine Io's $k_2$ Love number to be 0.125$\pm$0.047 which is inconsistent with a shallow magma ocean ($\le$ 500\,km from the surface).\\ 
Our simulation results of the latitude-dependent case support the theory of heating predominantly within the deep mantle and therefore higher heating rates in the polar regions. However, this work does not aim to address the existence of a subsurface magma ocean.

\section{Conclusions}
In this work we develop a surface temperature model to describe Io's sublimation atmosphere. It includes as heat sources time-dependent solar illumination, constant internal heat fluxes, the thermal radiation as well as the sunlight reflected from Jupiter, heat losses due to thermal radiation and considers time-dependent heat flux from the surface into subsurface layers and vice versa. Using SPICE Kernels, we calculated the subsurface temperatures as a function of latitude, longitude and time. We include assumptions about Io's thermal diffusivity as well as the exact celestial geometry to make sure that effects caused by the geometry such as the eclipse by Jupiter are also considered. Assuming initially a thermal diffusivity of $\alpha_0=2.41\times10^{-6}\,\text{m}^2\text{s}^{-1}$ (cf. \citeA{Leone_2011}), we perform a parameter study to determine a default thermal diffusivity by comparing our simulation results to observations. The large scale characteristics of Io's atmosphere that we describe in this work such as the latitudinal dependence (e.g., \citeA{Feldman2000}; \citeA{StrobelWolven2001}), the sub-anti Jovian hemisphere and day-night asymmetry (e.g., \citeA{Jessup2004}; \citeA{Feaga2009}) and the long-time variation (\citeA{Tsang_2012}) were discovered before and also partly attributed to the sublimation part of the atmosphere. In addition to identifying hemispheric season response to Io's inclination and heliocentric distance, our model of a purely sublimation driven atmosphere matches the cyclical cadence and order of magnitude of long-term spatial and temporal SO$_2$ abundance variability as well as variations of its global atmospheric structure on short time scales observed at Io.\\
We summarize our findings as follows: 
\begin{itemize}
    \item[1.] We found that a range of thermal diffusivities within $\frac{1}{12}\alpha_0\leq\alpha\leq\frac{1}{8}\alpha_0$ with $\alpha_0=2.41\times10^{-6}\,\text{m}^2\text{s}^{-1}$ (272-333 MKS in terms of thermal inertia)  is sufficient to explain most of the surface temperature and column density observations. We determine the default value of the thermal diffusivity, which fits the observed diurnal variation and latitudinal structure the best to be lower by a factor of 10 compared to our reference case $\alpha_0$. It corresponds to a thermal inertia value of 298 MKS.
    \item[2.] The atmosphere has been observed to be significantly denser on the anti-Jovian hemisphere when compared to the sub-Jovian hemisphere. This is consistent with missing eclipses on the anti-Jovian hemisphere, which causes this hemisphere to be exposed $\sim$2 hrs longer to sunlight every day and results in maximum SO$_2$ column densities that are larger by a factor of 4 according to our simulations. 
    \item[3.] The atmosphere is observed to be centered around an extended equatorial region and its column density decreases with latitudes. With our model we are able to reproduce a column density decrease from $1.5\times10^{16}$ cm$^{-2}$ at the equator to $3\times10^{14}$ cm$^{-2}$ at $\sim$50° latitude longitudinally averaged with our sublimation driven atmosphere.
    \item[4.] During the eclipse by Jupiter, Io's atmosphere is observed to collapse by almost one order of magnitude within $\sim$10 min. Our simulation found the atmosphere to collapse on similar timescales with decreasing equatorial column density values from $2.5\times10^{16}$ cm$^{-2}$ shortly before eclipse ingress to $1.7\times10^{15}$ cm$^{-2}$ shortly before egress. Therefore, we can consider 10 minutes to be the surface temperature's response time to the vanishing solar heat flux. 
    \item[5.] Due to the thermal inertia of the subsurface materials the diurnal maximum surface temperature occurs not at noon but at 15.00 LT on the sub-Jovian hemisphere and at 13.00 LT on the anti-Jovian.
    \item[6.] Since we  calculate subsurface temperatures as a function of time we investigate the influence of Io's diurnal temperature variation to determine the depth of solar influence (DOSI). We found the DOSI to be at $\sim$0.6 m depth. The skin depth, which requires the surface temperature to decrease by a factor of $\frac{1}{e}$ (at depth $z=d$ from the surface), is at 0.1 m for the thermal diffusivity value of $\frac{1}{10}\alpha_0$ with $\alpha_0=2.41\times10^{-6}\,\text{m}^2\text{s}^{-1}$.
    \item[7.] Simulating the surface temperature and corresponding column density over Jovian years, we found that the surface temperature varies by 5 K from perihelion to aphelion, with column densities decreasing/increasing by a factor of 7.
    \item[8.] We found that due to Io's inclination its atmosphere has hemispheric seasons, additional to the helio-distance driven surface temperature variability occuring on $\sim$decadal periodic timescales. The northern summer is near the perihelion and the northern winter is near the aphelion. We found the respective summer hemisphere to be on average 4 K hotter compared to the winter hemisphere. 
    \item[9.] In our simulation, the thermal radiation from Jupiter as well as the sunlight reflected from Jupiter were included but have a minor contribution of $\sim$3 K to the equatorial surface temperature at the anti-Jovian point. However, the contribution from Jupiter increases the corresponding $SO_2$ column density by a factor of $\sim$3.5 at times of maximum impact, which is during the night.
    \item[10.] We confirm the suggestion that Io's (active) volcanoes have only a small spatial scale perturbation to the atmosphere, e.g. modeled by \citeA{SaurStrobel2004}. The observed large scale spatial and temporal characteristics are well explained with the simulated sublimation driven atmosphere. The comparison of our simulation results to available observations suggests that volcanic contributions do not override the fundamental spatial and temporal patterns produced by a sublimation driven atmosphere.  
    \item[11.] Lastly, we found that Io's anomalously warm poles could be explained by an internal conductive heat flux that is higher by a factor of 3 compared to the initially assumed 0.48 Wm$^{-2}$, at least in the high latitude regions. According to our simulations, an enhanced globally averaged total internal heat flux of 3.4 Wm$^{-2}$ is not sufficient to explain Io's poles being hotter than expected by only internal heat sources.
\end{itemize}
With the model used in this work we are able to explain large scale spatial and temporal characteristics of Io's sublimation atmosphere simultaneously. Nevertheless, some features, such as the detailed physical properties of the surface and the near-surface materials remain unresolved with our model. To only mention a few, Io's subsurface probably has gradients in thermal inertia and density with depth \cite{dePater_2020}, which we treated to be homogeneous. Other than that, the model used in this work does not take into account any spatial variability of SO$_2$ ice properties which could affect the surface albedo as well as the sublimation rate itself. And as mentioned throughout the paper, the model does not include any volcanic impact at all at this point. All these and additional extensions remain part of possible future studies.

\newpage

\appendix 
\section{}\label{Appendix_A}
\subsection{Evolution equation for the surface temperature}\label{A1-Evolutionequation}
Here we derive the evolution equation for the surface temperature (Equation \ref{energy_equation}) from the general form of the internal energy equation in a solid.\\
The change of heat can be expressed as the divergence of the thermal diffusion flux and is given as 
\begin{equation}\label{A1}
    \frac{\partial}{\partial t}\left(\rho c_p T(r,t,\theta,\varphi)\right)=\nabla \cdot \left(\kappa\nabla T(r,t,\theta,\varphi)\right) + P - L
\end{equation}
with $P$ and $L$ production and loss terms. The general heat equation describes the change of a materials temperature $T$ and includes the mass density $\rho$, the specific heat capacity $c_p$ and the thermal conductivity $\kappa$. In our case, $T$ is a function of time $t$ and spherical coordinates $r$, latitude $\theta$ and longitude $\varphi$.
Assuming  $\rho$, $c_p$ and $\kappa$ to be constant and writing the resulting Laplace operator in spherical coordinates leads to:
\begin{eqnarray}\label{A2}
    \frac{\partial}{\partial t}T(r,t,\theta,\varphi)=\frac{\kappa}{\rho c_p}&\Biggl[&\frac{1}{r^2}\frac{\partial}{\partial r}\left(r^2\frac{\partial}{\partial r}T(r,t,\theta,\varphi)\right)+\frac{1}{r\sin\theta}\frac{\partial}{\partial\theta}\left(\frac{1}{r}\sin\theta\frac{\partial}{\partial\theta}T(r,t,\theta,\varphi)\right)\nonumber\\
    &+&\frac{1}{r\sin\theta}\frac{\partial}{\partial\varphi}\left(\frac{1}{r\sin\theta}\frac{\partial}{\partial\varphi}T(r,t,\theta,\varphi)\right)\Biggr]+\frac{1}{\rho c_p}\left(P-L\right).
\end{eqnarray}
Now we transform radial distance $r$ to depth $z$ with
$z=R_{Io}-r$ and $R_{Io}$ is Io's radius.
Consequently, it holds that $\frac{\partial}{\partial r}T(r,t,\theta,\varphi)=-\frac{\partial}{\partial z}T(z,t,\theta,\varphi)$. Furthermore, there is no significant surface temperature heat production and loss variability on scales of a few meters in horizontal direction. Therefore the resulting lateral conductive heat fluxes are orders of magnitude lower compared to the vertical and can be neglected the $\theta$- and $\varphi$-direction in equation (\ref{A2}). The evolution equation for the temperature is now only depending derivatives of the depth $z$:
\begin{eqnarray}\label{A3}
    \frac{\partial}{\partial t}T(z,t,\theta,\varphi)=\frac{\kappa}{\rho c_p}&\Biggl[&\frac{1}{(R_{Io}-z)^2}\left(-\frac{\partial}{\partial z}\right)\biggl[(R_{Io}-z)^2\left(-\frac{\partial}{\partial z}T(z,t,\theta,\varphi)\right)\biggr]\Biggr]\nonumber\\
    &+&\frac{1}{\rho c_p}\left(P-L\right).
\end{eqnarray}
For reasonable values of the thermal diffusivity and the temporal evolution of the sources and sinks within one Io day, the heat flux is spatially dependent only within depth $|z|\ll R_{Io}$. Thus we can approximate $R_{Io}-z\approx R_{Io}$ and (\ref{A3}) reduces to
\begin{equation}
    \frac{\partial}{\partial t}T(z,t,\theta,\varphi)=\frac{\kappa}{\rho c_p}\left(\frac{\partial^2}{\partial z^2}T(z,t,\theta,\varphi)\right)+\frac{1}{\rho c_p}\left(P-L\right).
\end{equation}
Lastly, we introduce the thermal diffusivity $\alpha=\frac{\kappa}{\rho c_p}$ which yields the final form of the energy equation we use in our simulations:
\begin{equation*}
    \frac{\partial}{\partial t}T(z,t,\theta,\varphi)=\alpha\frac{\partial^2}{\partial z^2}T(z,t,\theta,\varphi) + \frac{1}{\rho c_p} \left(P - L\right).
\end{equation*}
The production and loss terms are specified in detail in Section \ref{Model}.
\subsection{Boundary conditions}\label{A2boundary_conditions}
To obtain a complete set of equations that can be solved numerically we need boundary conditions for both, the outer boundary at Io's surface ($z=0$) and the inner boundary at a certain depth ($z=z_{low}$). Considering that the most of the heat production and the heat loss processes only act on Io's surface, the outer boundary condition is given as\\
\begin{equation}\label{innerbc}
    \frac{\partial}{\partial z}T(z,t,\theta,\varphi)=-\frac{1}{\kappa}\left(P_{sol}+P_{rad,Jup}+P_{ref,Jup}\right)+\frac{1}{\kappa}L
\end{equation}
with $P$ and $L$ production and loss terms, temperature $T$, time $t$, latitude $\theta$, longitude $\varphi$ and thermal conductivity $\kappa$. The heat production due to Io's tidal heating is only acting at an inner boundary $z=z_{low}$ such that the inner boundary condition reads as
\begin{equation}\label{outerbc}
    \frac{\partial}{\partial z}T(z,t,\theta,\varphi)=\frac{1}{\kappa}P_{int}.
\end{equation}
The production and loss terms are specified in detail in Section \ref{Model}.
\subsection{Model limitations}\label{A3-Limitations}
The main goal of this work is to reproduce the temporal variability and spatial structure of Io's atmosphere using a simple 1-D heat diffusion equation that includes the most important heat production and loss processes and has only very few free parameters principally the thermal inertia and later in the study also the internal heat flux. Therefore, the model neglects effects contributing to Io's surface temperature and column densities of its atmosphere, which we discuss in the following.
\begin{itemize}
    \item Latent heat:\\ Latent heat due to sublimation and condensation is not included in our model. \citeA{Walker_2012} determined these to be negligible compared to other heat production and loss processes and to not contribute significantly to Io's surface temperature and column density variation.
    \item Lateral heat fluxes:\\ In our 1-D model, we only include vertical heat fluxes from Io's surface to deeper structures (and vice versa) and neglect any heat fluxes in lateral direction. This assumption is justified based on the  small vertical compared to the large horizontal spatial scales (see also Appendix \ref{A1-Evolutionequation}). Therefore, the 1-D heat transfer model can be applied to each modeled point on Io's surface individually.
    \item Volcanic component:\\ Io's volcanoes are assumed to contribute to the atmosphere only locally and only if there is an eruption present \cite{dePater_2020}. Eclipse observations suggest that the averaged volcanic column density is on the order of $\sim$10$^{14}$ cm$^{-2}$ and therefore approximately one order of magnitude lower when compared to the sublimation driven column density of Io's dayside atmosphere \cite{SaurStrobel2004}. The proportion of volcanic and sublimation support may also vary over one Jovian year, with sublimation being the dominant source close to perihelion and the volcanic component to be dominant during aphelion \cite{giles_2024}. \citeA{Tsang2013} concluded that the volcanic component is in the range of $\sim$5$\times$10$^{16}$ cm$^{-2}$ and therefore may become the dominant source of SO$_2$ also on shorter timescales when the insolation decreases, e.g., on the nightside or during eclipse. 
    \item Surface sputtering:\\ Surface sputtering is considered to be a significant source for alkali salts NaCl and KCl \cite{roth_2025}, but it is generally not considered to significantly add to the mass density of Io's SO$_2$ atmosphere, except possibly on the nightside or during eclipse and also at high latitudes with vanishing solar illumination or/and and at locations where active plumes are absent and therefore the sublimation support or the volcanic impact gets small (\citeA{lellouch_2005}; \citeA{dePater2023IoBook}). While sputtering plays an important role in driving the atmospheres of other Galilean moons such as Europa and Ganymede, for Io a sputtering atmosphere is self-limited to a column density of $\sim$10$^{16}$ cm$^{-2}$ and is not in accordance with observations from, e.g., Voyager (e.g., \citeA{lellouch_2005}).
    \item Winds:\\ The influence of winds on the structure of the atmosphere is complex and difficult to assess quantitatively without numerical modelling. Pressure driven winds are expected to smear out vertical density variations, but day night and equatorial vs polar gradients are still maintained (e.g., \citeA{ingersoll_1985}; \citeA{moreno_1991}; \citeA{austin_2000}). In addition to that, \citeA{saur_2002} suggest that the magnitude of the drag forces due to the fast moving plasma constantly overtaking Io to be on the same order as those of the pressure gradients. \cite{thelen_2024} support that theory by showing clear effects of plasma flows on the SO$_2$ and NaCl winds in Io's atmosphere. However, since the aim of this work is to keep the underlying model as simple as possible to explain the observed large scale spatiotemporal structure and variability of Io's atmosphere, this effect exceeds the limits of our model.
    \item Surface properties:\\ The surface is not homogeneous with respect to albedo, 
    surface roughness, rock and ice patterns and the composition of the (sub)surface materials. In general, Io's albedo is expected to have a patchy structure due to the geology being composed of plains, lava flows and volcanoes (\citeA{Williams2023IoBook}). However, the surface frost bond albedo is estimated to have only low variation on the order of $\sim$0.7 (\citeA{veeder_1994}; \citeA{kerton_1996}; \citeA{Rathbun2004}). On large scales, albedo variations show the largest gradient from a bright, extended equatorial belt to darker polar regions (e.g., \citeA{Simonelli2001}). However, as discussed in Section \ref{LatVariation}, at high latitudes the internal heat becomes the dominant heat source to determine the surface temperature, which is independent of the albedo and the reason for us to neglect that latitudinal variability. In our study, we want to focus on the description of global-scale surface temperature and column density variations, which is why we use globally averaged albedo values in our simulations and do not account for the small scale albedo variations. We also assume Io's surface and subsurface to be homogeneous with respect to surface roughness or material composition and assume Io to be uniformly covered with SO$_2$ surface frost. Local deviations from these spatially averaged values are expected to locally modify surface temperatures and column densities. As such the averaged values of surface properties are intended to derive spatially averaged surface temperatures and column density which we compare with hemispheric observations.
    \item Vapor pressure equilibrium assumption: \\ Finally we assume the atmosphere to be in vapor pressure equilibrium with the underlying surface ice to calculate the sublimation atmosphere's column density according to \citeA{Wagman_1979}. In accordance with other items of \ref{A3-Limitations}, this assumption only holds true if the contribution to the atmosphere is mainly due to sublimation, there are no winds due to pressure gradients and one considers sputtering of SO$_2$ surface frost and chemical reactions to be absent \cite{moore_2011}. Furthermore, the atmospheric temperature is expected to react on timescales of several minutes to rapid changes of the surface temperatures (see main text Section \ref{Basics} for a more detailed discussion). Therefore, the simulation results presented in this work do not fully capture the atmospheric perturbations at locations that experience rapid temperature variations at a time due to ingress/egress in/out of eclipse or at the terminators. However, our work does not aim to resolve the detailed atmospheric variations at small timescales of only a few minutes.

\end{itemize}

\newpage

\section*{Open Research Section}
The model used in this work has been implemented with Python, with which the results were obtained as well. The source code is publicly available at GitHub (\url{https://github.com/ACDott/Software}) and in a repository \cite{dott_zenodo} .

\acknowledgments
IdP acknowledges funding from NASA SSW grant 80NSSC24K0306, as a subawardee of the lead campus, University of Texas at Austin.

%
%

\bibliography{io_ref.bib}

\begin{thebibliography}{}

\bibitem [\protect \citeauthoryear {%
Acton%
}{%
Acton%
}{%
{\protect \APACyear {1996}}%
}]{%
SPICE}
\APACinsertmetastar {%
SPICE}%
\begin{APACrefauthors}%
Acton, C\BPBI H.%
\end{APACrefauthors}%
\unskip\
\newblock
\APACrefYearMonthDay{1996}{}{}.
\newblock
{\BBOQ}\APACrefatitle {Ancillary data services of NASA's Navigation and
  Ancillary Information Facility} {Ancillary data services of nasa's navigation
  and ancillary information facility}.{\BBCQ}
\newblock
\APACjournalVolNumPages{Planetary and Space Science}{44}{1}{65-70}.
\newblock
\APACrefnote{Planetary data system}
\newblock
\begin{APACrefDOI} \doi{https://doi.org/10.1016/0032-0633(95)00107-7}
  \end{APACrefDOI}
\PrintBackRefs{\CurrentBib}

\bibitem [\protect \citeauthoryear {%
Antu{\~n}ano%
\ \protect \BOthers {.}}{%
Antu{\~n}ano%
\ \protect \BOthers {.}}{%
{\protect \APACyear {2019}}%
}]{%
antunano_2019}
\APACinsertmetastar {%
antunano_2019}%
\begin{APACrefauthors}%
Antu{\~n}ano, A.%
, Fletcher, L\BPBI N.%
, Orton, G\BPBI S.%
, Melin, H.%
, Milan, S.%
, Rogers, J.%
\BDBL {}Giles, R.%
\end{APACrefauthors}%
\unskip\
\newblock
\APACrefYearMonthDay{2019}{}{}.
\newblock
{\BBOQ}\APACrefatitle {Jupiter’s atmospheric variability from long-term
  ground-based observations at 5 $\mu$m} {Jupiter’s atmospheric variability
  from long-term ground-based observations at 5 $\mu$m}.{\BBCQ}
\newblock
\APACjournalVolNumPages{The Astronomical Journal}{158}{3}{130}.
\PrintBackRefs{\CurrentBib}

\bibitem [\protect \citeauthoryear {%
Austin%
\ \BBA {} Goldstein%
}{%
Austin%
\ \BBA {} Goldstein%
}{%
{\protect \APACyear {2000}}%
}]{%
austin_2000}
\APACinsertmetastar {%
austin_2000}%
\begin{APACrefauthors}%
Austin, J\BPBI V.%
\BCBT {}\ \BBA {} Goldstein, D\BPBI B.%
\end{APACrefauthors}%
\unskip\
\newblock
\APACrefYearMonthDay{2000}{}{}.
\newblock
{\BBOQ}\APACrefatitle {Rarefied gas model of Io's sublimation-driven
  atmosphere} {Rarefied gas model of io's sublimation-driven
  atmosphere}.{\BBCQ}
\newblock
\APACjournalVolNumPages{Icarus}{148}{2}{370--383}.
\PrintBackRefs{\CurrentBib}

\bibitem [\protect \citeauthoryear {%
Bagenal%
\ \BBA {} Dols%
}{%
Bagenal%
\ \BBA {} Dols%
}{%
{\protect \APACyear {2020}}%
}]{%
bagenaldols2020}
\APACinsertmetastar {%
bagenaldols2020}%
\begin{APACrefauthors}%
Bagenal, F.%
\BCBT {}\ \BBA {} Dols, V.%
\end{APACrefauthors}%
\unskip\
\newblock
\APACrefYearMonthDay{2020}{}{}.
\newblock
{\BBOQ}\APACrefatitle {The Space Environment of $\text{Io}$ and
  $\text{Europa}$} {The space environment of $\text{Io}$ and
  $\text{Europa}$}.{\BBCQ}
\newblock
\APACjournalVolNumPages{Journal of Geophysical Research: Space
  Physics}{125}{5}{}.
\newblock
\begin{APACrefDOI} \doi{https://doi.org/10.1029/2019JA027485} \end{APACrefDOI}
\PrintBackRefs{\CurrentBib}

\bibitem [\protect \citeauthoryear {%
Blöcker%
, Saur%
, Roth%
\BCBL {}\ \BBA {} Strobel%
}{%
Blöcker%
\ \protect \BOthers {.}}{%
{\protect \APACyear {2018}}%
}]{%
Bloecker_2018}
\APACinsertmetastar {%
Bloecker_2018}%
\begin{APACrefauthors}%
Blöcker, A.%
, Saur, J.%
, Roth, L.%
\BCBL {}\ \BBA {} Strobel, D\BPBI F.%
\end{APACrefauthors}%
\unskip\
\newblock
\APACrefYearMonthDay{2018}{}{}.
\newblock
{\BBOQ}\APACrefatitle {$\text{MHD}$ Modeling of the Plasma Interaction With
  $\text{Io's}$ Asymmetric Atmosphere} {$\text{MHD}$ modeling of the plasma
  interaction with $\text{Io's}$ asymmetric atmosphere}.{\BBCQ}
\newblock
\APACjournalVolNumPages{Journal of Geophysical Research: Space
  Physics}{123}{11}{9286-9311}.
\newblock
\begin{APACrefDOI} \doi{https://doi.org/10.1029/2018JA025747} \end{APACrefDOI}
\PrintBackRefs{\CurrentBib}

\bibitem [\protect \citeauthoryear {%
Cruikshank%
, Emery%
, Kornei%
, Bellucci%
\BCBL {}\ \BBA {} d’Aversa%
}{%
Cruikshank%
\ \protect \BOthers {.}}{%
{\protect \APACyear {2010}}%
}]{%
cruikshank_2010}
\APACinsertmetastar {%
cruikshank_2010}%
\begin{APACrefauthors}%
Cruikshank, D\BPBI P.%
, Emery, J\BPBI P.%
, Kornei, K\BPBI A.%
, Bellucci, G.%
\BCBL {}\ \BBA {} d’Aversa, E.%
\end{APACrefauthors}%
\unskip\
\newblock
\APACrefYearMonthDay{2010}{}{}.
\newblock
{\BBOQ}\APACrefatitle {Eclipse reappearances of Io: Time-resolved spectroscopy
  (1.9--4.2 $\mu$m)} {Eclipse reappearances of io: Time-resolved spectroscopy
  (1.9--4.2 $\mu$m)}.{\BBCQ}
\newblock
\APACjournalVolNumPages{Icarus}{205}{2}{516--527}.
\PrintBackRefs{\CurrentBib}

\bibitem [\protect \citeauthoryear {%
Davies%
, Perry%
, Williams%
\BCBL {}\ \BBA {} Nelson%
}{%
Davies%
, Perry%
, Williams%
\BCBL {}\ \BBA {} Nelson%
}{%
{\protect \APACyear {2024}}%
}]{%
Davies2023}
\APACinsertmetastar {%
Davies2023}%
\begin{APACrefauthors}%
Davies, A\BPBI G.%
, Perry, J\BPBI E.%
, Williams, D\BPBI A.%
\BCBL {}\ \BBA {} Nelson, D\BPBI M.%
\end{APACrefauthors}%
\unskip\
\newblock
\APACrefYearMonthDay{2024}{}{}.
\newblock
{\BBOQ}\APACrefatitle {Io’s polar volcanic thermal emission indicative of
  magma ocean and shallow tidal heating models} {Io’s polar volcanic thermal
  emission indicative of magma ocean and shallow tidal heating models}.{\BBCQ}
\newblock
\APACjournalVolNumPages{Nature Astronomy}{8}{1}{94--100}.
\newblock
\begin{APACrefDOI} \doi{https://doi.org/10.1038/s41550-023-02123-5}
  \end{APACrefDOI}
\PrintBackRefs{\CurrentBib}

\bibitem [\protect \citeauthoryear {%
Davies%
, Perry%
, Williams%
, Veeder%
\BCBL {}\ \BBA {} Nelson%
}{%
Davies%
, Perry%
, Williams%
, Veeder%
\BCBL {}\ \BBA {} Nelson%
}{%
{\protect \APACyear {2024}}%
}]{%
davies_2024}
\APACinsertmetastar {%
davies_2024}%
\begin{APACrefauthors}%
Davies, A\BPBI G.%
, Perry, J\BPBI E.%
, Williams, D\BPBI A.%
, Veeder, G\BPBI J.%
\BCBL {}\ \BBA {} Nelson, D\BPBI M.%
\end{APACrefauthors}%
\unskip\
\newblock
\APACrefYearMonthDay{2024}{}{}.
\newblock
{\BBOQ}\APACrefatitle {New Global Map of $\text{Io’s}$ Volcanic Thermal
  Emission and Discovery of Hemispherical Dichotomies} {New global map of
  $\text{Io’s}$ volcanic thermal emission and discovery of hemispherical
  dichotomies}.{\BBCQ}
\newblock
\APACjournalVolNumPages{The Planetary Science Journal}{5}{5}{121}.
\newblock
\begin{APACrefDOI} \doi{10.3847/PSJ/ad4346} \end{APACrefDOI}
\PrintBackRefs{\CurrentBib}

\bibitem [\protect \citeauthoryear {%
{de Pater}%
, {Goldstein}%
\BCBL {}\ \BBA {} {Lellouch}%
}{%
{de Pater}%
\ \protect \BOthers {.}}{%
{\protect \APACyear {2023}}%
}]{%
dePater2023IoBook}
\APACinsertmetastar {%
dePater2023IoBook}%
\begin{APACrefauthors}%
{de Pater}, I.%
, {Goldstein}, D.%
\BCBL {}\ \BBA {} {Lellouch}, E.%
\end{APACrefauthors}%
\unskip\
\newblock
\APACrefYearMonthDay{2023}{{\APACmonth{01}}}{}.
\newblock
{\BBOQ}\APACrefatitle {{The Plumes and Atmosphere of Io}} {{The Plumes and
  Atmosphere of Io}}.{\BBCQ}
\newblock
\BIn{} R\BPBI M\BPBI C.~{Lopes}, K.~{de Kleer}\BCBL {}\ \BBA {} J\BPBI
  T.~{Keane}\ (\BEDS), \APACrefbtitle {Io: A New View of Jupiter's Moon} {Io: A
  new view of jupiter's moon}\ (\BVOL~468, \BPG~233-290).
\newblock
\begin{APACrefDOI} \doi{10.1007/978-3-031-25670-7_8} \end{APACrefDOI}
\PrintBackRefs{\CurrentBib}

\bibitem [\protect \citeauthoryear {%
de Kleer%
, de Pater%
\BCBL {}\ \protect \BOthers {.}}{%
de Kleer%
, de Pater%
\BCBL {}\ \protect \BOthers {.}}{%
{\protect \APACyear {2019}}%
}]{%
de_Kleer_2019_AJ}
\APACinsertmetastar {%
de_Kleer_2019_AJ}%
\begin{APACrefauthors}%
de Kleer, K.%
, de Pater, I.%
, Molter, E\BPBI M.%
, Banks, E.%
, Davies, A\BPBI G.%
, Alvarez, C.%
\BDBL {}Tollefson, J.%
\end{APACrefauthors}%
\unskip\
\newblock
\APACrefYearMonthDay{2019}{}{}.
\newblock
{\BBOQ}\APACrefatitle {$\text{Io’s}$ Volcanic Activity from Time Domain
  Adaptive Optics Observations: 2013–2018} {$\text{Io’s}$ volcanic activity
  from time domain adaptive optics observations: 2013–2018}.{\BBCQ}
\newblock
\APACjournalVolNumPages{The Astronomical Journal}{158}{1}{29}.
\newblock
\begin{APACrefDOI} \doi{10.3847/1538-3881/ab2380} \end{APACrefDOI}
\PrintBackRefs{\CurrentBib}

\bibitem [\protect \citeauthoryear {%
de Kleer%
, McEwen%
\BCBL {}\ \protect \BOthers {.}}{%
de Kleer%
, McEwen%
\BCBL {}\ \protect \BOthers {.}}{%
{\protect \APACyear {2019}}%
}]{%
deKleer2019}
\APACinsertmetastar {%
deKleer2019}%
\begin{APACrefauthors}%
de Kleer, K.%
, McEwen, A\BPBI S.%
, Park, R\BPBI S.%
, Bierson, C\BPBI J.%
, Davies, A\BPBI G.%
, DellaGustina, D\BPBI N.%
\BDBL {}others%
\end{APACrefauthors}%
\unskip\
\newblock
\APACrefYearMonthDay{2019}{}{}.
\newblock
{\BBOQ}\APACrefatitle {Tidal heating: Lessons from $\text{Io}$ and the jovian
  system-final report} {Tidal heating: Lessons from $\text{Io}$ and the jovian
  system-final report}.{\BBCQ}
\newblock
\APACjournalVolNumPages{Keck Institute for Space Studies}{}{}{}.
\newblock
\begin{APACrefURL}
  \url{https://www.kiss.caltech.edu/final_reports/Tidal_Heating_final_report.pdf}
  \end{APACrefURL}
\PrintBackRefs{\CurrentBib}

\bibitem [\protect \citeauthoryear {%
de Pater%
, Keane%
, de Kleer%
\BCBL {}\ \BBA {} Davies%
}{%
de Pater%
\ \protect \BOthers {.}}{%
{\protect \APACyear {2021}}%
}]{%
depater2020review}
\APACinsertmetastar {%
depater2020review}%
\begin{APACrefauthors}%
de Pater, I.%
, Keane, J\BPBI T.%
, de Kleer, K.%
\BCBL {}\ \BBA {} Davies, A\BPBI G.%
\end{APACrefauthors}%
\unskip\
\newblock
\APACrefYearMonthDay{2021}{}{}.
\newblock
{\BBOQ}\APACrefatitle {A 2020 observational perspective of $\text{Io}$} {A 2020
  observational perspective of $\text{Io}$}.{\BBCQ}
\newblock
\APACjournalVolNumPages{Annual Review of Earth and Planetary
  Sciences}{49}{1}{643--678}.
\newblock
\begin{APACrefDOI} \doi{https://doi.org/10.1146/annurev-earth-082420-095244}
  \end{APACrefDOI}
\PrintBackRefs{\CurrentBib}

\bibitem [\protect \citeauthoryear {%
de Pater%
\ \BBA {} Lissauer%
}{%
de Pater%
\ \BBA {} Lissauer%
}{%
{\protect \APACyear {2015}}%
}]{%
dePater_2015_plansci}
\APACinsertmetastar {%
dePater_2015_plansci}%
\begin{APACrefauthors}%
de Pater, I.%
\BCBT {}\ \BBA {} Lissauer, J\BPBI J.%
\end{APACrefauthors}%
\unskip\
\newblock
\APACrefYear{2015}.
\newblock
\APACrefbtitle {Planetary sciences} {Planetary sciences}.
\newblock
\APACaddressPublisher{}{Cambridge University Press}.
\PrintBackRefs{\CurrentBib}

\bibitem [\protect \citeauthoryear {%
de Pater%
\ \protect \BOthers {.}}{%
de Pater%
\ \protect \BOthers {.}}{%
{\protect \APACyear {2020b}}%
}]{%
dePater_2020}
\APACinsertmetastar {%
dePater_2020}%
\begin{APACrefauthors}%
de Pater, I.%
, Luszcz-Cook, S.%
, Rojo, P.%
, Redwing, E.%
, de Kleer, K.%
\BCBL {}\ \BBA {} Moullet, A.%
\end{APACrefauthors}%
\unskip\
\newblock
\APACrefYearMonthDay{2020b}{}{}.
\newblock
{\BBOQ}\APACrefatitle {ALMA Observations of $\text{Io}$ Going into and Coming
  out of Eclipse} {Alma observations of $\text{Io}$ going into and coming out
  of eclipse}.{\BBCQ}
\newblock
\APACjournalVolNumPages{The Planetary Science Journal}{1}{3}{60}.
\newblock
\begin{APACrefDOI} \doi{10.3847/PSJ/abb93d} \end{APACrefDOI}
\PrintBackRefs{\CurrentBib}

\bibitem [\protect \citeauthoryear {%
de Pater%
, Roe%
, Graham%
, Strobel%
\BCBL {}\ \BBA {} Bernath%
}{%
de Pater%
\ \protect \BOthers {.}}{%
{\protect \APACyear {2002}}%
}]{%
dePater_2002}
\APACinsertmetastar {%
dePater_2002}%
\begin{APACrefauthors}%
de Pater, I.%
, Roe, H.%
, Graham, J\BPBI R.%
, Strobel, D\BPBI F.%
\BCBL {}\ \BBA {} Bernath, P.%
\end{APACrefauthors}%
\unskip\
\newblock
\APACrefYearMonthDay{2002}{}{}.
\newblock
{\BBOQ}\APACrefatitle {Detection of the Forbidden SO a1$\Delta$→ X3$\Sigma$-
  Rovibronic Transition on $\text{Io}$ at 1.7 $\mu$m} {Detection of the
  forbidden so a1$\delta$→ x3$\sigma$- rovibronic transition on $\text{Io}$
  at 1.7 $\mu$m}.{\BBCQ}
\newblock
\APACjournalVolNumPages{Icarus}{156}{1}{296--301}.
\newblock
\begin{APACrefDOI} \doi{https://doi.org/10.1006/icar.2001.6787}
  \end{APACrefDOI}
\PrintBackRefs{\CurrentBib}

\bibitem [\protect \citeauthoryear {%
Dott%
}{%
Dott%
}{%
{\protect \APACyear {2024}}%
}]{%
dott_zenodo}
\APACinsertmetastar {%
dott_zenodo}%
\begin{APACrefauthors}%
Dott, A\BHBI C.%
\end{APACrefauthors}%
\unskip\
\newblock
\APACrefYearMonthDay{2024}{}{}.
\newblock
{\BBOQ}\APACrefatitle {Implicit solver for 1-D heat equation to calculate
  $\text{Io's}$ surface temperature (assuming a homogenous subsurface
  structure)[Software].} {Implicit solver for 1-d heat equation to calculate
  $\text{Io's}$ surface temperature (assuming a homogenous subsurface
  structure)[software].}{\BBCQ}
\newblock

\newblock
\begin{APACrefDOI} \doi{https://doi.org/10.5281/zenodo.14199914}
  \end{APACrefDOI}
\PrintBackRefs{\CurrentBib}

\bibitem [\protect \citeauthoryear {%
Feaga%
, McGrath%
\BCBL {}\ \BBA {} Feldman%
}{%
Feaga%
\ \protect \BOthers {.}}{%
{\protect \APACyear {2009}}%
}]{%
Feaga2009}
\APACinsertmetastar {%
Feaga2009}%
\begin{APACrefauthors}%
Feaga, L\BPBI M.%
, McGrath, M.%
\BCBL {}\ \BBA {} Feldman, P\BPBI D.%
\end{APACrefauthors}%
\unskip\
\newblock
\APACrefYearMonthDay{2009}{}{}.
\newblock
{\BBOQ}\APACrefatitle {Io's dayside $\text{SO}_{2}$ atmosphere} {Io's dayside
  $\text{SO}_{2}$ atmosphere}.{\BBCQ}
\newblock
\APACjournalVolNumPages{Icarus}{201}{2}{570-584}.
\newblock
\begin{APACrefDOI} \doi{https://doi.org/10.1016/j.icarus.2009.01.029}
  \end{APACrefDOI}
\PrintBackRefs{\CurrentBib}

\bibitem [\protect \citeauthoryear {%
Feldman%
\ \protect \BOthers {.}}{%
Feldman%
\ \protect \BOthers {.}}{%
{\protect \APACyear {2000}}%
}]{%
Feldman2000}
\APACinsertmetastar {%
Feldman2000}%
\begin{APACrefauthors}%
Feldman, P\BPBI D.%
, Strobel, D\BPBI F.%
, Moos, H\BPBI W.%
, Retherford, K\BPBI D.%
, Wolven, B\BPBI C.%
, McGrath, M\BPBI A.%
\BDBL {}Ballester, G\BPBI E.%
\end{APACrefauthors}%
\unskip\
\newblock
\APACrefYearMonthDay{2000}{}{}.
\newblock
{\BBOQ}\APACrefatitle {Lyman-$\alpha$ imaging of the $\text{SO}_2$ distribution
  on $\text{Io}$} {Lyman-$\alpha$ imaging of the $\text{SO}_2$ distribution on
  $\text{Io}$}.{\BBCQ}
\newblock
\APACjournalVolNumPages{Geophysical Research Letters}{27}{12}{1787-1790}.
\newblock
\begin{APACrefDOI} \doi{https://doi.org/10.1029/1999GL011067} \end{APACrefDOI}
\PrintBackRefs{\CurrentBib}

\bibitem [\protect \citeauthoryear {%
Giles%
\ \protect \BOthers {.}}{%
Giles%
\ \protect \BOthers {.}}{%
{\protect \APACyear {2024}}%
}]{%
giles_2024}
\APACinsertmetastar {%
giles_2024}%
\begin{APACrefauthors}%
Giles, R\BPBI S.%
, Spencer, J\BPBI R.%
, Tsang, C\BPBI C.%
, Greathouse, T\BPBI K.%
, Lellouch, E.%
\BCBL {}\ \BBA {} L{\'o}pez-Valverde, M\BPBI A.%
\end{APACrefauthors}%
\unskip\
\newblock
\APACrefYearMonthDay{2024}{}{}.
\newblock
{\BBOQ}\APACrefatitle {Seasonal and longitudinal variability in Io’s SO2
  atmosphere from 22 years of IRTF/TEXES observations} {Seasonal and
  longitudinal variability in io’s so2 atmosphere from 22 years of irtf/texes
  observations}.{\BBCQ}
\newblock
\APACjournalVolNumPages{Icarus}{418}{}{116151}.
\PrintBackRefs{\CurrentBib}

\bibitem [\protect \citeauthoryear {%
{Giono}%
\ \BBA {} {Roth}%
}{%
{Giono}%
\ \BBA {} {Roth}%
}{%
{\protect \APACyear {2021}}%
}]{%
Giono_Roth2021}
\APACinsertmetastar {%
Giono_Roth2021}%
\begin{APACrefauthors}%
{Giono}, G.%
\BCBT {}\ \BBA {} {Roth}, L.%
\end{APACrefauthors}%
\unskip\
\newblock
\APACrefYearMonthDay{2021}{}{}.
\newblock
{\BBOQ}\APACrefatitle {{Io's SO$_{2}$ atmosphere from HST
  Lyman-{\ensuremath{\alpha}} images: 1997 to 2018}} {{Io's SO$_{2}$ atmosphere
  from HST Lyman-{\ensuremath{\alpha}} images: 1997 to 2018}}.{\BBCQ}
\newblock
\APACjournalVolNumPages{Icarus}{359}{}{114212}.
\newblock
\begin{APACrefDOI} \doi{10.1016/j.icarus.2020.114212} \end{APACrefDOI}
\PrintBackRefs{\CurrentBib}

\bibitem [\protect \citeauthoryear {%
Guillot%
, Stevenson%
, Hubbard%
\BCBL {}\ \BBA {} Saumon%
}{%
Guillot%
\ \protect \BOthers {.}}{%
{\protect \APACyear {2004}}%
}]{%
guillot_2004}
\APACinsertmetastar {%
guillot_2004}%
\begin{APACrefauthors}%
Guillot, T.%
, Stevenson, D\BPBI J.%
, Hubbard, W\BPBI B.%
\BCBL {}\ \BBA {} Saumon, D.%
\end{APACrefauthors}%
\unskip\
\newblock
\APACrefYearMonthDay{2004}{}{}.
\newblock
{\BBOQ}\APACrefatitle {The interior of Jupiter} {The interior of
  jupiter}.{\BBCQ}
\newblock
\APACjournalVolNumPages{Jupiter: The planet, satellites and
  magnetosphere}{35}{}{57}.
\PrintBackRefs{\CurrentBib}

\bibitem [\protect \citeauthoryear {%
Hamilton%
\ \protect \BOthers {.}}{%
Hamilton%
\ \protect \BOthers {.}}{%
{\protect \APACyear {2013}}%
}]{%
Hamilton2013}
\APACinsertmetastar {%
Hamilton2013}%
\begin{APACrefauthors}%
Hamilton, C\BPBI W.%
, Beggan, C\BPBI D.%
, Still, S.%
, Beuthe, M.%
, Lopes, R\BPBI M.%
, Williams, D\BPBI A.%
\BDBL {}Wright, W.%
\end{APACrefauthors}%
\unskip\
\newblock
\APACrefYearMonthDay{2013}{}{}.
\newblock
{\BBOQ}\APACrefatitle {Spatial distribution of volcanoes on $\text{Io}$:
  Implications for tidal heating and magma ascent} {Spatial distribution of
  volcanoes on $\text{Io}$: Implications for tidal heating and magma
  ascent}.{\BBCQ}
\newblock
\APACjournalVolNumPages{Earth and Planetary Science Letters}{361}{}{272-286}.
\newblock
\begin{APACrefDOI} \doi{https://doi.org/10.1016/j.epsl.2012.10.032}
  \end{APACrefDOI}
\PrintBackRefs{\CurrentBib}

\bibitem [\protect \citeauthoryear {%
Ingersoll%
, Summers%
\BCBL {}\ \BBA {} Schlipf%
}{%
Ingersoll%
\ \protect \BOthers {.}}{%
{\protect \APACyear {1985}}%
}]{%
ingersoll_1985}
\APACinsertmetastar {%
ingersoll_1985}%
\begin{APACrefauthors}%
Ingersoll, A\BPBI P.%
, Summers, M\BPBI E.%
\BCBL {}\ \BBA {} Schlipf, S\BPBI G.%
\end{APACrefauthors}%
\unskip\
\newblock
\APACrefYearMonthDay{1985}{}{}.
\newblock
{\BBOQ}\APACrefatitle {Supersonic meteorology of Io: Sublimation-driven flow of
  SO2} {Supersonic meteorology of io: Sublimation-driven flow of so2}.{\BBCQ}
\newblock
\APACjournalVolNumPages{Icarus}{64}{3}{375--390}.
\PrintBackRefs{\CurrentBib}

\bibitem [\protect \citeauthoryear {%
K.~Jessup%
\ \protect \BOthers {.}}{%
K.~Jessup%
\ \protect \BOthers {.}}{%
{\protect \APACyear {2004}}%
}]{%
Jessup2004}
\APACinsertmetastar {%
Jessup2004}%
\begin{APACrefauthors}%
Jessup, K.%
, Spencer, J.%
, Ballester, G.%
, Vigel, M.%
, Howell, R.%
, Roesler, F.%
\BCBL {}\ \BBA {} Yelle, R.%
\end{APACrefauthors}%
\unskip\
\newblock
\APACrefYearMonthDay{2004}{}{}.
\newblock
{\BBOQ}\APACrefatitle {The Atmospheric Signature of $\text{Io's}$
  $\text{Prometheus}$ Plume and Anti-$\text{Jovian}$ Hemisphere: Evidence for a
  Sublimation Atmosphere} {The atmospheric signature of $\text{Io's}$
  $\text{Prometheus}$ plume and anti-$\text{Jovian}$ hemisphere: Evidence for a
  sublimation atmosphere}.{\BBCQ}
\newblock
\APACjournalVolNumPages{Icarus}{169}{1}{197-215}.
\newblock
\begin{APACrefDOI} \doi{https://doi.org/10.1016/j.icarus.2003.11.015}
  \end{APACrefDOI}
\PrintBackRefs{\CurrentBib}

\bibitem [\protect \citeauthoryear {%
K\BPBI L.~Jessup%
\ \BBA {} Spencer%
}{%
K\BPBI L.~Jessup%
\ \BBA {} Spencer%
}{%
{\protect \APACyear {2015}}%
}]{%
Jessup2014}
\APACinsertmetastar {%
Jessup2014}%
\begin{APACrefauthors}%
Jessup, K\BPBI L.%
\BCBT {}\ \BBA {} Spencer, J\BPBI R.%
\end{APACrefauthors}%
\unskip\
\newblock
\APACrefYearMonthDay{2015}{}{}.
\newblock
{\BBOQ}\APACrefatitle {Spatially resolved $\text{HST/STIS}$ observations of
  $\text{Io’s}$ dayside equatorial atmosphere} {Spatially resolved
  $\text{HST/STIS}$ observations of $\text{Io’s}$ dayside equatorial
  atmosphere}.{\BBCQ}
\newblock
\APACjournalVolNumPages{Icarus}{248}{}{165-189}.
\newblock
\begin{APACrefDOI} \doi{https://doi.org/10.1016/j.icarus.2014.10.020}
  \end{APACrefDOI}
\PrintBackRefs{\CurrentBib}

\bibitem [\protect \citeauthoryear {%
Kerton%
, Fanale%
\BCBL {}\ \BBA {} Salvail%
}{%
Kerton%
\ \protect \BOthers {.}}{%
{\protect \APACyear {1996}}%
}]{%
kerton_1996}
\APACinsertmetastar {%
kerton_1996}%
\begin{APACrefauthors}%
Kerton, C.%
, Fanale, F.%
\BCBL {}\ \BBA {} Salvail, J.%
\end{APACrefauthors}%
\unskip\
\newblock
\APACrefYearMonthDay{1996}{}{}.
\newblock
{\BBOQ}\APACrefatitle {The state of SO2 on Io's surface} {The state of so2 on
  io's surface}.{\BBCQ}
\newblock
\APACjournalVolNumPages{Journal of Geophysical Research:
  Planets}{101}{E3}{7555--7563}.
\PrintBackRefs{\CurrentBib}

\bibitem [\protect \citeauthoryear {%
Kervazo%
, Tobie%
, Choblet%
, Dumoulin%
\BCBL {}\ \BBA {} B{\v{e}}hounkov{\'a}%
}{%
Kervazo%
\ \protect \BOthers {.}}{%
{\protect \APACyear {2022}}%
}]{%
kervazo_2022}
\APACinsertmetastar {%
kervazo_2022}%
\begin{APACrefauthors}%
Kervazo, M.%
, Tobie, G.%
, Choblet, G.%
, Dumoulin, C.%
\BCBL {}\ \BBA {} B{\v{e}}hounkov{\'a}, M.%
\end{APACrefauthors}%
\unskip\
\newblock
\APACrefYearMonthDay{2022}{}{}.
\newblock
{\BBOQ}\APACrefatitle {Inferring Io’s interior from tidal monitoring}
  {Inferring io’s interior from tidal monitoring}.{\BBCQ}
\newblock
\APACjournalVolNumPages{Icarus}{373}{}{114737}.
\PrintBackRefs{\CurrentBib}

\bibitem [\protect \citeauthoryear {%
Khurana%
\ \protect \BOthers {.}}{%
Khurana%
\ \protect \BOthers {.}}{%
{\protect \APACyear {2011}}%
}]{%
Khurana_2011}
\APACinsertmetastar {%
Khurana_2011}%
\begin{APACrefauthors}%
Khurana, K\BPBI K.%
, Jia, X.%
, Kivelson, M\BPBI G.%
, Nimmo, F.%
, Schubert, G.%
\BCBL {}\ \BBA {} Russell, C\BPBI T.%
\end{APACrefauthors}%
\unskip\
\newblock
\APACrefYearMonthDay{2011}{}{}.
\newblock
{\BBOQ}\APACrefatitle {Evidence of a Global Magma Ocean in $\text{Io’s}$
  Interior} {Evidence of a global magma ocean in $\text{Io’s}$
  interior}.{\BBCQ}
\newblock
\APACjournalVolNumPages{Science}{332}{6034}{1186-1189}.
\newblock
\begin{APACrefDOI} \doi{10.1126/science.1201425} \end{APACrefDOI}
\PrintBackRefs{\CurrentBib}

\bibitem [\protect \citeauthoryear {%
Lellouch%
}{%
Lellouch%
}{%
{\protect \APACyear {2005}}%
}]{%
lellouch_2005}
\APACinsertmetastar {%
lellouch_2005}%
\begin{APACrefauthors}%
Lellouch, E.%
\end{APACrefauthors}%
\unskip\
\newblock
\APACrefYearMonthDay{2005}{}{}.
\newblock
{\BBOQ}\APACrefatitle {Io’s atmosphere and surface-atmosphere interactions}
  {Io’s atmosphere and surface-atmosphere interactions}.{\BBCQ}
\newblock
\APACjournalVolNumPages{Space science reviews}{116}{}{211--224}.
\PrintBackRefs{\CurrentBib}

\bibitem [\protect \citeauthoryear {%
Lellouch%
\ \protect \BOthers {.}}{%
Lellouch%
\ \protect \BOthers {.}}{%
{\protect \APACyear {2015}}%
}]{%
lellouch_2015}
\APACinsertmetastar {%
lellouch_2015}%
\begin{APACrefauthors}%
Lellouch, E.%
, Ali-Dib, M.%
, Jessup, K\BHBI L.%
, Smette, A.%
, K{\"a}ufl, H\BHBI U.%
\BCBL {}\ \BBA {} Marchis, F.%
\end{APACrefauthors}%
\unskip\
\newblock
\APACrefYearMonthDay{2015}{}{}.
\newblock
{\BBOQ}\APACrefatitle {Detection and characterization of Io’s atmosphere from
  high-resolution 4-$\mu$m spectroscopy} {Detection and characterization of
  io’s atmosphere from high-resolution 4-$\mu$m spectroscopy}.{\BBCQ}
\newblock
\APACjournalVolNumPages{Icarus}{253}{}{99--114}.
\PrintBackRefs{\CurrentBib}

\bibitem [\protect \citeauthoryear {%
Lellouch%
, Belton%
, de Pater%
, Gulkis%
\BCBL {}\ \BBA {} Encrenaz%
}{%
Lellouch%
\ \protect \BOthers {.}}{%
{\protect \APACyear {1990}}%
}]{%
lellouch_1990}
\APACinsertmetastar {%
lellouch_1990}%
\begin{APACrefauthors}%
Lellouch, E.%
, Belton, M.%
, de Pater, I.%
, Gulkis, S.%
\BCBL {}\ \BBA {} Encrenaz, T.%
\end{APACrefauthors}%
\unskip\
\newblock
\APACrefYearMonthDay{1990}{}{}.
\newblock
{\BBOQ}\APACrefatitle {Io's atmosphere from microwave detection of
  $\text{SO}_2$} {Io's atmosphere from microwave detection of
  $\text{SO}_2$}.{\BBCQ}
\newblock
\APACjournalVolNumPages{Nature}{346}{6285}{639--641}.
\newblock
\begin{APACrefDOI} \doi{https://doi.org/10.1038/346639a0} \end{APACrefDOI}
\PrintBackRefs{\CurrentBib}

\bibitem [\protect \citeauthoryear {%
Lellouch%
, McGrath%
\BCBL {}\ \BBA {} Jessup%
}{%
Lellouch%
\ \protect \BOthers {.}}{%
{\protect \APACyear {2007}}%
}]{%
lellouch2007}
\APACinsertmetastar {%
lellouch2007}%
\begin{APACrefauthors}%
Lellouch, E.%
, McGrath, M\BPBI A.%
\BCBL {}\ \BBA {} Jessup, K\BPBI L.%
\end{APACrefauthors}%
\unskip\
\newblock
\APACrefYearMonthDay{2007}{}{}.
\newblock
{\BBOQ}\APACrefatitle {$\text{Io’s}$ atmosphere} {$\text{Io’s}$
  atmosphere}.{\BBCQ}
\newblock
\APACjournalVolNumPages{Io After Galileo}{}{}{231--264}.
\PrintBackRefs{\CurrentBib}

\bibitem [\protect \citeauthoryear {%
Leone%
, Wilson%
\BCBL {}\ \BBA {} Davies%
}{%
Leone%
\ \protect \BOthers {.}}{%
{\protect \APACyear {2011}}%
}]{%
Leone_2011}
\APACinsertmetastar {%
Leone_2011}%
\begin{APACrefauthors}%
Leone, G.%
, Wilson, L.%
\BCBL {}\ \BBA {} Davies, A.%
\end{APACrefauthors}%
\unskip\
\newblock
\APACrefYearMonthDay{2011}{}{}.
\newblock
{\BBOQ}\APACrefatitle {The geothermal gradient of $\text{Io}$: Consequences for
  lithosphere structure and volcanic eruptive activity} {The geothermal
  gradient of $\text{Io}$: Consequences for lithosphere structure and volcanic
  eruptive activity}.{\BBCQ}
\newblock
\APACjournalVolNumPages{Icarus}{211}{}{623-635}.
\newblock
\begin{APACrefDOI} \doi{10.1016/j.icarus.2010.10.016} \end{APACrefDOI}
\PrintBackRefs{\CurrentBib}

\bibitem [\protect \citeauthoryear {%
Matsuyama%
, Steinke%
\BCBL {}\ \BBA {} Nimmo%
}{%
Matsuyama%
\ \protect \BOthers {.}}{%
{\protect \APACyear {2022}}%
}]{%
Matsyama_2022}
\APACinsertmetastar {%
Matsyama_2022}%
\begin{APACrefauthors}%
Matsuyama, I\BPBI N.%
, Steinke, T.%
\BCBL {}\ \BBA {} Nimmo, F.%
\end{APACrefauthors}%
\unskip\
\newblock
\APACrefYearMonthDay{2022}{}{}.
\newblock
{\BBOQ}\APACrefatitle {{Tidal Heating in Io}} {{Tidal Heating in Io}}.{\BBCQ}
\newblock
\APACjournalVolNumPages{Elements}{18}{6}{374-378}.
\newblock
\begin{APACrefDOI} \doi{10.2138/gselements.18.6.374} \end{APACrefDOI}
\PrintBackRefs{\CurrentBib}

\bibitem [\protect \citeauthoryear {%
McEwen%
, Keszthelyi%
, Lopes%
, Schenk%
\BCBL {}\ \BBA {} Spencer%
}{%
McEwen%
\ \protect \BOthers {.}}{%
{\protect \APACyear {2004}}%
}]{%
McEwen2004}
\APACinsertmetastar {%
McEwen2004}%
\begin{APACrefauthors}%
McEwen, A.%
, Keszthelyi, L.%
, Lopes, R.%
, Schenk, P.%
\BCBL {}\ \BBA {} Spencer, J.%
\end{APACrefauthors}%
\unskip\
\newblock
\APACrefYearMonthDay{2004}{}{}.
\newblock
{\BBOQ}\APACrefatitle {The lithosphere and surface of $\text{Io}$} {The
  lithosphere and surface of $\text{Io}$}.{\BBCQ}
\newblock
\APACjournalVolNumPages{Jupiter. The Planet, Satellites and
  Magnetosphere}{}{}{}.
\PrintBackRefs{\CurrentBib}

\bibitem [\protect \citeauthoryear {%
McGrath%
, Belton%
, Spencer%
\BCBL {}\ \BBA {} Sartoretti%
}{%
McGrath%
\ \protect \BOthers {.}}{%
{\protect \APACyear {2000}}%
}]{%
McGrath2000}
\APACinsertmetastar {%
McGrath2000}%
\begin{APACrefauthors}%
McGrath, M\BPBI A.%
, Belton, M\BPBI J.%
, Spencer, J\BPBI R.%
\BCBL {}\ \BBA {} Sartoretti, P.%
\end{APACrefauthors}%
\unskip\
\newblock
\APACrefYearMonthDay{2000}{}{}.
\newblock
{\BBOQ}\APACrefatitle {Spatially Resolved Spectroscopy of $\text{Io's}$
  $\text{Pele}$ Plume and $\text{SO}_2$ Atmosphere} {Spatially resolved
  spectroscopy of $\text{Io's}$ $\text{Pele}$ plume and $\text{SO}_2$
  atmosphere}.{\BBCQ}
\newblock
\APACjournalVolNumPages{Icarus}{146}{2}{476-493}.
\newblock
\begin{APACrefDOI} \doi{https://doi.org/10.1006/icar.1999.6412}
  \end{APACrefDOI}
\PrintBackRefs{\CurrentBib}

\bibitem [\protect \citeauthoryear {%
Moore%
}{%
Moore%
}{%
{\protect \APACyear {2011}}%
}]{%
moore_2011}
\APACinsertmetastar {%
moore_2011}%
\begin{APACrefauthors}%
Moore, C\BPBI H.%
\end{APACrefauthors}%
\unskip\
\newblock
\APACrefYearMonthDay{2011}{}{}.
\newblock
{\BBOQ}\APACrefatitle {Monte Carlo simulation of the Jovian plasma torus
  interaction with Io’s atmosphere and the resultant aurora during eclipse}
  {Monte carlo simulation of the jovian plasma torus interaction with io’s
  atmosphere and the resultant aurora during eclipse}.{\BBCQ}
\newblock

\PrintBackRefs{\CurrentBib}

\bibitem [\protect \citeauthoryear {%
Moreno%
, Schubert%
, Baumgardner%
, Kivelson%
\BCBL {}\ \BBA {} Paige%
}{%
Moreno%
\ \protect \BOthers {.}}{%
{\protect \APACyear {1991}}%
}]{%
moreno_1991}
\APACinsertmetastar {%
moreno_1991}%
\begin{APACrefauthors}%
Moreno, M\BPBI A.%
, Schubert, G.%
, Baumgardner, J.%
, Kivelson, M\BPBI G.%
\BCBL {}\ \BBA {} Paige, D\BPBI A.%
\end{APACrefauthors}%
\unskip\
\newblock
\APACrefYearMonthDay{1991}{}{}.
\newblock
{\BBOQ}\APACrefatitle {Io's volcanic and sublimation atmospheres} {Io's
  volcanic and sublimation atmospheres}.{\BBCQ}
\newblock
\APACjournalVolNumPages{Icarus}{93}{1}{63--81}.
\PrintBackRefs{\CurrentBib}

\bibitem [\protect \citeauthoryear {%
{Morrison}%
\ \BBA {} {Telesco}%
}{%
{Morrison}%
\ \BBA {} {Telesco}%
}{%
{\protect \APACyear {1980}}%
}]{%
Morrison1980}
\APACinsertmetastar {%
Morrison1980}%
\begin{APACrefauthors}%
{Morrison}, D.%
\BCBT {}\ \BBA {} {Telesco}, C.%
\end{APACrefauthors}%
\unskip\
\newblock
\APACrefYearMonthDay{1980}{}{}.
\newblock
{\BBOQ}\APACrefatitle {{Io: Observational constraints on internal energy and
  thermophysics of the surface}} {{Io: Observational constraints on internal
  energy and thermophysics of the surface}}.{\BBCQ}
\newblock
\APACjournalVolNumPages{Icarus}{44}{2}{226-233}.
\newblock
\begin{APACrefDOI} \doi{10.1016/0019-1035(80)90018-4} \end{APACrefDOI}
\PrintBackRefs{\CurrentBib}

\bibitem [\protect \citeauthoryear {%
Moullet%
, Gurwell%
, Lellouch%
\BCBL {}\ \BBA {} Moreno%
}{%
Moullet%
\ \protect \BOthers {.}}{%
{\protect \APACyear {2010}}%
}]{%
moullet_2010}
\APACinsertmetastar {%
moullet_2010}%
\begin{APACrefauthors}%
Moullet, A.%
, Gurwell, M\BPBI A.%
, Lellouch, E.%
\BCBL {}\ \BBA {} Moreno, R.%
\end{APACrefauthors}%
\unskip\
\newblock
\APACrefYearMonthDay{2010}{}{}.
\newblock
{\BBOQ}\APACrefatitle {Simultaneous mapping of $\text{SO}_2$, $\text{SO}$,
  $\text{NaCl}$ in $\text{Io’s}$ atmosphere with the $\text{Submillimeter}$
  $\text{Array}$} {Simultaneous mapping of $\text{SO}_2$, $\text{SO}$,
  $\text{NaCl}$ in $\text{Io’s}$ atmosphere with the $\text{Submillimeter}$
  $\text{Array}$}.{\BBCQ}
\newblock
\APACjournalVolNumPages{Icarus}{208}{1}{353--365}.
\newblock
\begin{APACrefDOI} \doi{https://doi.org/10.1016/j.icarus.2010.02.009}
  \end{APACrefDOI}
\PrintBackRefs{\CurrentBib}

\bibitem [\protect \citeauthoryear {%
Nash%
\ \BBA {} Nelson%
}{%
Nash%
\ \BBA {} Nelson%
}{%
{\protect \APACyear {1979}}%
}]{%
Nash1979}
\APACinsertmetastar {%
Nash1979}%
\begin{APACrefauthors}%
Nash, D\BPBI B.%
\BCBT {}\ \BBA {} Nelson, R\BPBI M.%
\end{APACrefauthors}%
\unskip\
\newblock
\APACrefYearMonthDay{1979}{}{}.
\newblock
{\BBOQ}\APACrefatitle {Spectral evidence for sublimates and adsorbates on
  $\text{Io}$} {Spectral evidence for sublimates and adsorbates on
  $\text{Io}$}.{\BBCQ}
\newblock
\APACjournalVolNumPages{Nature}{280}{}{763--766}.
\newblock
\begin{APACrefDOI} \doi{https://doi.org/10.1038/280763a0} \end{APACrefDOI}
\PrintBackRefs{\CurrentBib}

\bibitem [\protect \citeauthoryear {%
R.~Park%
\ \protect \BOthers {.}}{%
R.~Park%
\ \protect \BOthers {.}}{%
{\protect \APACyear {2025}}%
}]{%
park_2025}
\APACinsertmetastar {%
park_2025}%
\begin{APACrefauthors}%
Park, R.%
, Jacobson, R.%
, Gomez~Casajus, L.%
, Nimmo, F.%
, Ermakov, A.%
, Keane, J.%
\BDBL {}others%
\end{APACrefauthors}%
\unskip\
\newblock
\APACrefYearMonthDay{2025}{}{}.
\newblock
{\BBOQ}\APACrefatitle {Io’s tidal response precludes a shallow magma ocean}
  {Io’s tidal response precludes a shallow magma ocean}.{\BBCQ}
\newblock
\APACjournalVolNumPages{Nature}{638}{8049}{69--73}.
\PrintBackRefs{\CurrentBib}

\bibitem [\protect \citeauthoryear {%
R\BPBI S.~Park%
, Folkner%
, Williams%
\BCBL {}\ \BBA {} Boggs%
}{%
R\BPBI S.~Park%
\ \protect \BOthers {.}}{%
{\protect \APACyear {2021}}%
}]{%
park_2021}
\APACinsertmetastar {%
park_2021}%
\begin{APACrefauthors}%
Park, R\BPBI S.%
, Folkner, W\BPBI M.%
, Williams, J\BPBI G.%
\BCBL {}\ \BBA {} Boggs, D\BPBI H.%
\end{APACrefauthors}%
\unskip\
\newblock
\APACrefYearMonthDay{2021}{}{}.
\newblock
{\BBOQ}\APACrefatitle {The $\text{JPL}$ planetary and lunar ephemerides
  $\text{DE440}$ and $\text{DE441}$} {The $\text{JPL}$ planetary and lunar
  ephemerides $\text{DE440}$ and $\text{DE441}$}.{\BBCQ}
\newblock
\APACjournalVolNumPages{The Astronomical Journal}{161}{3}{105}.
\newblock
\begin{APACrefDOI} \doi{10.3847/1538-3881/abd414} \end{APACrefDOI}
\PrintBackRefs{\CurrentBib}

\bibitem [\protect \citeauthoryear {%
Peale%
, Cassen%
\BCBL {}\ \BBA {} Reynolds%
}{%
Peale%
\ \protect \BOthers {.}}{%
{\protect \APACyear {1979}}%
}]{%
Peale1979}
\APACinsertmetastar {%
Peale1979}%
\begin{APACrefauthors}%
Peale, S.%
, Cassen, P.%
\BCBL {}\ \BBA {} Reynolds, R.%
\end{APACrefauthors}%
\unskip\
\newblock
\APACrefYearMonthDay{1979}{}{}.
\newblock
{\BBOQ}\APACrefatitle {Melting of $\text{Io}$ by Tidal Dissipation} {Melting of
  $\text{Io}$ by tidal dissipation}.{\BBCQ}
\newblock
\APACjournalVolNumPages{Science (New York, N.Y.)}{203}{}{892-4}.
\newblock
\begin{APACrefDOI} \doi{10.1126/science.203.4383.892} \end{APACrefDOI}
\PrintBackRefs{\CurrentBib}

\bibitem [\protect \citeauthoryear {%
Rathbun%
\ \protect \BOthers {.}}{%
Rathbun%
\ \protect \BOthers {.}}{%
{\protect \APACyear {2004}}%
}]{%
Rathbun2004}
\APACinsertmetastar {%
Rathbun2004}%
\begin{APACrefauthors}%
Rathbun, J.%
, Spencer, J.%
, Tamppari, L.%
, Martin, T.%
, Barnard, L.%
\BCBL {}\ \BBA {} Travis, L.%
\end{APACrefauthors}%
\unskip\
\newblock
\APACrefYearMonthDay{2004}{}{}.
\newblock
{\BBOQ}\APACrefatitle {Mapping of $\text{Io's}$ thermal radiation by the
  $\text{Galileo}$ photopolarimeter–radiometer $\text{(PPR)}$ instrument}
  {Mapping of $\text{Io's}$ thermal radiation by the $\text{Galileo}$
  photopolarimeter–radiometer $\text{(PPR)}$ instrument}.{\BBCQ}
\newblock
\APACjournalVolNumPages{Icarus}{169}{1}{127-139}.
\newblock
\begin{APACrefDOI} \doi{https://doi.org/10.1016/j.icarus.2003.12.021}
  \end{APACrefDOI}
\PrintBackRefs{\CurrentBib}

\bibitem [\protect \citeauthoryear {%
Retherford%
\ \protect \BOthers {.}}{%
Retherford%
\ \protect \BOthers {.}}{%
{\protect \APACyear {2019}}%
}]{%
retherford2019}
\APACinsertmetastar {%
retherford2019}%
\begin{APACrefauthors}%
Retherford, K.%
, Roth, L.%
, Becker, T.%
, Feaga, L.%
, Tsang, C.%
, Jessup, K.%
\BCBL {}\ \BBA {} Grava, C.%
\end{APACrefauthors}%
\unskip\
\newblock
\APACrefYearMonthDay{2019}{}{}.
\newblock
{\BBOQ}\APACrefatitle {Io’s Atmosphere Silhouetted by Jupiter Ly-$\alpha$}
  {Io’s atmosphere silhouetted by jupiter ly-$\alpha$}.{\BBCQ}
\newblock
\APACjournalVolNumPages{The Astronomical Journal}{158}{4}{154}.
\newblock
\begin{APACrefDOI} \doi{10.3847/1538-3881/ab3c4c} \end{APACrefDOI}
\PrintBackRefs{\CurrentBib}

\bibitem [\protect \citeauthoryear {%
Roth%
\ \protect \BOthers {.}}{%
Roth%
\ \protect \BOthers {.}}{%
{\protect \APACyear {2025}}%
}]{%
roth_2025}
\APACinsertmetastar {%
roth_2025}%
\begin{APACrefauthors}%
Roth, L.%
, Bl{\"o}cker, A.%
, de Kleer, K.%
, Goldstein, D.%
, Lellouch, E.%
, Saur, J.%
\BDBL {}others%
\end{APACrefauthors}%
\unskip\
\newblock
\APACrefYearMonthDay{2025}{}{}.
\newblock
{\BBOQ}\APACrefatitle {Mass supply from Io to Jupiter’s magnetosphere} {Mass
  supply from io to jupiter’s magnetosphere}.{\BBCQ}
\newblock
\APACjournalVolNumPages{Space Science Reviews}{221}{1}{13}.
\PrintBackRefs{\CurrentBib}

\bibitem [\protect \citeauthoryear {%
Roth%
\ \protect \BOthers {.}}{%
Roth%
\ \protect \BOthers {.}}{%
{\protect \APACyear {2020}}%
}]{%
roth_2020}
\APACinsertmetastar {%
roth_2020}%
\begin{APACrefauthors}%
Roth, L.%
, Boissier, J.%
, Moullet, A.%
, S{\'a}nchez-Monge, {\'A}.%
, de Kleer, K.%
, Yoneda, M.%
\BDBL {}others%
\end{APACrefauthors}%
\unskip\
\newblock
\APACrefYearMonthDay{2020}{}{}.
\newblock
{\BBOQ}\APACrefatitle {An attempt to detect transient changes in
  $\text{Io’s}$ $\text{SO}_2$ and $\text{NaCl}$ atmosphere} {An attempt to
  detect transient changes in $\text{Io’s}$ $\text{SO}_2$ and $\text{NaCl}$
  atmosphere}.{\BBCQ}
\newblock
\APACjournalVolNumPages{Icarus}{350}{}{113925}.
\newblock
\begin{APACrefDOI} \doi{https://doi.org/10.1016/j.icarus.2020.113925}
  \end{APACrefDOI}
\PrintBackRefs{\CurrentBib}

\bibitem [\protect \citeauthoryear {%
Roth%
\ \protect \BOthers {.}}{%
Roth%
\ \protect \BOthers {.}}{%
{\protect \APACyear {2017}}%
}]{%
Roth_2017}
\APACinsertmetastar {%
Roth_2017}%
\begin{APACrefauthors}%
Roth, L.%
, Saur, J.%
, Retherford, K\BPBI D.%
, Blöcker, A.%
, Strobel, D\BPBI F.%
\BCBL {}\ \BBA {} Feldman, P\BPBI D.%
\end{APACrefauthors}%
\unskip\
\newblock
\APACrefYearMonthDay{2017}{}{}.
\newblock
{\BBOQ}\APACrefatitle {Constraints on $\text{Io's}$ interior from auroral spot
  oscillations} {Constraints on $\text{Io's}$ interior from auroral spot
  oscillations}.{\BBCQ}
\newblock
\APACjournalVolNumPages{Journal of Geophysical Research: Space
  Physics}{122}{2}{1903-1927}.
\newblock
\begin{APACrefDOI} \doi{https://doi.org/10.1002/2016JA023701} \end{APACrefDOI}
\PrintBackRefs{\CurrentBib}

\bibitem [\protect \citeauthoryear {%
Roth%
, Saur%
, Retherford%
, Strobel%
\BCBL {}\ \BBA {} Spencer%
}{%
Roth%
\ \protect \BOthers {.}}{%
{\protect \APACyear {2011}}%
}]{%
Roth2011}
\APACinsertmetastar {%
Roth2011}%
\begin{APACrefauthors}%
Roth, L.%
, Saur, J.%
, Retherford, K\BPBI D.%
, Strobel, D\BPBI F.%
\BCBL {}\ \BBA {} Spencer, J\BPBI R.%
\end{APACrefauthors}%
\unskip\
\newblock
\APACrefYearMonthDay{2011}{}{}.
\newblock
{\BBOQ}\APACrefatitle {Simulation of $\text{Io’s}$ auroral emission:
  Constraints on the atmosphere in eclipse} {Simulation of $\text{Io’s}$
  auroral emission: Constraints on the atmosphere in eclipse}.{\BBCQ}
\newblock
\APACjournalVolNumPages{Icarus}{214}{2}{495-509}.
\newblock
\begin{APACrefDOI} \doi{10.1016/j.icarus.2011.05.014} \end{APACrefDOI}
\PrintBackRefs{\CurrentBib}

\bibitem [\protect \citeauthoryear {%
Saur%
, Neubauer%
, Strobel%
\BCBL {}\ \BBA {} Summers%
}{%
Saur%
\ \protect \BOthers {.}}{%
{\protect \APACyear {2002}}%
}]{%
saur_2002}
\APACinsertmetastar {%
saur_2002}%
\begin{APACrefauthors}%
Saur, J.%
, Neubauer, F\BPBI M.%
, Strobel, D\BPBI F.%
\BCBL {}\ \BBA {} Summers, M\BPBI E.%
\end{APACrefauthors}%
\unskip\
\newblock
\APACrefYearMonthDay{2002}{}{}.
\newblock
{\BBOQ}\APACrefatitle {Interpretation of Galileo's Io plasma and field
  observations: I0, I24, and I27 flybys and close polar passes} {Interpretation
  of galileo's io plasma and field observations: I0, i24, and i27 flybys and
  close polar passes}.{\BBCQ}
\newblock
\APACjournalVolNumPages{Journal of Geophysical Research: Space
  Physics}{107}{A12}{SMP--5}.
\PrintBackRefs{\CurrentBib}

\bibitem [\protect \citeauthoryear {%
Saur%
\ \BBA {} Strobel%
}{%
Saur%
\ \BBA {} Strobel%
}{%
{\protect \APACyear {2004}}%
}]{%
SaurStrobel2004}
\APACinsertmetastar {%
SaurStrobel2004}%
\begin{APACrefauthors}%
Saur, J.%
\BCBT {}\ \BBA {} Strobel, D\BPBI F.%
\end{APACrefauthors}%
\unskip\
\newblock
\APACrefYearMonthDay{2004}{}{}.
\newblock
{\BBOQ}\APACrefatitle {Relative contributions of sublimation and volcanoes to
  $\text{Io's}$ atmosphere inferred from its plasma interaction during solar
  eclipse} {Relative contributions of sublimation and volcanoes to
  $\text{Io's}$ atmosphere inferred from its plasma interaction during solar
  eclipse}.{\BBCQ}
\newblock
\APACjournalVolNumPages{Icarus}{171}{2}{411-420}.
\newblock
\begin{APACrefDOI} \doi{https://doi.org/10.1016/j.icarus.2004.05.010}
  \end{APACrefDOI}
\PrintBackRefs{\CurrentBib}

\bibitem [\protect \citeauthoryear {%
Segatz%
, Spohn%
, Ross%
\BCBL {}\ \BBA {} Schubert%
}{%
Segatz%
\ \protect \BOthers {.}}{%
{\protect \APACyear {1988}}%
}]{%
Segatz_1988}
\APACinsertmetastar {%
Segatz_1988}%
\begin{APACrefauthors}%
Segatz, M.%
, Spohn, T.%
, Ross, M.%
\BCBL {}\ \BBA {} Schubert, G.%
\end{APACrefauthors}%
\unskip\
\newblock
\APACrefYearMonthDay{1988}{}{}.
\newblock
{\BBOQ}\APACrefatitle {Tidal dissipation, surface heat flow, and figure of
  viscoelastic models of $\text{Io}$} {Tidal dissipation, surface heat flow,
  and figure of viscoelastic models of $\text{Io}$}.{\BBCQ}
\newblock
\APACjournalVolNumPages{Icarus}{75}{2}{187-206}.
\newblock
\begin{APACrefDOI} \doi{https://doi.org/10.1016/0019-1035(88)90001-2}
  \end{APACrefDOI}
\PrintBackRefs{\CurrentBib}

\bibitem [\protect \citeauthoryear {%
Simonelli%
, Dodd%
\BCBL {}\ \BBA {} Veverka%
}{%
Simonelli%
\ \protect \BOthers {.}}{%
{\protect \APACyear {2001}}%
}]{%
Simonelli2001}
\APACinsertmetastar {%
Simonelli2001}%
\begin{APACrefauthors}%
Simonelli, D\BPBI P.%
, Dodd, C.%
\BCBL {}\ \BBA {} Veverka, J.%
\end{APACrefauthors}%
\unskip\
\newblock
\APACrefYearMonthDay{2001}{}{}.
\newblock
{\BBOQ}\APACrefatitle {Regolith variations on Io: Implications for bolometric
  albedos} {Regolith variations on io: Implications for bolometric
  albedos}.{\BBCQ}
\newblock
\APACjournalVolNumPages{Journal of Geophysical Research:
  Planets}{106}{E12}{33241-33252}.
\newblock
\begin{APACrefDOI} \doi{https://doi.org/10.1029/2000JE001350} \end{APACrefDOI}
\PrintBackRefs{\CurrentBib}

\bibitem [\protect \citeauthoryear {%
Sinclair%
\ \protect \BOthers {.}}{%
Sinclair%
\ \protect \BOthers {.}}{%
{\protect \APACyear {2023}}%
}]{%
sinclair_2023}
\APACinsertmetastar {%
sinclair_2023}%
\begin{APACrefauthors}%
Sinclair, J.%
, West, R.%
, Barbara, J.%
, Tao, C.%
, Orton, G.%
, Greathouse, T.%
\BDBL {}Irwin, P.%
\end{APACrefauthors}%
\unskip\
\newblock
\APACrefYearMonthDay{2023}{}{}.
\newblock
{\BBOQ}\APACrefatitle {Long-term variability of Jupiter’s northern auroral
  8-$\mu$m CH4 emissions} {Long-term variability of jupiter’s northern
  auroral 8-$\mu$m ch4 emissions}.{\BBCQ}
\newblock
\APACjournalVolNumPages{Icarus}{406}{}{115740}.
\PrintBackRefs{\CurrentBib}

\bibitem [\protect \citeauthoryear {%
Smoluchowski%
}{%
Smoluchowski%
}{%
{\protect \APACyear {1967}}%
}]{%
smoluchowski_1967}
\APACinsertmetastar {%
smoluchowski_1967}%
\begin{APACrefauthors}%
Smoluchowski, R.%
\end{APACrefauthors}%
\unskip\
\newblock
\APACrefYearMonthDay{1967}{}{}.
\newblock
{\BBOQ}\APACrefatitle {Internal structure and energy emission of Jupiter}
  {Internal structure and energy emission of jupiter}.{\BBCQ}
\newblock
\APACjournalVolNumPages{Nature}{215}{5102}{691--695}.
\PrintBackRefs{\CurrentBib}

\bibitem [\protect \citeauthoryear {%
Spencer%
\ \protect \BOthers {.}}{%
Spencer%
\ \protect \BOthers {.}}{%
{\protect \APACyear {2005}}%
}]{%
Spencer2005}
\APACinsertmetastar {%
Spencer2005}%
\begin{APACrefauthors}%
Spencer, J\BPBI R.%
, Lellouch, E.%
, Richter, M\BPBI J.%
, López-Valverde, M\BPBI A.%
, Jessup, K\BPBI L.%
, Greathouse, T\BPBI K.%
\BCBL {}\ \BBA {} Flaud, J\BHBI M.%
\end{APACrefauthors}%
\unskip\
\newblock
\APACrefYearMonthDay{2005}{}{}.
\newblock
{\BBOQ}\APACrefatitle {Mid-infrared detection of large longitudinal asymmetries
  in $\text{Io's}$ SO$_{2}$ atmosphere} {Mid-infrared detection of large
  longitudinal asymmetries in $\text{Io's}$ so$_{2}$ atmosphere}.{\BBCQ}
\newblock
\APACjournalVolNumPages{Icarus}{176}{2}{283-304}.
\newblock
\begin{APACrefDOI} \doi{https://doi.org/10.1016/j.icarus.2005.01.019}
  \end{APACrefDOI}
\PrintBackRefs{\CurrentBib}

\bibitem [\protect \citeauthoryear {%
Steinke%
, van Sliedregt%
, Vilella%
, van~der Wal%
\BCBL {}\ \BBA {} Vermeersen%
}{%
Steinke%
\ \protect \BOthers {.}}{%
{\protect \APACyear {2020}}%
}]{%
Steinke2020}
\APACinsertmetastar {%
Steinke2020}%
\begin{APACrefauthors}%
Steinke, T.%
, van Sliedregt, D.%
, Vilella, K.%
, van~der Wal, W.%
\BCBL {}\ \BBA {} Vermeersen, B.%
\end{APACrefauthors}%
\unskip\
\newblock
\APACrefYearMonthDay{2020}{}{}.
\newblock
{\BBOQ}\APACrefatitle {Can a Combination of Convective and Magmatic Heat
  Transport in the Mantle Explain $\text{Io's}$ Volcanic Pattern?} {Can a
  combination of convective and magmatic heat transport in the mantle explain
  $\text{Io's}$ volcanic pattern?}{\BBCQ}
\newblock
\APACjournalVolNumPages{Journal of Geophysical Research: Planets}{125}{12}{}.
\newblock
\begin{APACrefDOI} \doi{https://doi.org/10.1029/2020JE006521} \end{APACrefDOI}
\PrintBackRefs{\CurrentBib}

\bibitem [\protect \citeauthoryear {%
Strobel%
\ \BBA {} Wolven%
}{%
Strobel%
\ \BBA {} Wolven%
}{%
{\protect \APACyear {2001}}%
}]{%
StrobelWolven2001}
\APACinsertmetastar {%
StrobelWolven2001}%
\begin{APACrefauthors}%
Strobel, D\BPBI F.%
\BCBT {}\ \BBA {} Wolven, B\BPBI C.%
\end{APACrefauthors}%
\unskip\
\newblock
\APACrefYearMonthDay{2001}{}{}.
\newblock
{\BBOQ}\APACrefatitle {The atmosphere of $\text{Io}$: Abundances and sources of
  sulfur dioxide and atomic hydrogen} {The atmosphere of $\text{Io}$:
  Abundances and sources of sulfur dioxide and atomic hydrogen}.{\BBCQ}
\newblock
\APACjournalVolNumPages{Astrophysics and Space Science}{277}{}{271--287}.
\newblock
\begin{APACrefDOI} \doi{https://doi.org/10.1023/A:1012261209678}
  \end{APACrefDOI}
\PrintBackRefs{\CurrentBib}

\bibitem [\protect \citeauthoryear {%
Thelen%
\ \protect \BOthers {.}}{%
Thelen%
\ \protect \BOthers {.}}{%
{\protect \APACyear {2024}}%
}]{%
thelen_2024}
\APACinsertmetastar {%
thelen_2024}%
\begin{APACrefauthors}%
Thelen, A\BPBI E.%
, de Kleer, K.%
, Cordiner, M\BPBI A.%
, de Pater, I.%
, Moullet, A.%
\BCBL {}\ \BBA {} Luszcz-Cook, S.%
\end{APACrefauthors}%
\unskip\
\newblock
\APACrefYearMonthDay{2024}{}{}.
\newblock
{\BBOQ}\APACrefatitle {Io’s SO2 and NaCl Wind Fields from ALMA} {Io’s so2
  and nacl wind fields from alma}.{\BBCQ}
\newblock
\APACjournalVolNumPages{The Astrophysical Journal Letters}{978}{1}{L1}.
\PrintBackRefs{\CurrentBib}

\bibitem [\protect \citeauthoryear {%
Tsang%
, Spencer%
\BCBL {}\ \BBA {} Jessup%
}{%
Tsang%
\ \protect \BOthers {.}}{%
{\protect \APACyear {2015}}%
}]{%
tsang_2015}
\APACinsertmetastar {%
tsang_2015}%
\begin{APACrefauthors}%
Tsang, C\BPBI C.%
, Spencer, J\BPBI R.%
\BCBL {}\ \BBA {} Jessup, K\BPBI L.%
\end{APACrefauthors}%
\unskip\
\newblock
\APACrefYearMonthDay{2015}{}{}.
\newblock
{\BBOQ}\APACrefatitle {Non-detection of post-eclipse changes in Io’s
  Jupiter-facing atmosphere: Evidence for volcanic support?} {Non-detection of
  post-eclipse changes in io’s jupiter-facing atmosphere: Evidence for
  volcanic support?}{\BBCQ}
\newblock
\APACjournalVolNumPages{Icarus}{248}{}{243--253}.
\PrintBackRefs{\CurrentBib}

\bibitem [\protect \citeauthoryear {%
{Tsang}%
\ \protect \BOthers {.}}{%
{Tsang}%
\ \protect \BOthers {.}}{%
{\protect \APACyear {2012}}%
}]{%
Tsang_2012}
\APACinsertmetastar {%
Tsang_2012}%
\begin{APACrefauthors}%
{Tsang}, C\BPBI C\BPBI C.%
, {Spencer}, J\BPBI R.%
, {Lellouch}, E.%
, {L{\'o}pez-Valverde}, M\BPBI A.%
, {Richter}, M\BPBI J.%
\BCBL {}\ \BBA {} {Greathouse}, T\BPBI K.%
\end{APACrefauthors}%
\unskip\
\newblock
\APACrefYearMonthDay{2012}{}{}.
\newblock
{\BBOQ}\APACrefatitle {{Io's atmosphere: Constraints on sublimation support
  from density variations on seasonal timescales using NASA IRTF/TEXES
  observations from 2001 to 2010}} {{Io's atmosphere: Constraints on
  sublimation support from density variations on seasonal timescales using NASA
  IRTF/TEXES observations from 2001 to 2010}}.{\BBCQ}
\newblock
\APACjournalVolNumPages{Icarus}{217}{1}{277-296}.
\newblock
\begin{APACrefDOI} \doi{10.1016/j.icarus.2011.11.005} \end{APACrefDOI}
\PrintBackRefs{\CurrentBib}

\bibitem [\protect \citeauthoryear {%
{Tsang}%
\ \protect \BOthers {.}}{%
{Tsang}%
\ \protect \BOthers {.}}{%
{\protect \APACyear {2013}}%
}]{%
Tsang2013}
\APACinsertmetastar {%
Tsang2013}%
\begin{APACrefauthors}%
{Tsang}, C\BPBI C\BPBI C.%
, {Spencer}, J\BPBI R.%
, {Lellouch}, E.%
, {L{\'o}pez-Valverde}, M\BPBI A.%
, {Richter}, M\BPBI J.%
, {Greathouse}, T\BPBI K.%
\BCBL {}\ \BBA {} {Roe}, H.%
\end{APACrefauthors}%
\unskip\
\newblock
\APACrefYearMonthDay{2013}{}{}.
\newblock
{\BBOQ}\APACrefatitle {{Io{\textquoteright}s contracting atmosphere post 2011
  perihelion: Further evidence for partial sublimation support on the
  anti-Jupiter hemisphere}} {{Io{\textquoteright}s contracting atmosphere post
  2011 perihelion: Further evidence for partial sublimation support on the
  anti-Jupiter hemisphere}}.{\BBCQ}
\newblock
\APACjournalVolNumPages{Icarus}{226}{1}{1177-1181}.
\newblock
\begin{APACrefDOI} \doi{10.1016/j.icarus.2013.06.032} \end{APACrefDOI}
\PrintBackRefs{\CurrentBib}

\bibitem [\protect \citeauthoryear {%
Tsang%
, Spencer%
, Lellouch%
, Lopez-Valverde%
\BCBL {}\ \BBA {} Richter%
}{%
Tsang%
\ \protect \BOthers {.}}{%
{\protect \APACyear {2016}}%
}]{%
Tsang2016}
\APACinsertmetastar {%
Tsang2016}%
\begin{APACrefauthors}%
Tsang, C\BPBI C\BPBI C.%
, Spencer, J\BPBI R.%
, Lellouch, E.%
, Lopez-Valverde, M\BPBI A.%
\BCBL {}\ \BBA {} Richter, M\BPBI J.%
\end{APACrefauthors}%
\unskip\
\newblock
\APACrefYearMonthDay{2016}{}{}.
\newblock
{\BBOQ}\APACrefatitle {The collapse of $\text{Io's}$ primary atmosphere in
  Jupiter eclipse} {The collapse of $\text{Io's}$ primary atmosphere in jupiter
  eclipse}.{\BBCQ}
\newblock
\APACjournalVolNumPages{Journal of Geophysical Research:
  Planets}{121}{8}{1400-1410}.
\newblock
\begin{APACrefDOI} \doi{https://doi.org/10.1002/2016JE005025} \end{APACrefDOI}
\PrintBackRefs{\CurrentBib}

\bibitem [\protect \citeauthoryear {%
Tyler%
, Henning%
\BCBL {}\ \BBA {} Hamilton%
}{%
Tyler%
\ \protect \BOthers {.}}{%
{\protect \APACyear {2015}}%
}]{%
Tyler_2015}
\APACinsertmetastar {%
Tyler_2015}%
\begin{APACrefauthors}%
Tyler, R\BPBI H.%
, Henning, W\BPBI G.%
\BCBL {}\ \BBA {} Hamilton, C\BPBI W.%
\end{APACrefauthors}%
\unskip\
\newblock
\APACrefYearMonthDay{2015}{}{}.
\newblock
{\BBOQ}\APACrefatitle {TIDAL HEATING IN A MAGMA OCEAN WITHIN JUPITER’S MOON
  $\text{Io}$} {Tidal heating in a magma ocean within jupiter’s moon
  $\text{Io}$}.{\BBCQ}
\newblock
\APACjournalVolNumPages{The Astrophysical Journal Supplement
  Series}{218}{2}{22}.
\newblock
\begin{APACrefDOI} \doi{10.1088/0067-0049/218/2/22} \end{APACrefDOI}
\PrintBackRefs{\CurrentBib}

\bibitem [\protect \citeauthoryear {%
Veeder%
\ \protect \BOthers {.}}{%
Veeder%
\ \protect \BOthers {.}}{%
{\protect \APACyear {2015}}%
}]{%
Veeder_2015}
\APACinsertmetastar {%
Veeder_2015}%
\begin{APACrefauthors}%
Veeder, G\BPBI J.%
, Davies, A\BPBI G.%
, Matson, D\BPBI L.%
, Johnson, T\BPBI V.%
, Williams, D\BPBI A.%
\BCBL {}\ \BBA {} Radebaugh, J.%
\end{APACrefauthors}%
\unskip\
\newblock
\APACrefYearMonthDay{2015}{}{}.
\newblock
{\BBOQ}\APACrefatitle {Io: Heat flow from small volcanic features} {Io: Heat
  flow from small volcanic features}.{\BBCQ}
\newblock
\APACjournalVolNumPages{Icarus}{245}{}{379-410}.
\newblock
\begin{APACrefDOI} \doi{https://doi.org/10.1016/j.icarus.2014.07.028}
  \end{APACrefDOI}
\PrintBackRefs{\CurrentBib}

\bibitem [\protect \citeauthoryear {%
Veeder%
, Matson%
, Johnson%
, Blaney%
\BCBL {}\ \BBA {} Goguen%
}{%
Veeder%
\ \protect \BOthers {.}}{%
{\protect \APACyear {1994}}%
}]{%
veeder_1994}
\APACinsertmetastar {%
veeder_1994}%
\begin{APACrefauthors}%
Veeder, G\BPBI J.%
, Matson, D\BPBI L.%
, Johnson, T\BPBI V.%
, Blaney, D\BPBI L.%
\BCBL {}\ \BBA {} Goguen, J\BPBI D.%
\end{APACrefauthors}%
\unskip\
\newblock
\APACrefYearMonthDay{1994}{}{}.
\newblock
{\BBOQ}\APACrefatitle {Io's heat flow from infrared radiometry: 1983--1993}
  {Io's heat flow from infrared radiometry: 1983--1993}.{\BBCQ}
\newblock
\APACjournalVolNumPages{Journal of Geophysical Research:
  Planets}{99}{E8}{17095--17162}.
\PrintBackRefs{\CurrentBib}

\bibitem [\protect \citeauthoryear {%
Veeder%
, Matson%
, Johnson%
, Davies%
\BCBL {}\ \BBA {} Blaney%
}{%
Veeder%
\ \protect \BOthers {.}}{%
{\protect \APACyear {2004}}%
}]{%
Veeder2004_polar}
\APACinsertmetastar {%
Veeder2004_polar}%
\begin{APACrefauthors}%
Veeder, G\BPBI J.%
, Matson, D\BPBI L.%
, Johnson, T\BPBI V.%
, Davies, A\BPBI G.%
\BCBL {}\ \BBA {} Blaney, D\BPBI L.%
\end{APACrefauthors}%
\unskip\
\newblock
\APACrefYearMonthDay{2004}{}{}.
\newblock
{\BBOQ}\APACrefatitle {The polar contribution to the heat flow of $\text{Io}$}
  {The polar contribution to the heat flow of $\text{Io}$}.{\BBCQ}
\newblock
\APACjournalVolNumPages{Icarus}{169}{1}{264-270}.
\newblock
\begin{APACrefDOI} \doi{https://doi.org/10.1016/j.icarus.2003.11.016}
  \end{APACrefDOI}
\PrintBackRefs{\CurrentBib}

\bibitem [\protect \citeauthoryear {%
Wagman%
}{%
Wagman%
}{%
{\protect \APACyear {1979}}%
}]{%
Wagman_1979}
\APACinsertmetastar {%
Wagman_1979}%
\begin{APACrefauthors}%
Wagman, D.%
\end{APACrefauthors}%
\unskip\
\newblock
\APACrefYearMonthDay{1979}{}{}.
\newblock
{\BBOQ}\APACrefatitle {Sublimation pressure and enthalpy of sublimation of
  $\text{SO}_2$, Chem} {Sublimation pressure and enthalpy of sublimation of
  $\text{SO}_2$, chem}.{\BBCQ}
\newblock
\APACjournalVolNumPages{Thermodyn. Data Cent. Rep., Natl. Bur. of Standards,
  Washington, DC}{}{}{}.
\PrintBackRefs{\CurrentBib}

\bibitem [\protect \citeauthoryear {%
A.~Walker%
\ \protect \BOthers {.}}{%
A.~Walker%
\ \protect \BOthers {.}}{%
{\protect \APACyear {2010}}%
}]{%
Walker2010}
\APACinsertmetastar {%
Walker2010}%
\begin{APACrefauthors}%
Walker, A.%
, Gratiy, S.%
, Goldstein, D.%
, Moore, C.%
, Varghese, P.%
, Trafton, L.%
\BDBL {}Stewart, B.%
\end{APACrefauthors}%
\unskip\
\newblock
\APACrefYearMonthDay{2010}{}{}.
\newblock
{\BBOQ}\APACrefatitle {A comprehensive numerical simulation of $\text{Io’s}$
  sublimation-driven atmosphere} {A comprehensive numerical simulation of
  $\text{Io’s}$ sublimation-driven atmosphere}.{\BBCQ}
\newblock
\APACjournalVolNumPages{Icarus}{207}{}{409-432}.
\newblock
\begin{APACrefDOI} \doi{https://doi.org/10.1016/j.icarus.2010.01.012}
  \end{APACrefDOI}
\PrintBackRefs{\CurrentBib}

\bibitem [\protect \citeauthoryear {%
A\BPBI C.~Walker%
, Moore%
, Goldstein%
, Varghese%
\BCBL {}\ \BBA {} Trafton%
}{%
A\BPBI C.~Walker%
\ \protect \BOthers {.}}{%
{\protect \APACyear {2012}}%
}]{%
Walker_2012}
\APACinsertmetastar {%
Walker_2012}%
\begin{APACrefauthors}%
Walker, A\BPBI C.%
, Moore, C\BPBI H.%
, Goldstein, D\BPBI B.%
, Varghese, P\BPBI L.%
\BCBL {}\ \BBA {} Trafton, L\BPBI M.%
\end{APACrefauthors}%
\unskip\
\newblock
\APACrefYearMonthDay{2012}{}{}.
\newblock
{\BBOQ}\APACrefatitle {A parametric study of $\text{Io’s}$ thermophysical
  surface properties and subsequent numerical atmospheric simulations based on
  the best fit parameters} {A parametric study of $\text{Io’s}$
  thermophysical surface properties and subsequent numerical atmospheric
  simulations based on the best fit parameters}.{\BBCQ}
\newblock
\APACjournalVolNumPages{Icarus}{220}{1}{225--253}.
\newblock
\begin{APACrefDOI} \doi{https://doi.org/10.1016/j.icarus.2012.05.001}
  \end{APACrefDOI}
\PrintBackRefs{\CurrentBib}

\bibitem [\protect \citeauthoryear {%
{Williams}%
, {Schenk}%
\BCBL {}\ \BBA {} {Radebaugh}%
}{%
{Williams}%
\ \protect \BOthers {.}}{%
{\protect \APACyear {2023}}%
}]{%
Williams2023IoBook}
\APACinsertmetastar {%
Williams2023IoBook}%
\begin{APACrefauthors}%
{Williams}, D\BPBI A.%
, {Schenk}, P\BPBI M.%
\BCBL {}\ \BBA {} {Radebaugh}, J.%
\end{APACrefauthors}%
\unskip\
\newblock
\APACrefYearMonthDay{2023}{{\APACmonth{01}}}{}.
\newblock
{\BBOQ}\APACrefatitle {{Geology of Io}} {{Geology of Io}}.{\BBCQ}
\newblock
\BIn{} R\BPBI M\BPBI C.~{Lopes}, K.~{de Kleer}\BCBL {}\ \BBA {} J\BPBI
  T.~{Keane}\ (\BEDS), \APACrefbtitle {Io: A New View of Jupiter's Moon} {Io: A
  new view of jupiter's moon}\ (\BVOL~468, \BPG~233-290).
\newblock
\begin{APACrefDOI} \doi{10.1007/978-3-031-25670-7_8} \end{APACrefDOI}
\PrintBackRefs{\CurrentBib}

\bibitem [\protect \citeauthoryear {%
Williams%
}{%
Williams%
}{%
{\protect \APACyear {1995}}%
{\protect \APACexlab {{\protect \BCnt {1}}}}}]{%
IoFactsheet}
\APACinsertmetastar {%
IoFactsheet}%
\begin{APACrefauthors}%
Williams, D\BPBI R.%
\end{APACrefauthors}%
\unskip\
\newblock
\APACrefYearMonthDay{1995{\protect \BCnt {1}}}{}{}.
\newblock
\APACrefbtitle {Jovian Satellite Fact Sheet.} {Jovian satellite fact sheet.}
\newblock
\APAChowpublished
  {\url{https://nssdc.gsfc.nasa.gov/planetary/factsheet/joviansatfact.html}}.
\newblock
\APACrefnote{Accessed 2023-09-05}
\PrintBackRefs{\CurrentBib}

\bibitem [\protect \citeauthoryear {%
Williams%
}{%
Williams%
}{%
{\protect \APACyear {1995}}%
{\protect \APACexlab {{\protect \BCnt {2}}}}}]{%
JupiterFactsheet}
\APACinsertmetastar {%
JupiterFactsheet}%
\begin{APACrefauthors}%
Williams, D\BPBI R.%
\end{APACrefauthors}%
\unskip\
\newblock
\APACrefYearMonthDay{1995{\protect \BCnt {2}}}{}{}.
\newblock
\APACrefbtitle {Jupiter Fact Sheet.} {Jupiter fact sheet.}
\newblock
\APAChowpublished
  {\url{https://nssdc.gsfc.nasa.gov/planetary/factsheet/jupiterfact.html}}.
\newblock
\APACrefnote{Accessed 2023-09-05}
\PrintBackRefs{\CurrentBib}

\end{thebibliography}

%
%
%
%
%

\end{document}